\documentclass[11pt]{article}
\usepackage{amscd,amsmath,amsxtra,amssymb,theorem,latexsym,amsfonts,color,graphics}

\newtheorem{Thm}{Theorem}[section]
\newtheorem{Lemma}[Thm]{Lemma}
\newtheorem{Prop}[Thm]{Proposition}

\newtheorem{Rmk}[Thm]{Remark}
\newtheorem{Remark}[Thm]{Remark}

\newcommand{\R}{{\mathbb R}}

\newcommand{\C}{{\mathbb C}}

\newcommand{\N}{{\mathbb N}}

\newcommand{\Mbb}{{\mathbb M}}
\newcommand{\Gbb}{{\mathbb G}}
\newcommand{\BBB}{{\mathbb B}}

\newcommand{\Tr}{{\rm Tr \ }}

\newcommand{\diag}{{\rm diag \, }}

\newcommand{\arccosh}{{\rm arccosh \ }}

\newcommand{\Ln}{{\rm Ln \ }}
\renewcommand{\log}{{\rm ln \ }}

\newcommand{\Arg}{{\rm Arg \ }}
\newcommand{\Log}{{\rm Ln \ }}

\begin{document}

\title{ Resurgent analysis of the  Witten Laplacian \\ in one dimension -- II. 
} 
\author{Alexander GETMANENKO\\ 
\footnotesize Max Planck Institute for Mathematics in the Sciences, D-04103 Leipzig, Germany; \\
\footnotesize Kavli Institute for Physics and Mathematics of the Universe, \\
\footnotesize The University of Tokyo, 5-1-5 Kashiwanoha, Kashiwa, 277-8568, Japan\\
\footnotesize  {\tt Alexander.Getmanenko@ipmu.jp}.}
\maketitle

\begin{abstract} The Witten Laplacian in one dimension is studied further by methods of resurgent analysis in order to approach Fukaya's conjectures relating WKB asymptotics and disc instantons. We carry out explicit computations of exponential asymptotic expansions of exponentially small (i.e. $O(e^{-c/h})$, $c>0$, $h\to 0+$) eigenvalues and of corresponding eigenfunctions of the Witten Laplacian; a general algorithm as well as two examples are discussed. \end{abstract}

\section{Introduction}

We are continuing the project started in ~\cite{G}, where we proposed to study the Witten Laplacian by methods of resurgent analysis in order to prove conjectures by Fukaya ~\cite[\S 5.2]{Fuk} relating WKB asymptotics and disc instantons. The reader is referred to the introductory section of ~\cite{G} for philosophy and motivation, as well as for a brief review of resurgent analysis. 

The present paper is purely computational; it relies on techniques of resurgent analysis and the theory of exponential asymptotics whose rigorous mathematical  justification is a subject of current research. 

{\bf Resurgent functions and exponential asymptotics.} We will continue to use the terminology of ~\cite{G} which is a blend of ~\cite{CNP} and ~\cite{ShSt}. In particular,  by a resurgent function of the semiclassical parameter $h\to 0+$ we will mean an equivalence class of holomorphic functions $\phi(h)$ representable as a Laplace transform
$$ \phi(h) \ = \ \int_\gamma e^{-\frac{\xi}{h}} \Phi(\xi) d\xi $$
of an endlessly continuable function $\Phi(\xi)$ (``major of $\phi$") along an infinite contour $\gamma$.  See ~\cite[Sec.2]{G} for the precise definition of the equivalence relation used, the meaning of the words ``endlessly continuable", and the choice of $\gamma$. 

Under favorable conditions on singularities of $\Phi(\xi)$, a resurgent function $\phi(h)$ can be developed into an exponential asymptotic series
\begin{equation} \phi(h) \ \sim \ \sum_j e^{-\frac{c_j}{h}} (a_{0j} + a_{1j}h+ a_{2j}h^2 + ... ). \label{expasym} \end{equation}
 In the present paper we will also see resurgent  functions where the RHS of \eqref{expasym} contains series in $h$ and $h\log h$. 

If $\phi(h)$ depends on any additional parameters, then so does $\Phi(\xi)$, and the RHS may exhibit apparent discontinuities; this is known as a Stokes phenomenon, cf. ~\cite[Sec.3]{G}. 

We think of resurgent analysis as a way of making mathematical sense of exponential asymptotics and describing the Stokes phenomenon. In the present paper we will work mostly with expansions as in the RHS of \eqref{expasym} and use a description of the Stokes phenomena available from the previous literature; we will cite analytic statements without discussing their current rigor status.

{\bf Spectrum of the Witten Laplacian.} Consider a generic enough, in the sense of ~\cite{G}, real trigonometric polynomial $f(q)\in \R[\sin 2\pi q, \cos 2\pi q]$ with $n$ real local minima and $n$ real local maxima on $[0,1)$, and associate to it the following $h$-differential operator called the {\it Witten Laplacian }
\begin{equation}  P = -h^2 \partial_q^2 + (f')^2 - hf'', \ \ \ \ h\to 0+ \label{WL0}. \end{equation}
Motivated by Morse theory, we are interested in {\it exponentially small}, or {\it low-lying} eigenvalues $E(h)$ of $P$ with periodic boundary conditions, namely those eigenvalues that can be estimated $0\le E(h) \le e^{-\frac{c}{h}}$ for $h\to 0+$ and $c>0$. 

In ~\cite{G} we have shown, modulo standard black boxes in resurgent analysis, that $P$ has $n$ exponentially small resurgent eigenvalues $E_k(h)$ and that the corresponding eigenfunctions $\phi_k(q,h)$ are resurgent with respect to $h$ for $q\not\in (f')^{-1}(0)$. There we also sketched a method for calculating $E_k(h)$ which we recall in section \ref{SecOutline}. For technical reasons, in ~\cite{G} we worked with the semiclassical parameter $h$ satisfying $0<\arg h<< 1$, $|h|\to 0$. 

Because of the Stokes phenomenon, we cannot hope to write asymptotic expansions of eigenfunctions of $P$ valid for all $q\in [0,1)$; for every line segment between two consecutive critical points of $f(q)$ a different exponential asymptotic expansion is valid. Those asymptotic expansions were not discussed in ~\cite{G} and will be calculated here.

{\bf Statement of results.} This purely computational paper has two main results, both of which contribute to our computational dexterity when it comes to explicit exponential asymptotics of eigenvalues and eigenfunctions of the Witten Laplacian corresponding to a trigonometric polynomial $f(q)$  with $n$ real local minima $q_1,...,q_{2n-1}$ and $n$ real local maxima $q_2,..,q_{2n}$ on the period, where $0\ge q_1<q_2<...<q_{2n-1}<q_{2n}<1$.

%; {\it post factum} they turned out to be somewhat disconnected but we did not expect them to be when we started our work.

The first result is a computation of the connection coefficients for the equation $P\psi=hE_r\psi$, $E_r\in \C$, that matches exponential asymptotic expansions of its solutions on the intervals $(q_{j-1},q_j)$ and $(q_{j},q_{j+1})$ (called  {\it connection coefficients across the point $q_j$}) to order $O(h)$. Less precise results and a somewhat shortcut treatment of ~\cite{G} was stopping us from confidently proceeding to the calculation of periodic eigenfunctions of the Witten Laplacian, see below.

\begin{Thm} Let $q_1$ be a real local minimum and $q_2$ be a real local maximum of $f(q)$. Then: \\
a) with respect to the basis of formal WKB solutions \eqref{mar4f1} the connection coefficient, defined by \eqref{mar4f2}, across the point $q_1$  is expressed by the formula \eqref{mar4f3}; \\
b) with respect to the basis of formal WKB solutions \eqref{mar4f4} the connection coefficient, defined by \eqref{mar4f5}, across the point $q_2$ is expressed by the formula \eqref{mar4f6}. 
\end{Thm}

We warn the reader again that this theorem is proven modulo standard black boxes in the complex WKB method; here we present a calculation whose complete justification is a topic for future research. 

In finite time, the formulas \eqref{mar4f3}, \eqref{mar4f6} results could have been made more precise both with respect to $h$ and $E_r$, see Sec.\ref{FiniteTime}.

The intrigue of this calculation is the conjectured analyticity of the reduced connection coefficients $(c'_1)^{red}(E,h)$, cf.\eqref{c1redDefd}, and $c_2^{red}(E,h)$, cf.\eqref{c2redDefd}, with respect to $E$. See remark \ref{SatoConj} about the current mathematical status of this analyticity statement known as the Sato's conjecture, and sec.\ref{SomeProofs} for a partial argument in the case of the Witten Laplacian. Thus, our calculation is a nice explicit example of the Sato's conjecture.

The second result consists in explicit calculation of the low-lying eigenvalues and the corresponding eigenfunction of the Witten Laplacian in two specific examples. 

A bit of notation: let ${\cal E}^{c}$ mean terms of exponential asymptotics whose exponential prefactors are $\le e^{\frac{c}{h}}$; let us write \label{hatsimN} $\phi(h)\hat \sim e^{\frac{a}{h}}$ for $a\in \R$ if the leading exponential term in the exponential asymptotic expansion of $\phi(h)$ is $e^{\frac{a}{h}}$ times an expansion in terms of $h$ and $h\ln h$.

{\it Example 1 (section \ref{SectionEx1})}. If $f(q)=\frac{1}{2\pi}\left[ \sin 2\pi q + \cos 4\pi q\right]$, then the operator \eqref{WL0} has two exponentially small eigenvalues
$$ E^{(1)} = 0 \ \ \  \text{and} \ \ \  E^{(2)} \sim e^{-\frac{9}{8\pi h}} (6\sqrt{5}h+ o(h^1)) +   e^{-\frac{18}{8\pi h}} (-\frac{27}{\pi}h\log h + O(h^1)) + {\cal E}^{-\frac{25}{8\pi }}.$$
The eigenfunction corresponding to $E^{(1)}=0$ is $e^{-\frac{f(q)}{h}}$, and the eigenfunction corresponding to $E^{(2)}$ has the following asymptotics (actually, more precise information on the asymptotics of the eigenfunction is easy to obtain from our calculations in section \ref{SectionEx1}):
\begin{equation} \begin{array}{ccc} 
q\in (0,\frac{1}{4}-\frac{1}{2\pi}\arcsin \frac{1}{4}) && (1+{\cal E}^{-\frac{9}{8\pi }})\tilde \psi_+(q,h) + ie^{\frac{9}{8\pi h}}(\frac{2}{\sqrt{5}}+O(h))\tilde \psi_-(q,h); \\
q\in (\frac{1}{4}-\frac{1}{2\pi}\arcsin\frac{1}{4},\frac{1}{2}) && (1+{\cal E}^{-\frac{9}{8\pi }})\tilde \psi_+(q,h) + ie^{-\frac{7}{8\pi h}}(\frac{2}{\sqrt{3}}+O(h))\tilde \psi_-(q,h);\\
q\in (\frac{1}{2},\frac{3}{4}+\frac{1}{2\pi}\arcsin\frac{1}{4}) && (-1+{\cal E}^{-\frac{9}{8\pi }})\tilde \psi_+(q,h) + ie^{-\frac{7}{8\pi h}}(\frac{2}{\sqrt{3}}+O(h))\tilde \psi_-(q,h); \\
q\in (\frac{3}{4}+\frac{1}{2\pi}\arcsin\frac{1}{4}, 1) && (-1+{\cal E}^{-\frac{9}{8\pi }})\tilde \psi_+(q,h) + ie^{\frac{9}{8\pi h}}(\frac{2}{\sqrt{5}}+O(h))\tilde \psi_-(q,h), 
\end{array} \label{ansEx1} \end{equation}
where the basis $\tilde\psi_{\pm}(x,h)$ of formal WKB solutions is introduced on page \pageref{tildebasis} (please keep in mind that $f_{\rm here}(q+\frac{1}{8})=f_{\rm p.\pageref{tildebasis}}(q)$). 

Post factum it turned out that we did not need $O(h)$-level contributions to the connection coefficients to obtain these formulas; perhaps, this is the case for all sufficiently generic examples of $f(q)$. Using the full precision of the formulas \eqref{mar4f3} and \eqref{mar4f6} would have given us one more term in the $(h,\ln h)$-expansions multiplying $e^{-\frac{9}{8\pi h}}$ and $e^{-\frac{18}{8\pi h}}$ in the expression for $E^{(2)}$.

{\it Example 2 (section \ref{SectionEx2}).} If $f(q)$ is a real trigonometric polynomial with two local minima $q_1,q_3$ and two local maxima $q_2,q_4$, $0<q_1<q_2<q_3<q_4<1$, such that 
$$ f(q_1)=0, \ \ \ f(q_2) =\frac{1}{2}, \ \ \ f(q_3)=\frac{b}{2}, \ \ \ f(q_4)=\frac{a}{2}, $$
where 
$$ 0< b<a<\frac{1}{2} \ \ \ \text{and} \ \ \ 2a<3b, $$
 then the operator \eqref{WL0} has two exponentially small eigenvalues
$$ E^{(1)} = 0 \ \ \  \text{and} \ \ \  E^{(2)} \hat\sim e^{\frac{b-a}{h}}.$$
The eigenfunction corresponding to $E^{(1)}=0$ is $e^{-\frac{f(q)}{h}}$, and the eigenfunction corresponding to $E^{(2)}$ has on $(q_j,q_{j+1})$, $q_5:=q_1+1$, $j=1,..,4$, the exponential asymptotic expansions $\tilde D_+^{(j)}\tilde \psi_{+}+\tilde D_-^{(j)}\tilde \psi_{-}$ where the basis $\tilde\psi_{\pm}(x,h)$ of formal WKB solutions is introduced on page \pageref{tildebasis} and approximations to $\tilde D^{(j)}_\pm$ are given by \eqref{efunEx2}.

In remarks \ref{ResurgImp1}, \ref{ResurgImp2}  we put our finger on the specific algebraic reason why methods of complex WKB resurgent analysis are essential for such a calculation and why they look more powerful than $C^\infty$ methods of, e.g., ~\cite{HeKlNi}. 

In conclusion, the ability to perform explicit calculations developed in this paper will be needed in our future work towards Fukaya's conjecture. Remarks \ref{ResurgImp1}, \ref{ResurgImp2} and computations leading to them may be of independent pedagogical interest.

The {\bf structure of the paper} follows the outline given in section \ref{Method}. 

%{\bf Structure of the paper.} 

%The structure of this paper, that is a continuation of ~\cite{G} and uses its material freely, is as follows. In the section \ref{FormalWKBSolus} we recall the notation and calculate various monodromies of formal solution of the Witten Laplacian \eqref{WLcsb}. In the section \ref{ConnDouble} we perform a more precise calculation of the connection coefficients and of the tunneling cycle monodromies $\tau_j$ than we did in ~\cite{G}. A general procedure of calculating asymptotic expansions of eigenfunctions is recalled in section  \ref{procedure} and applied to two examples in sections \ref{SectionEx1} and \ref{SectionEx2}. In addition, for the example of \ref{SectionEx1}, we have calculated the first subdominant exponential in the hyperasymptotic expansion of the nonzero low-lying eigenvalue. The paper concludes with an appendix containing a list of elementary formulae used in this text.
 
% \newpage

\section{Method for calculation of the spectrum of the Witten Laplacian.} \label{Method}

\subsection{Eigenvalues} \label{SecOutline}

We will now review a method to calculate exponentially small eigenvalues of the Witten Laplacian proposed in ~\cite{G}. 

The semiclassical parameter $h$ is assumed to satisfy $0<\arg h<< 1$, $|h|\to 0$.

Let $f(q)$ be a real polynomial in $\sin 2\pi q$ and $\cos 2\pi q$,  with $n$ real local minima $q_1,...,q_{2n-1}$ and $n$ real local maxima $q_2,..,q_{2n}$ on the period, where $0<q_1<q_2<...<q_{2n-1}<q_{2n}<1$. We will assume that $f'(0)\ne 0$. 

\vskip2.3pc

{\it Step 1. Formal solutions of the Witten Laplacian.} Let $E$ be a complex number, $|E|$ sufficiently small. We choose two formal WKB solutions 
\begin{equation} \psi_{\pm}(E,q,h) = e^{\pm \frac{S(E,q)}{h}} (a^{\pm}_0(E,q) + a^{\pm}_1(E,q) h + a^{\pm}_2(E,q)h^2 + ... ) \label{psipmdefd} \end{equation}
of the differential equation 
\begin{equation} P\psi(E,q,h) \ = \ E\psi(E,q,h). \label{PpsiEpsi} \end{equation}
The words {\it ``formal WKB solution"} mean that $\psi_{\pm}$ satisfy \eqref{PpsiEpsi} in the sense of formal power series in $h$. Dependence on $E$ will be often suppressed in our notation. Actually, different choices of the formal solutions will be convenient in different chapters, so the notation $\psi_{\pm}(E,q,h)$ will not be kept beyond this Sec.\ref{Method}.

The equation \eqref{PpsiEpsi} is a stationary Schr\"odinger equation with the potential $[f'(q)]^2 -E -hf''(q)$. 
In the standard terminology of the WKB analysis, points $q$ for which the $h$-independent part $[f'(q)]^2-E$ of the potential vanishes are called {\it turning points} of the equation \eqref{PpsiEpsi}. They are called {\it simple} or {\it double} turning points according to the multiplicity of the zero of the function $[f'(q)]^2-E$. Higher-order turning points are also studied in resurgent analysis but they will not be relevant for our discussion.

When $E=0$, the equation \eqref{PpsiEpsi} has $2n$ double turning points on $[0,1)$ located at the critical points of $f(q)$. When we deform $E$ away from zero and make $0<|E|<<1$, each of these double turning points gives rise to a pair of simple turning points. This is the origin of several delicate features of our analysis.  
In particular, the ingredients $S(E,q)$, $a^{\pm}_j(E,q)$ on the RHS of \eqref{psipmdefd} are ramified analytic functions of $E,q$. 

Section \ref{FormalWKBSolus} is devoted to the explicit form of the formal solutions \ref{psipmdefd} and to their properties as multivalued functions of $E$ and $q$.

\vskip2.3pc

{\it Step 2. Connection problem; limit $E=hE_r$.} This next step crucially requires the study of the equation  \eqref{PpsiEpsi} for complex values of $E$ and $q$, but its outcome can be explained in term of $q$ confined to the real axis and $E\ge 0$.

The critical points $0<q_1<...<q_{2n}<1$ of $f(q)$ are double turning points of the equation \eqref{PpsiEpsi} for $E=0$. For $0<E<<1$, they split into pairs of simple turning points $0<q_1^-(E)<q_1^+(E)<q_2^-(E) <... <q_{2n}^+(E)<1$. 

The content of the Stokes phenomenon is that actual (as opposed to formal) resurgent solutions $\psi(E,q,h)$ of \eqref{PpsiEpsi} have different asymptotic expansions in terms of the formal WKB solutions $\psi_{\pm}(E,q,h)$ on different intervals between the turning points; see ~\cite[\S 3]{G} for definitions. E.g.,
$$ \psi(E,q,h) \sim A_+(E,h) \psi_+(E,q,h) + A_-(E,h) \psi_-(E,q,h) \ \ \text{for} \ q_{j-1}^{-}<q<q_j^{-}(E), $$
and
$$ \psi(E,q,h) \sim B_+(E,h) \psi_+(E,q,h) + B_-(E,h) \psi_-(E,q,h) \ \ \text{for} \ q_{j}^{+}(E)<q<q_{j+1}^{-}(E);  $$
here we let $q_{0}^{+}(E)=q_{2n}^{+}(E)-1$ and $q_{2n+1}^{-}(E)=q_{1}^{-}(E)+1$.

In ~\cite{G} we conceptualize coefficients $A_\pm(E),B_\pm(E)$ as resurgent symbols, see ~\cite[\S 2.2]{G}. For $0<|E|<<1$, they correspond to exponential asymptotics 
as on the RHS of \eqref{expasym}. 

The connection problem consists in finding a $2\times 2$ {\it connection matrix} $C^{(q_j)}(E,h)$ of $q$-independent resurgent symbols such that 
$$ \left( \begin{array}{c} B_+(E,h) \\ B_-(E,h) \end{array} \right) = C^{(q_j)}(E,h) \left( \begin{array}{c} A_+(E,h) \\ A_-(E,h) \end{array} \right). $$ 

The delicate and, to our knowledge, only partially understood on the rigorous level step is to replace a number $E$ by an expression $hE_r$, $E_r\in \C$; here we follow the notation of ~\cite{DDP97}, ~\cite{DP99}, where ``r" stands for ``reduced".
The $h$-independent part of the potential $(f')^2-hf''-hE_r$ is now $(f')^2$, and so the equation 
\begin{equation} P\psi(q,h)=hE_r\psi(q,h), \ \ \ \ E_r\in \C \label{WLEr} \end{equation}
 has double turning points at $q_j$, $j=1,...,2n$. A careful calculation of the limit yields connection matrices $C^{(q_j)}(hE_r, h)$.

We discussed this calculation in ~\cite[\S 7]{G}, but now we would like to have more accurate results -- more terms in the asymptotic expansions. This is accomplished in section \ref{secE2hEr}.

\vskip2.3pc

{\it Step 3. Quantization condition.} We will now write down conditions on the complex number $E_r$ for the equation \eqref{WLEr} to admit a periodic resurgent function solution. This requirement is equivalent to the existence of coefficients (resurgent symbols) $A_+(h),A_-(h)$, not simultaneously zero, satisfying
$$ (\psi_+(hE_r,q+1,h), \psi_-(hE_r,q+1,h) ) C^{(q_{2n})}(hE_r,h) .... C^{(q_1)}(hE_r,h) \left( \begin{array}{c} A_+(h) \\ A_-(h) \end{array} \right) \ = \ \ \ \ $$ \begin{equation} \ \ \ \  \ = \ (\psi_+(hE_r,q,h), \psi_-(hE_r,q,h) ) \left( \begin{array}{c} A_+(h) \\ A_-(h) \end{array} \right). \label{periodicity} \end{equation}
We call the $2\times 2$ matrix 
\begin{equation} F(E_r,h) = \diag\left(\frac{\psi_+(hE_r,q+1,h)}{\psi_+(hE_r,q,h)}, \frac{\psi_-(hE_r,q+1,h)}{\psi_-(hE_r,q,h)} \right)C^{(q_{2n})}(hE_r,h) .... C^{(q_1)}(hE_r,h) \label{transferMtxDefd} \end{equation}
the {\it transfer matrix}.

A condition on $E_r$ that there exists a nonzero pair $(A_+(h), A_-(h))$ satisfying \eqref{periodicity} is called {\it the quantization condition}; it can be written in the form
\begin{equation} \det(F(E_r,h)-Id)\ = \ 0. \label{detZero} \end{equation}

The algebraic structure of the quantization condition was made explicit in ~\cite[\S 8]{G} using the specific form of the entries in $C^{(q_j)}(hE_r,h)$. The quantization condition can be written in term of the following $4n+1$ quantities:\\ 
i) $\mu_j(hE_r,h)$, $j=1,...,2n$ -- monodromies of one of the formal solutions $\psi_{\pm}(hE_r,q,h)$ around the turning points $q_j$; \\
ii) $\tau_j(hE_r,h)$ for $j=1,...,2n$ -- limits of monodromies of one of $\psi_{\pm}(hE_r,q,h)$ along the ``tunneling cycles" connecting $q_j$ and $q_{j+1}$; \\ 
iii) $1+\kappa(E_r,h)E_r \ = \ \psi_{\pm}(hE_r,q+1,h)/\psi_{\pm}(hE_r,q,h)$ for one of the choices $\psi_+$ or $\psi_-$, cf.\eqref{coeffk}.  \\
 We will recall precise statements in sec.\ref{QCPs}, cf.\eqref{QCondMtxForm}.

\vskip2.3pc

{\it Step 4. Solving quantization conditions by means of a Newton polygon.} The equation \eqref{detZero} is satisfied for $E_r=0$, but otherwise the requirement that  $E_r$ should be an $h$-independent complex number is too restrictive. We will now solve the quantization condition \eqref{detZero} allowing $E_r$ to be any exponentially small resurgent function, $E_r=E_r(h)=O(e^{-\frac{c}{h}})$, $c>0$.

As in ~\cite[\S 9]{G}, we write the quantization condition in the form
\begin{equation} {\cal F}(E_r,h) := \sum_{k,c} E_r^k e^{\frac{c}{h}} a_{k,c}(h) \ = \ 0, \label{NPGenForm} \end{equation}
where $k$ ranges over positive integers and $c$ over a subset of $\R$ bounded from above and without accumulation points.

\begin{figure} \includegraphics{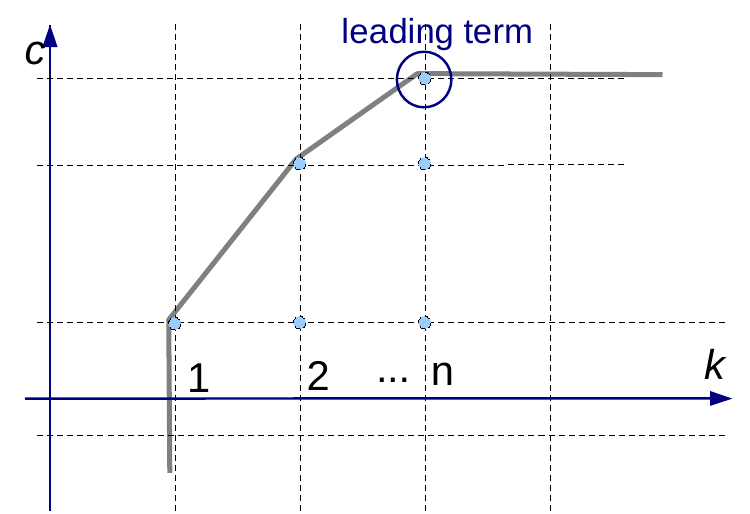} \caption{An example of a Newton polygon representing the quantization condition} \label{Paper4p1}
\end{figure}

We plot the points $(k,c)$ corresponding to nonzero $a_{k,c}(h)$ on a coordinate plane, and consider the Newton polygon ${\cal N}$ which is the convex hull of
$$ \bigcup_{(k,c): a_{k,c}(h)\ne 0} [k;+\infty) \times (-\infty;c] , $$
fig.\ref{Paper4p1}. 

By the {\it leading term} of ${\cal N}$ we will mean the term $E_r^k e^{\frac{c}{h}} a_{k,c}(h)$ corresponding to the leftmost vertex of the horizontal edge of ${\cal N}$; such a term necessarily has degree $n$ in $E_r$, ~\cite[\S 10]{G}. In ~\cite[\S 9]{G} we have written up an iterative procedure for finding exponentially small solutions of ${\cal F}(E_r(h),h)=0$. In a nondegenerate case there will be $n$ such solutions $E_r^{(1)}=0, E_r^{(2)},...,E_r^{(n)}$, and for $k=2,..,n$
$$ E_r^{(k)} \sim e^{-t_k/h} b(h,\log h) + e.l.o.t., $$
where $t_k$ is one of the slopes of an edge of ${\cal N}$, $b(h,\log h)$ is an expansion in terms of $h$ and $\log h$, and e.l.o.t. means terms of exponentially lower order.

In sec.\ref{SolvingQCex1} we will perform this calculation in details for a specific example.

\vskip2.3pc

We finally note that substituting $E=hE_r(h)$ into the original equation \eqref{detZero} and into the resurgent solutions $\psi(E,q,h)$ gives resurgent eigenfunctions of $P$ corresponding to this resurgent eigenvalues. Rigorous results partially justifying this fact are given in ~\cite{G12}.

\subsection{Eigenfunctions}

Eigenfunctions of the Witten Laplacian corresponding to an eigenvalue $hE_r(h)$ can be calculated as follows.

If $E_r(h)$ satisfies \eqref{detZero}, then the kernel of the matrix $F(E_r(h),h)-Id$, where $F$ is the transfer matrix \eqref{transferMtxDefd}, is nontrivial. Since this is a $2\times 2$ matrix, an element of its kernel can be read from the coefficients of the matrix.

Clearly, $A_+(h)\psi_+(hE_r(h),q,h) + A_-(h)\psi_-(hE_r(h),q,h)$ will be an asymptotic representation of the eigenfunction for $0<q<q_1$. 

Using the connection formulas to compare asymptotic representations of eigenfunctions on different intervals $(q_j,q_{j+1})$, we conclude that on an interval $(q_j,q_{j+1})$ the asymptotic representation of the same eigenfunction is given by
$$ (\psi_+(hE_r,q,h), \psi_-(hE_r,q,h) ) C^{(q_{j})}(hE_r,h) .... C^{(q_1)}(hE_r,h) \left( \begin{array}{c} A_+(h) \\ A_-(h) \end{array} \right). $$ 

We present results of this kind of computation in two examples, sections \ref{eigenfunEx1} and \ref{SectionEx2}. 

\begin{Remark} {\rm In order to obtain nontrivial results on all intervals $(q_j,q_{j+1})$, we have to work with exponential asymptotic expansions of $A_{\pm}(h)$ {\it beyond the leading exponential order} which in turn depend on the subleading exponential terms in $E_r(h)$. This suggests that an analogous computation would be hard to perform by $C^\infty$ methods, i.e. without complex WKB.}
\end{Remark}

% \newpage 

\section{Notation and formal WKB solutions.} \label{FormalWKBSolus}

\subsection{Notation, cuts, signs, and branches.} \label{CutsSignsBranches}

For the purposes of calculation performed in this paper, it is enough to limit our considerations to a neighborhood of the real axis in the $q$-plane.

Let us recall the notation of ~\cite{G}.  Let $f(q)\in \R[\sin(2\pi q),\cos(2\pi q)]$ be a real trigonometric polynomial,  with $n$ real local minima $q_1,...,q_{2n-1}$ and $n$ real local maxima $q_2,..,q_{2n}$ on the period $[0,1)$, where $0<q_1<q_2<...<q_{2n-1}<q_{2n}<1$. We require $f''(q_j)\ne 0$. 

In this section \ref{FormalWKBSolus} we will discuss  formal WKB solutions of 
\begin{equation} P\psi \ := \ \left[ -h^2 \partial^2_q + (f')^2 - hf''\right]\psi \ = \ E \psi, \label{WLcsb} \end{equation}
where $E$ is a complex number and $\psi=\psi(E,q,h)$.

For $E\ne 0$ and $|E|$ sufficiently small, the classical momentum $p(q)=\sqrt{E-(f'(q))^2}$ is defined on a two sheeted cover of the complex plane of $q$. For $E=0$, the two determinations of $p(q)$ are $\pm f'(q)$, and one can think of the Riemann surface of $p(q)$ as of two separate sheets having contact at points $q_j$ where $f'(q_j)=0$.

Our formulas will be written for $E>0$; analytic continuation to other values of $E$ will be implicit.

For every fixed $E$, the ramification points of $p(q)$ coincide with solutions of the equation $E=(f'(q))^2$ which we call {\it turning points} of the equation $P\psi(q,h)=E\psi(q,h)$. If $E=0$, we speak of {\it double turning points} because $(f'(q))^2$ has double zeros at the critical points $q_j$, $1\le j\le 2n$, of $f(q)$. If $0<E<<1$, the double turning points $q_j$ split into pairs of simple turning points $q_j^-(E)<q_j<q_j^+(E)$.

\begin{figure}[h]\includegraphics{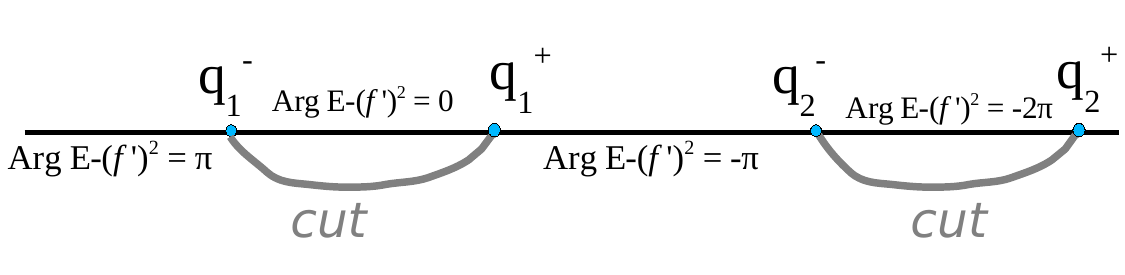} \caption{Choice of $\Arg E-(f')^2$ on the first sheet of the Riemann surface of $p(q)$ for a fixed $E$, $0<E<<1$.} \label{nuthesisp5}
\end{figure}

 The Riemann surface of $p(q)$ can be described as the plane with cut connecting $q_j^-$ to $q_j^+$ and going a little below the real axis. To specify the determination of $p(q)$ on the first sheet, we define $\Arg (E-(f')^2)$ for real values of $q$ on figure \ref{nuthesisp5}. As $E\to 0$, on the first sheet $i p(q,E)\to f'(q)$.

On our pictures we will draw contours (or their parts) on the first sheet of $p(q)$ as solid curves, and contours on the second sheet of $p(q)$ as dashed curves.

\subsection{Formal solutions of $P\psi=E\psi$.} \label{FormalSolSec}

In order to find a formal WKB solution of (\ref{WLcsb}), we will be looking for a series 
\begin{equation} y(q,h) = y_0(q) + hy_1(q) + h^2 y_2(q) + ... \label{yjdefd} \end{equation}
solving the equation
\begin{equation} (P-E)\left\{ \exp \left\{ \int^q \frac{i}{h}\sqrt{E-(f'(q))^2} + y(q',h)dq'\right\} \right\} = 0 \label{mar4eq7} \end{equation}
in the sense of formal power series. The equivalent condition on $y(q,h)$ is given by the Riccati equation
$$  2y(q) i\sqrt{E-(f')^2} = - h y(q,h)^2 +  
i\frac{f'f''}{\sqrt{E-(f')^2}} - h y'(q)   - f''  $$
which yields a recursive procedure for calculating $y_j$. In particular,
$$  y_0(q) =  \frac{f'f''}{2(E-(f')^2)}    - \frac{f'' }{2i\sqrt{E-(f')^2}}, $$ 
\small 
$$ y_1(q) \ = \  - \frac{5(f')^2(f'')^2}{8i (E-(f')^2)^{5/2}}  - \frac{f'(f'')^2}{2(E-(f')^2)^2}  
-  \frac{(f'')^2 }{8i(E-(f')^2)^{3/2}}  -   
\frac{f'f^{(3)}}{4i(E-(f')^2)^{3/2}}  - \frac{f^{(3)}}{4 (E-(f')^2)}, $$
\normalsize
etc. 

We have two choices of $p(q)=\sqrt{E-(f'(q))^2}$ corresponding to the first and the second sheet of the Riemann surface of $p(q)$ introduced in Sec.\ref{CutsSignsBranches}. Accordingly, we have a two formal solutions 
$$ \phi_{\pm}(E,q,h) = \exp \left\{ \int_{q_0}^q \frac{i}{h}\sqrt{E-(f')^2} + y(q')dq'\right\} $$
where the sign ``$+$" corresponds to the first sheet and the sign ``$-$" corresponds to the second sheet. For definiteness, we put $q_0=0$, although any other point in the interval $(q_{2n}-1,q_1)$ would work just as well. It is customary to say that $\phi_{\pm}(E,q,h)$ are {\it normalized} in such a way that $\phi_{\pm}(q_0)=1$.

\begin{figure}[h] \includegraphics{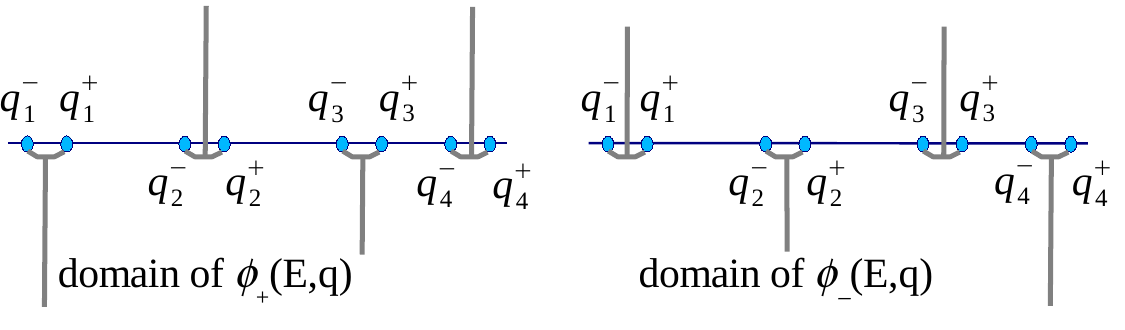} \caption{Domains with respect to the $q$ variable of $\phi_+(E,q,h)$ and $\phi_-(E,q,h)$, $0<E<<1$} \label{Paper4p2} \end{figure}

For $0<E<<1$, $\phi_+(E,q,h)$ becomes $\phi_-(E,q,h)$ times a $q$-independent factor $C_j(E,h)$, and vice versa, when we cross a cut between $q_j^-(E)$ and $q_j^+(E)$. Further, to make $\phi_{\pm}(E,q,h)$ univalued functions of $q$, let us introduce additional cuts as shown on fig.\ref{Paper4p2}.

\subsection{Formal solutions of $P\psi=hE_r\psi$.} \label{FormalhEr}

Inserting $hE_r$, $E_r\in \C$, into the formal solutions $\phi_{\pm}(E,q,h)$ and regrouping the terms accordingly to the powers of $h$ yields formal solutions $\phi_\pm(hE_r,q,h)$ of the equation 
\begin{equation} (-h^2 \partial^2_q + [f']^2 - hf'' )\phi = h E_r \phi. \label{ReducedSchroe} \end{equation}
Their domains will be taken the same as on fig.\ref{Paper4p2} except cuts between $q_j^-$ and $q_j^+$ are now shrunk to punctures. For the equation $P\psi=hE_r\psi$, we will take $p(q)=\pm f'(q)$ and speak of two sheets of the Riemann surface of $p(q)$: upper sheet with $p(q)=f'(q)$ and lower sheet with $p(q)=-f'(q)$.

%By inserting $hE_r$, $E_r\in \C$, in the power series expansions of $\psi_{\pm}(E,q,h)$ , we obtain formal WKB solutions $\psi_{\pm}(hE_r,q,h)$ of the equation $P\psi=hE_r\psi$.

%corresponding to the first and second sheet of the Riemann surface, normalized in such a way that $\psi_{+}(q_0)=\psi_{-}(q_0)=1$ and defined on the domains (complex plane with vertical cuts starting at $q_j$ ) shown on fig.\ref{RDRW2p32}.

%\begin{figure}[h] \includegraphics{RDRW2p32.pdf} \caption{Domains of $\phi_+$ and $\phi_-$} \label{RDRW2p32} \end{figure}

%In terms of $\phi_+, \phi_-$ the transfer matrix and the quantization condition will be written, in the same way as in ~\cite[section VIII]{G}.

% \newpage

\section{Monodromies of formal solutions.} \label{MonoFS}

For a fixed $E$, let $\rho(t)$, $0\le t\le 1$, be a path on the Riemann surface of $p(q)$, and let $\phi(q,h)$ be a formal WKB solution as in Sec.\ref{FormalSolSec}, but now understood as a multivalued analytic function on the Riemann surface of $p(q)$. We call $\phi(\rho(1))/\phi(\rho(0))$ the {\it formal monodromy}, or the {\it monodromy of a formal solution} along the path $\rho$, and we call $s_\rho(h)$ satisfying $\phi(\rho(1))/\phi(\rho(0))= \exp [2\pi i s_\rho]$, the {\it monodromy exponent} along the path $\rho$.

\begin{figure}[h]\includegraphics{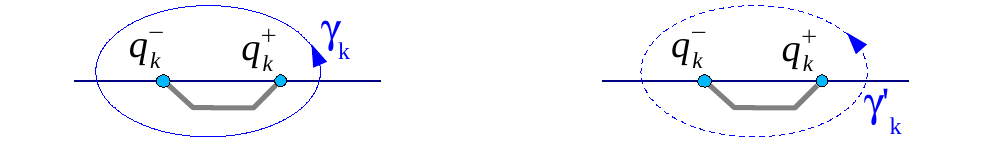}  \caption{Paths $\gamma_k$ and $\gamma'_k$.} \label{Paper4p3}
\end{figure}

\begin{figure}[h]\includegraphics{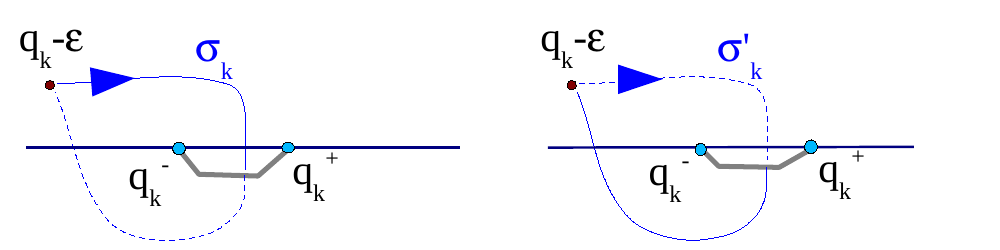}  \caption{Paths $\sigma_k$ and $\sigma'_k$.} \label{Paper4p4}
\end{figure}

Definition for $0<E<<1$ of the paths whose monodromies we would like to compute is given on fig.\ref{Paper4p3} and fig.\ref{Paper4p4}. On fig.\ref{Paper4p4}, $\varepsilon$ will be taken as a small complex number with $0\le\arg \varepsilon \le\pi$; this restriction will be useful later in order to specify which Stokes region the point $q_k-\varepsilon$ belongs to. 

Notice that $\gamma_k$ and $\gamma'_k$ also make sense when $E=0$ whereas $\sigma_k$ and $\sigma'_k$ get ``pinched" when $E\to 0$. 

\subsection{Some Taylor series} \label{TaylorSer}

In this Sec.\ref{TaylorSer} we collect various Taylor series and simple formulas among them that will be relevant for the calculations later on.

Let $q_\ell\in \R$ and $f'(q_\ell)=0$. 
%We will need to calculate various integrals along paths passing around or near the point $q_\ell$, and in this subsection we will set up the appropriate notation. 
For $q$ near $q_\ell$ the substitution  $$ u=-f'(q) $$ is one-to-one, and thus $u$ can be taken as a local coordinate near $q_\ell$. 

Let us introduce the numbers $a_j(q_\ell)$ (sometimes written as $a_j$ if the index $\ell$ is clear) by 
$$ -(q-q_\ell) = a_0(q_\ell) f'(q) + \frac{1}{2}a_1(q_\ell)  [f'(q)]^2 + \frac{1}{3} a_2(q_\ell) [f'(q)]^3+... .$$
%where 
%$$f'(q)=f''(q_1)(q-q_1)+\frac{f^{(3)}(q_1)}{2}(q-q_1)^2+\frac{f^{(4)}(q_1)}{3!}(q-q_1)^3+...$$
In particular,  
\begin{equation}  a_0 = -\frac{1}{f''(q_\ell)}, \ \   a_1 = \frac{f^{(3)}(q_\ell)}{[f''(q_\ell)]^3}, \ \  a_2 = \frac{f^{(4)}(q_\ell)}{2[f''(q_\ell)]^4} - \frac{3[f^{(3)}(q_\ell)]^2}{2[f''(q_\ell)]^5}, ... . \label{a0a1a2} \end{equation}
In general, the Lagrange inversion formula allows us to write a general expression for $a_n$. \footnote{We thank Prof.S.Garoufalidis for this remark.}
% the calculation is contained in RDRW2.Appx5 and was checked on Maple.

It follows then that 
\begin{equation}  -\frac{1}{f''(q)} = a_0(q_\ell)+a_1(q_\ell) f'(q)+a_2(q_\ell) [f'(q)]^2 +... = \sum_{j=0}^\infty a_j (-1)^j u^j \label{OneOverDDF} \end{equation}
and 
$$ f''(q)-f''(q_\ell) = f''(q_\ell) \sum_{j=1}^\infty a_j [f'(q)]^j  f''(q) \ = \ f''(q_\ell) \sum_{j=1}^\infty a_j (-1)^j u^j  f''. $$

We similarly introduce coefficients $b_j=b_j(q_\ell)$ by the requirement that 
\begin{equation} -f''(q) = b_0(q_\ell) + b_1(q_\ell) u + b_2(q_\ell) u^2+... . \label{DDfB} \end{equation}
should hold near $q_\ell$. In particular, %cf RDRW2.maple3
$$ b_0 = -f''(q_\ell); \ \ b_1=\frac{f^{(3)}(q_\ell)}{f''(q_\ell)}; \ \ b_2= \frac{[f^{(3)}(q_\ell)]^2-f''(q_\ell)f^{(4)}(q_\ell)}{2[f''(q_\ell)]^3}. $$
We obtain by differentiation 
$$ -f^{(3)}(q) = \sum_{j=0}^\infty j b_j [-f'(q)]^{j-1} [-f''(q)]. $$
% $$ \frac{f^{(3)}}{f''} = \sum j b_j u^{j-1} \ = \ \sum_{j=0}^\infty (j+1)b_{j+1} u^j.$$

Finally, for $A=-f'(q_\ell-\varepsilon)$, we have
%$$  f(q_\ell)-f(q_\ell-\varepsilon) = \int_{q_\ell-\varepsilon}^{q_\ell} f'(q)dq = $$ 
% $$ \ = \ \int_{A}^0 (-u)\frac{dq}{du} du =$$
%(Change the variable to $u=-f'$,  $\frac{du}{dq}=-f''$, so $\frac{dq}{du}=-\frac{1}{f''}=a_0-a_1u+...+(-1)^n a_n u^n+...$ where the series is convergent in some neighborhood of $u=0$)
\begin{equation} f(q_\ell)-f(q_\ell-\varepsilon) = \int_{q_\ell-\varepsilon}^{q_\ell} f'(q)dq = \int_0^A u [\sum_{j=0}^\infty (-1)^j a_j u^j]du = \sum_{j=0}^\infty (-1)^j a_j \frac{A^{j+2}}{j+2}. \label{fqthruA} \end{equation} 

\subsection{Formal monodromy along $\sigma_k$, $k$ odd.} \label{sigmaOdd}

In order to have fewer indices in the notation, we will treat the representative cases of $k=1$. 

We have:
$$ 2\pi i s_{\sigma_1} (E,h) = \frac{1}{h} (\Delta S)(E) + \sum_{j=0}^\infty h^j (\Delta y_j)(E), $$
where $(\Delta S)(E)$, $(\Delta y_j)(E)$ are analytic functions of $E$ at least for small positive $E$ defined by
\begin{equation}\Delta S (E) \ = \ i\int_{\sigma_1} \sqrt{E-(f')^2} dq',  \ \ \ \Delta y_j(E) \ = \ \int_{\sigma_1} y_j(q) dq,  \label{DeltaSDeltay} \end{equation}

As we will be taking limit for $E\to 0$, we will be interested in approximate values of $\Delta S(E)$, $\Delta y_j(E)$ for $E\to 0+$. 

\subsubsection{Calculation of the  $\Delta S$ summand in $2\pi i s_{\sigma_1}$. }

We have 
$$ \Delta S = \int_{\sigma_1} i\sqrt{E-(f'(q))^2} dq = 2\int_{q_1-\varepsilon}^{q_1^-} [-\sqrt{(f')^2-E}]dq = $$
(change of variables $u=-f'(q)$, $u=\sqrt{E} \cosh t$, $A=-f'(q_1-\varepsilon)$; $a_j=a_j(q_1)$ were defined in section \ref{TaylorSer})
$$ \ = \ -2\sum_{j=0}^\infty (-1)^j a_j \int_{A}^{\sqrt{E}}u^j \sqrt{u^2-E} du  \ = \ 
-2\sum_{j=0}^\infty (-1)^j a_j \int_{\arccosh (A/\sqrt{E})}^0 E^{\frac{j}{2}+1} \cosh^j t \sinh^2 t dt \ = \ $$
%$$ \ = \ a_0(A^2-\frac{E}{2}-\frac{E^2}{8A^2} - E\arccosh(A/\sqrt{E}))
% \ + \ a_2(\frac{A^4}{2} +\frac{A^2 E}{2} -  \frac{E^2}{16} +  \frac{E^2}{4}\arccosh(A/\sqrt{E})) + $$
%$$ \ + \ \sum_{\text{$j=1$ or $\ge 3$}} (-1)^j a_j \left( 2\frac{A^{j+2}}{j+2} - \frac{A^j E}{j} - \frac{A^{j-2}E^2}{4(j-2)} \right) + o(E^2) $$
(use \eqref{IntSinhCosh}, (\ref{fqthruA}) and (\ref{ArccoshAs}) )
$$ \ = \ 2[f(q_1)-f(q_1-\varepsilon)] + a_0(-\frac{E}{2}-\frac{E^2}{8A^2} - E(\Ln\frac{2A}{\sqrt{E}}-\frac{E}{4A^2}))
 \ + \ a_2(\frac{A^2 E}{2} -  \frac{E^2}{16} +  \frac{E^2}{4}\Ln(\frac{2A}{\sqrt{E}}) ) + $$
\begin{equation} \ + \ \sum_{\text{$j=1$ or $\ge 3$}} (-1)^j a_j \left(- \frac{A^j E}{j} - \frac{A^{j-2}E^2}{4(j-2)} \right) + o(E^2). \label{eq24feb} \end{equation}

\subsubsection{Calculation of the  $\Delta y_0$ summand in $2\pi i s_{\sigma_1}$. }

Recall that 
$$ y_0(q) \ = \  \frac{f' f''}{2(E-(f')^2)} - \frac{f''}{2i\sqrt{E-(f')^2}}. $$

\begin{Lemma} We have
$$ \Delta y_0 (E) \ = \ \int_{\sigma_1} y_0(q)dq \ = \ \arccosh \frac{(-f'(q_1-\varepsilon))}{\sqrt{E}} - \frac{\pi i}{2} , $$
where the branch of $\arccosh$ is chosen so as to coincide with the principal real value of of $\arccosh$ for $E>0$ and $q_1-\varepsilon$ on the real axis immediately to the left of $q_1^-$. 
\end{Lemma}
\textsc{Proof.}  Integrating the first summand in $y_0$, we have
$$ \int_{\sigma_1} \frac{f'f''}{2(E-(f')^2)} dq = \left. -\frac{1}{4} \Ln(E-(f')^2) \right|_{\partial \sigma_1} = -\frac{1}{4} \cdot 2\pi i = -\frac{\pi i}{2}. $$
To integrate the second summand, use a substitution $u=-f'(q)$ and $A=-f'(q)$:   
$$ \int_{\sigma_1} \frac{f''dq}{2i\sqrt{E-(f')^2}} \ = \  
- \int_{\sigma_1} \frac{f''dq}{2\sqrt{(f')^2-E} } \ = \ 
- \int_{A}^{\sqrt{E}} \frac{(-du)}{\sqrt{u^2-E}}  \ = \ -\arccosh \frac{A}{\sqrt{E}}. $$
(In the second term of this line the arithmetic square root is meant when $E>0$ and when $q$ is real immediately to the left of $q_1^-$.) 
%Performing the change of variables $u=\sqrt{E}\cosh t$ and using (\ref{IntCosh})
%\begin{equation} -\Ln 2A + \frac{1}{2}\Ln E + \frac{E}{4A^2} +o(E)  \label{Dec15} \end{equation}
Subtracting the latter value from the former, we obtain the statement. 
$\Box$

\subsubsection{Calculation of the  $\Delta y_1$ summand in $2\pi i s_{\sigma_1}$. }
\label{DeltaY1Q1} 

Recall that
\begin{equation}  y_1(q) \ = \  - \frac{5(f')^2(f'')^2}{8i (E-(f')^2)^{5/2}}  - \frac{f'(f'')^2}{2(E-(f')^2)^2}  
-  \frac{(f'')^2 }{8i(E-(f')^2)^{3/2}}  -   
\frac{f'f^{(3)}}{4i(E-(f')^2)^{3/2}}  - \frac{f^{(3)}}{4 (E-(f')^2)} . \label{y1is} \end{equation}

% $$ y_1 \ = \  -\frac{5(-f')^2(-f'')(-f'')}{8i(E-(f')^2)^{5/2}}    - 
% \frac{(-f')(-f'')(f'')}{2(E-(f')^2)^{2}} - \frac{(-f'')(-f'')}{8i(E-(f')^2)^{3/2}}    -\frac{f'f^{(3)}}{4i(E-(f')^2)^{3/2}}     - \frac{f^{(3)}}{4(E-(f')^2)}. $$ 

\begin{figure} \includegraphics{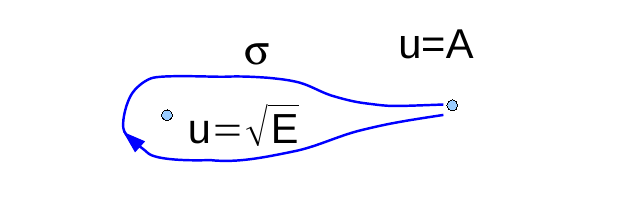} \caption{Contour $\sigma$ in the complex plane of $u$, sec.\ref{DeltaY1Q1}} \label{RDRW2p4} \end{figure}

In the integral $\int_{\sigma_1}y_1(q)dq$ let us make a substitution $u=-f'(q)$, write $ -f''(q) = b_0 + b_1 u + b_2u^2+...$ as in \eqref{DDfB} , and put $A=-f'(q_1-\varepsilon)$. Choose the contour $\sigma$ in the $u$-plane as on fig.\ref{RDRW2p4}; if $A$ is close to the real axis, in the formulas below $\arg (E-u^2)^{1/2}\approx \frac{\pi}{2}$ at the beginning of the path $\sigma$ and $\arg (E-u^2)^{1/2}\approx -\frac{\pi}{2}$ at the end of the path $\sigma$. 
%
%Note that $-(f''-f''(q_1))=b_1 u+b_2u^2+...$ since $u=0$ for $q=q_1$. Note further that $(f''-f''(q_1))^2=(f''+b_0)^2=[f'']^2+2f''b_0+b_0^2 = (-f'')(-f''-2b_0+\frac{b_0^2}{-f''})$ and $-1/f''=\sum_0^\infty a_j(-1)^j u^j$.
%
Then \small  
$$ \int_{\sigma_1} y_1(q)dq \ = \ \int_\sigma \left\{ -\frac{5 u^2 \sum_{j=0}^\infty b_j u^j}{8i(E-u^2)^{5/2}} + 
\frac{u\sum_{j=0}^\infty b_j u^j}{2(E-u^2)^{2}} -\frac{ \sum_{j=0}^\infty b_j u^j}{8i(E-u^2)^{3/2}} \right.$$
$$ \ \ \ \left.  - \frac{ u \sum_{j=0}^\infty (j+1)b_{j+1} u^j}{4i(E-u^2)^{3/2}}   
+ \frac{\sum_{j=0}^\infty (j+1)b_{j+1} u^j}{4(E-u^2)} \right\}du . $$ 
\normalsize

Since $\frac{u\sum_{j=0}^\infty b_j u^j}{2(E-u^2)^{2}} + \frac{\sum_{j=0}^\infty (j+1)b_{j+1} u^j}{4(E-u^2)} $ is a full differential of a function univalued in $u$, this part of the integrand can be dropped, and so
\small
$$ \int_{\sigma_1} y_1(q)dq \ = \ \int_\sigma \left\{ -\frac{5 u^2 \sum_{j=0}^\infty b_j u^j}{8i(E-u^2)^{5/2}} -\frac{ \sum_{j=0}^\infty b_j u^j}{8i(E-u^2)^{3/2}} 
 - \frac{ \sum_{j=1}^\infty j b_{j} u^{j}}{4i(E-u^2)^{3/2}}   
 \right\}du . $$ 
\normalsize
%The argument of $\sqrt{E-(f')^2}$ was chosen as $\pi/2$ to the left of $q_1^-$, so passing to the arithmetic square root for $q_1-\varepsilon$ to the left of $q_1^-$ and $q$ on the first sheet, obtain:
We prefer to rewrite the denominator in terms of $(u^2-E)^{1/2}$ which is positive for $u$ real, $u>\sqrt{E}$ and close to the beginning of the path $\sigma$,  and negative close for $u$ close to the end of $\sigma$:
$$ \int_{\sigma_1} y_1(q)dq \ = \ \int_\sigma \left\{ \frac{5 u^2 \sum_{j=0}^\infty b_j u^j}{8(u^2-E)^{5/2}} -\frac{  \sum_{j=0}^\infty (1+2j) b_j u^j}{8(u^2-E)^{3/2}} \right\}du . $$ 
Integrating by parts twice while using formulas \eqref{sqrt32deg0}-\eqref{sqrt52deg2}, we obtain
%$$  \ = \ -\left.\frac{\sum_{j=0}^\infty 5 b_j u^{j+1}}{3\cdot 8(u^2-E)^{3/2}}\right|_{\partial\sigma} + 
%\int_\sigma \left\{ \frac{5  \sum_{j=0}^\infty (j+1) b_j u^j}{3\cdot 8(u^2-E)^{3/2}} -\frac{  \sum_{j=0}^\infty (1+2j) b_j u^j}{8(u^2-E)^{3/2}} \right\} du \ = \ $$ 
%$$  \ = \ -\left.\frac{\sum_{j=0}^\infty 5 b_j u^{j+1}}{3\cdot 8(u^2-E)^{3/2}}\right|_{\partial\sigma} + 
%\int_\sigma  \frac{ \sum_{j=0}^\infty (2-j) b_j u^j}{3\cdot 8(u^2-E)^{3/2}}  du \ = \ $$ 
%$$  \ = \ -\left.\frac{\sum_{j=0}^\infty 5 b_j u^{j+1}}{3\cdot 8(u^2-E)^{3/2}}\right|_{\partial\sigma} 
%+ \left. \frac{2b_0}{3\cdot 8} \cdot \left( -\frac{1}{E}\frac{u}{\sqrt{u^2-E}}\right)\right|_{\partial\sigma}  
%+ \left. \frac{1\cdot b_1}{3\cdot 8} \cdot \left(-\frac{1}{u^2-E} \right)\right|_{\partial\sigma} + $$
%$$ + \left. \sum_{j=2}^\infty \frac{(2-j)b_j}{3\cdot 8} \cdot \left( -\frac{  b_j u^{j-1}}{(u^2-E)^{1/2}}\right)\right|_{\partial\sigma} 
%+ \int_\sigma  \frac{\sum_{j=2}^\infty (2-j)(j-1) b_j u^{j-2}}{3\cdot 8(u^2-E)^{1/2}}  du \ = \ $$ 
$$  \int_{\sigma_1} y_1(q)dq \ = \ -\left.\frac{\sum_{j=0}^\infty 5 b_j u^{j+1}}{24(u^2-E)^{3/2}}\right|_{\partial\sigma} 
- \left. \frac{b_0}{12E}\frac{u}{\sqrt{u^2-E}}\right|_{\partial\sigma}  
-   $$
$$ -  \left.\sum_{j=1}^\infty \frac{(2-j)b_j}{24} \frac{  b_j u^{j-1}}{(u^2-E)^{1/2}}\right|_{\partial\sigma}
+ \int_\sigma  \frac{\sum_{j=2}^\infty (2-j)(j-1) b_j u^{j-2}}{24(u^2-E)^{1/2}}  du. $$

The first three summands give
$$  \frac{5 \sum_{j=0}^{\infty} b_j A^{j+1} }{12 (A^2-E)^{3/2} } 
+  \frac{b_0}{6E} \frac{ A}{(A^2-E)^{1/2}} 
+   \frac{ \sum_{j=1}^\infty (2-j)b_j A^{j-1}}{12(A^2-E)^{1/2}}  .$$
Performing the change of variables $u=\sqrt{E}\cosh t$ in the fourth summand and using \eqref{IntCosh}, 
$$ \int_\sigma  \frac{\sum_{j=0}^\infty (j+1)[-jb_{j+2}] u^{j}}{24\sqrt{u^2-E}} du \ = \ % $$
%$$ \ = \ \frac{1}{12} \sum_{j=0}^\infty E^{\frac{j}{2}}(j+1) [ - j b_{j+2}] \int_{\arccosh(A/\sqrt{E})}^0    \cosh^j(t) dt \ = \ $$
%$$ \ = \ - \  \frac{1}{12} \sum_{j=1}^\infty \frac{(j+1)}{j}[-jb_{j+2}]  A^j  \ + \ o(E^0) $$
%$$ \ = \  \ 
 \frac{1}{12} \sum_{j=1}^\infty (j+1)b_{j+2}  A^j  \ + \ o(E^0). $$
Thus,
$$ \int_{\sigma_1} y_1(q)dq \ = \ \frac{5\sum_{j=0}^\infty b_j A^{j+1}}{12(A^2-E)^{3/2}}  + \frac{b_0}{6E} \frac{A}{(A^2-E)^{1/2}} + \frac{\sum_{j=1}^{\infty}(2-j)b_j A^{j-1}}{12(A^2-E)^{1/2}} + \frac{1}{12}\sum_{j=1}^\infty (j+1)b_{j+2}A^j + o(E^0) $$
%$$ \ = \ \frac{5}{12} \sum_{j=0}^\infty b_j A^{j-2}   + \frac{b_0}{6E} \left( 1 + \frac{E}{2A^2} \right)  + \frac{1}{12} \sum_{j=1}^{\infty}(2-j)b_j A^{j-2} + \frac{1}{12}\sum_{j=3}^\infty (j-1)b_{j}A^{j-2} + o(E^0) $$
%$$ \ = \ \frac{5}{12} \sum_{j=0}^2  b_j A^{j-2}   + \frac{b_0}{6E} +  \frac{b_0}{12 A^2}   + \frac{1}{12} \sum_{j=1}^{2}(2-j)b_j A^{j-2} + \frac{1}{12}\sum_{j=3}^\infty 6 b_{j}A^{j-2} + o(E^0) $$
%$$ \ = \ \frac{1}{2} b_0 A^{-2} +\frac{1}{2} b_1 A^{-1} + \frac{5}{12} b_2   + \frac{b_0}{6E}   + \frac{1}{2}\sum_{j=3}^\infty b_{j}A^{j-2} + o(E^0) $$
$$ \ = \ - \frac{1}{12} b_2   + \frac{b_0}{6E}   + \frac{1}{2}\sum_{j=0}^\infty b_{j}A^{j-2} + o(E^0). $$
Note that the error term $o(E^0)$ in the previous formula cannot be simply replaced by $O(E^1)$, as terms of order $E\ln E$ can also be present.

\subsection{Formal monodromy along $\sigma'_k$, $k$ even} \label{sigmaPEven}

The calculation is analogous to the one performed in Sec.\ref{sigmaOdd}; therefore, only the answers will be given. For definiteness let us work with $k=2$. 

In this section \ref{sigmaPEven}, denote:
$$ 2\pi i s_{\sigma'_2} (E,h) = \frac{1}{h} (\Delta S)(E) + \sum_{j=0}^\infty h^j (\Delta y_j)(E), $$
$$ \Delta S (E) \ = \ i\int_{\sigma'_2} \sqrt{E-(f')^2} dq',  \ \ \ \Delta y_j(E) \ = \ \int_{\sigma'_2} y_j(q) dq. $$

In this section \ref{sigmaPEven} we write $A=-f'(q_2-\varepsilon)$, $a_j=a_j(q_2)$, $b_j=b_j(q_2)$.

\subsubsection{The $\Delta S$ summand in $2\pi i s_{\sigma'_2}$. }

%%%%%%%%%%%%%%%%%%%% Calculation in comments is supposed to be correct
$$ \int_{\sigma'_2} i\sqrt{E-(f'(q))^2} dq \ = \ 
%2\int_{q_2-\varepsilon}^{q_2^{-}} [-\sqrt{(f')^2-E}]dq \ = \ $$
%(under the substitutions $-f'=u=-\sqrt{E}\cosh t$) 
%$$ \ = \ -2\int_A^{-\sqrt{E}} (\sum_{k=0}^\infty (-1)^k a_k u^k ) \sqrt{u^2-E} du \ = \ 2\sum_{k=0}^\infty  a_k E^{\frac{k}{2}+1} \int_{\arccosh(-\frac{A}{\sqrt{E}})}^0  \cosh^k t \sinh^2 t dt \ = \ $$
% $$ \ = \  
-2[f(q_2)-f(q_2-\varepsilon)] + a_0(\frac{E}{2}+\frac{E^2}{8A^2} + E ( \Ln\left( \frac{-2A}{\sqrt{E}} \right) - \frac{E}{2A^2} ) ) $$  
\begin{equation}  \ - \ a_2(
\frac{A^2E}{2} -  \frac{E^2}{16} +  \frac{E^2}{4}\Ln\left( \frac{-2A}{\sqrt{E}} \right)) \ - \ 
\sum_{\text{$k=1$ or $\ge 3$}} (-1)^k a_k \left( \frac{A^k E}{2k} + \frac{A^{k-2}E^2}{8(k-2)}\right) + O(E^2). \label{fe24f2} \end{equation} 
%see RDRW2.Appx136.tex for details

\subsubsection{The $\Delta y_0$ summand in $2\pi i s_{\sigma'_2}$. }

We have
$$ \Delta y_0 \ = \  \int_{\sigma'_2} y_0(q)dq \ = \ -\arccosh \frac{f'(q_2-\varepsilon)}{\sqrt{E}} - \frac{\pi i}{2} $$
where the branch of $\arccosh$ is chosen so as to coincide with the principal real value of of $\arccosh$ for $E>0$ and $q_2-\varepsilon$ on the real axis immediately to the left of $q_2^-$. 

% RDRW2.Appx109.tex contains the treatment of the even $j$ case

\subsubsection{The $\Delta y_1$ summand in $2\pi i s_{\sigma'_2}$. }

We have  
$$ \int_{\sigma'_2} y_1(q)dq \ = \ \frac{1}{12} b_2  -   \frac{b_0}{6E}   
-  \frac{1}{2} \sum_{j=0}^\infty b_{j}  A^{j-2}  \ + \ o(E^0) . $$

%%%%%%%%%%%%%%%%%%%%%%%%%%%%%%%%%%%%%%%%%%%%%%%%%%%%%%%%%%%%%%%%%%%%%%55

\subsection{Formal monodromy along $\gamma_k$} \label{gammamon}

The closed paths $\gamma_k$, $\gamma'_k$ were defined on fig.\ref{Paper4p3}. 

Once we calculate $s_{\gamma_k}$, we will automatically know $s_{\gamma'_k}$ since by ~\cite[Lemma 5.2]{G} 
\begin{equation}  s_{\gamma_k} \ + \ s_{\gamma'_k} \ = \ -1. \label{ggprim} \end{equation}

The following notation will be used in the rest of the paper:
$$ \omega_{\gamma_k}(E) = \oint_{\gamma_k} \sqrt{E-(f'(q))^2} dq. $$

\begin{Prop} \label{sgamma1calcd} We have %Let $\gamma_1$ be a counterclockwise loop around both $q_1^{-}$ and $q_1^{+}$ on the first sheet of the Riemann surface of the classical momentum. Then the monodromy of the corresponding formal WKB solution $e^{2\pi i s_{\gamma_1}}$ satisfies 
the following equality of $E$-dependent formal Laurent series in $h$:
$$ 2 \pi i s_{\gamma_k} = \frac{i}{h} \omega_{\gamma_k}(E) + \sum_{j=0}^\infty \Omega_{\gamma_k}^{(j)}(E) h^j, $$  
where for $E\to 0$
\begin{equation} \omega_{\gamma_k}(E) =  -\frac{\pi E}{f''(q_k)} + \frac{\pi a_2(q_k) E^2 }{4} + o(E^2) , \label{fe27f1} \end{equation} 
\begin{equation} \Omega_{\gamma_k}^{(0)} = -2\pi i, \label{fe27f2} \end{equation}
\begin{equation} \Omega_{\gamma_k}^{(1)} = O(E), \label{fe27f3} \end{equation} 
and where $$a_2(q_k)=\frac{f^{(4)}(q_k)}{2[f''(q_k)]^4} - \frac{3}{2}\frac{[f^{(3)}(q_k)]^2}{[f''(q_k)]^5} $$ 
as in \eqref{a0a1a2}.
\end{Prop}

%\begin{figure}
%\includegraphics{RDRW2p34n.pdf} \ \   \includegraphics{RDRW2p34a.pdf}
%\caption{Paths $\gamma_k$ and $\gamma'_k$.} \label{RDRWp20dash1a}
%\end{figure}

%Reminder on notation: $e^{\frac{i}{h}\omega_{\gamma_1}+O(1)} = e^{2\pi i s_{\gamma_1}}$, so $s_{\gamma_1} \approx \frac{\omega_{\gamma_1}}{2\pi h}$.

\textsc{Proof.} In order to have fewer indices, put $k=1$. We need to show \eqref{fe27f1}, \eqref{fe27f2}, and \eqref{fe27f3}.

{\it Formula \eqref{fe27f1}.} Performing the substitution $u=-f'(q)$ (so that $q=q_1^+$ corresponds to $u=-\sqrt{E}$ and  $q=q_1^-$ corresponds to $u=\sqrt{E}$) and using \eqref{OneOverDDF}, we have
$$\oint_{\gamma_1} \sqrt{E-(f')^2}dq = 2\int_{-\sqrt{E}}^{\sqrt{E}} \sqrt{E-u^2} (a_0-a_1u+a_2u^2-...)du \ = \ 
%\ = \ \oint \sqrt{E-u^2} \left( \sum_{j=0}^\infty a_j (-1)^j u^j \right) du 
  \sum_{j \ \text{even}} 2\frac{(j-1)!!}{(j+2)!!} E^{\frac{j}{2}+1}\pi a_j \ = \ $$
\begin{equation}  
 \ = \  2\frac{1}{2}  E \pi a_0 +  2\frac{1}{2\cdot 4} E^{2} \pi a_2 \ + \ o(E^2)  
  \ = \   -\frac{\pi E}{f''(q_1)} + \frac{\pi a_2 E^2 }{4} \ + \ o(E^2).  \label{Oct6} \end{equation}

%% {\it Step 1:} $\omega_{\gamma_1} := \int_{\gamma_1} \sqrt{E-(f')^2} dq = -\frac{\pi}{f''(q_1)}\cdot E + O(E^2)$.
%%   Write $f'(q) = f''(q_1)(q-q_1) + r(q-q_1)$, where $r(q-q_1) = O\left( (q-q_1)^2 \right)$. 
%% 
%% So, using the formula 
%% $$ \oint u^k \sqrt{E-u^2} du \ = \ \left\{ 
%% \begin{array}{ccc} 
%% 0, && \text{$k$ odd} \\
%% -2\frac{(k-1)!!}{(k+2)!!} E^{\frac{k}{2}+1}\pi, && \text{$k$ even}
%% \end{array}
%% \right. , $$
%% obtain (under the usual substitution $u=-f'$, $-\frac{1}{f''}=\sum_{k=0}^\infty a_k (-1)^k u^k$ and using that $f'(q_1^+)=\sqrt{E}$, so $q=q_1^+ \leftrightarrow u=-\sqrt{E}$, $q=q_1^- \leftrightarrow u=\sqrt{E}$.)

%\oint \sqrt{E} \sqrt{1- \left(\frac{f''(q_1)}{\sqrt{E}}(q-q_1) + \frac{1}{\sqrt{E}} r(q-q_1) \right)^2} dq  \ = \ $$ $$ \ = \  2\int_{1}^{-1} \frac{E}{f''(q_1)} \sqrt{1- x^2}dx  \ + \ O(E^2), $$ 
%so $\omega_\gamma<0$ if $E>0$ and  (according to our convention) $q_1$ is a local minimum. 

{\it Formula \eqref{fe27f2}.}% RDRW2.Appx133.tex contains a full version
The integration  
$$\oint_{\gamma_1} y_0 dq \ = \  \oint_{\gamma_1} dq \left\{\frac{f'f''}{2(E-(f')^2)} - \frac{f''}{2i\sqrt{E-(f')^2}} \right\} \ = \ -2\pi i $$
is performed by substitution $u=-f'(q)$ in the integral. 

{\it Formula \eqref{fe27f3}.}   Make a change of variables $u=-f'(q)$, choose a closed clockwise contour $\gamma$ in the $u$-plane containing inside it points $u=\pm\sqrt{E}$,  and proceed analogously to section \ref{DeltaY1Q1}.
$$ \int_{\gamma_1} y_1(q)dq \ = \ \int_\gamma \left\{ \frac{5 u^2 \sum_{j=0}^\infty b_j u^j}{8(u^2-E)^{5/2}} -\frac{  \sum_{j=0}^\infty (1+2j) b_j u^j}{8(u^2-E)^{3/2}} \right\} \ = \ $$
$$ \ = \  \int_\gamma  \frac{\sum_{j=2}^\infty (2-j)(j-1) b_j u^{j-2}}{3\cdot 8(u^2-E)^{1/2}}  du \ = \ $$
$$  \ = \  2\pi \sum_{2k=j>2} b_j E^{\frac{j-2}{2}} \frac{(j-2)(j-1)}{3\cdot 8}\frac{(j-3)!!}{(j-2)!!} \ = \ 
\ = \  \frac{\pi}{12} \sum_{k \ge 2}^\infty b_{2k} E^{k-1} \frac{(2k-1)!!}{(2k-4)!!} \ = \ 
O(E^0) $$
Here use the formula $\oint_\gamma \frac{u^k}{\sqrt{E-u^2}}du=-2\pi E^{k/2} \frac{(k-1)!!}{k!!}$ for even $k$.  $\Box$

% RDRW2.Appx60.tex contains $s_\gamma$ for the shifted potential. 

\subsection{Monodromies from $q_j-\varepsilon$ to $q_{j+1}-\varepsilon$.}

\begin{figure}[h] \includegraphics{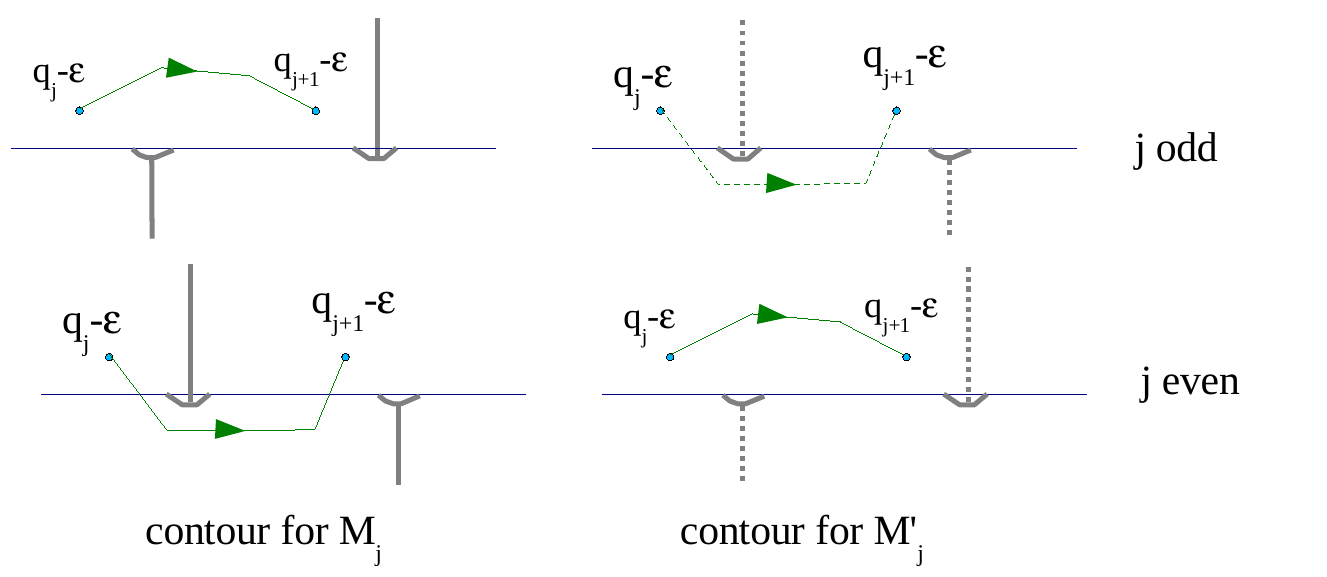} \caption{Integration contours defining $M_j$ and $M'_j$.} \label{Paper4p5} \end{figure}

%\begin{figure}[h] \includegraphics{RDRW2p5.pdf} \caption{Integration contours defining $M_j$ and $M'_j$, $j$ odd} \label{RDRW2p5} \end{figure} 

%\begin{figure}[h] \includegraphics{RDRW2p6.pdf} \caption{Integration contours defining $M_j$ and $M'_j$, $j$ even} \label{RDRW2p6} \end{figure} 

Define $M_j$, $M'_j$ to be the monodromies of the formal WKB solutions along the paths shown on fig.\ref{Paper4p5}, where $M_j$ are taken on the first sheet of the Riemann surface of $p(q)$ and $M'_j$ on the second. Note that in ~\cite{G} we denoted $M_j,M'_j$ by $A_j,A'_j$.

\begin{Lemma} \label{LemmaMj} We have:
$$ M_j \ = \ \exp \left\{ \frac{1}{h}\Mbb^{(-1)}_j + \sum_{k=0}^\infty h^k \Mbb^{(k)}_j \right\} $$
where $\Mbb_j^{(k)}=\Mbb_j^{(k)}(E)$, $k\ge -1$, are analytic functions of $E$ at least for $0<E<<1$, such that when  $E\to 0$,
$$ \Mbb^{(-1)}_j \ = \ [f(q_{j+1}-\varepsilon)-f(q_j-\varepsilon)] - E \int_{(q_j-\varepsilon)_I}^{(q_{j+1}-\varepsilon)_I} \frac{dq}{f'(q)} + O(E^2), $$
$$ \Mbb^{(0)}_j \ = \ (-1)^{j-1} \pi i \ + \ \frac{1}{2}\ln \left| \frac{f'(q_j-\varepsilon)}{f'(q_{j+1}-\varepsilon)} \right| \ + \ \frac{1}{4}\ln \frac{[f'(q_j-\varepsilon)]^2-E}{[f'(q_{j+1}-\varepsilon)]^2-E} \ + \ \frac{E}{8[f'(q_{j+1}-\varepsilon)]^2}-\frac{E}{8[f'(q_j-\varepsilon)]^2} + O(E^2), $$
$$ \Mbb^{(1)}_j \ = \ 
 \frac{f''(q_{j+1}-\varepsilon)}{2(f'(q_{j+1}-\varepsilon))^2} - \frac{f''(q_j-\varepsilon)}{2(f'(q_j-\varepsilon))^2} + O(E), $$
where $\int_{(q_j-\varepsilon)_I}^{(q_{j+1}-\varepsilon)_I}$ means that the integration path lies on the first sheet of the Riemann surface of $p(q)$ and $\ln$ stands for the branch of logarithm that is real for positive arguments.
\end{Lemma} 

%\begin{Lemma} \label{LemmaMj} We have:
%\small
%$$ M_j \ = \ -\exp\left\{\frac{[f(q_{j+1}-\varepsilon)-f(q_j-\varepsilon)]}{h}  - \frac{E}{2h} \int_{(q_j-\varepsilon)_I}^{(q_{j+1}-\varepsilon)_I} \frac{dq}{f'(q)} \right\} \sqrt{\frac{|f'(q_j-\varepsilon)|}{|f'(q_{j+1}-\varepsilon)|}} \sqrt[4]{\frac{[f'(q_j-\varepsilon)]^2-E}{[f'(q_{j+1}-\varepsilon)]^2-E}} \times \ \ \ \ $$ $$ \times \exp\left[  \frac{E}{8[f'(q_{j+1}-\varepsilon)]^2}-\frac{E}{8[f'(q_j-\varepsilon)]^2} + h( \frac{f''(q_{j+1}-\varepsilon)}{2(f'(q_{j+1}-\varepsilon))^2} - \frac{f''(q_j-\varepsilon)}{2(f'(q_j-\varepsilon))^2}) \right] (1+O(E^2/h)+O(E^2)+O(Eh)+O(h^2)),$$
%\normalsize
%where $\int_{(q_j-\varepsilon)_I}^{(q_{j+1}-\varepsilon)_I}$ means that the integration path lies on the first sheet of the Riemann surface of $p(q)$.
%\end{Lemma} 

\textsc{Proof.} The statement about $\Mbb^{(-1)}_j$ is obvious since 
$$\frac{i}{h}\int_{(q_j-\varepsilon)_I}^{(q_{j+1}-\varepsilon)_I} \sqrt{E-[f'(q)]^2}dq
 = \frac{i}{h}\int_{q_1-\varepsilon}^{q_2-\varepsilon} f'(q) (1-\frac{1}{2}\frac{E}{(f'(q))^2}) dq + O(E^2). $$

%cf RDRW2.Appx70.tex for the even case
%%  

%% \textsc{Proof.} We will present the argument for $M_1$, it will be the same for all odd $j$ and very analogous for even $j$. 
%% We have 
%% $$M_1 \ = \exp\left\{ \int_{(q_1-\varepsilon)_I}^{(q_2-\varepsilon)_I} \left[ \frac{i}{h}\sqrt{E-(f')^2} + y_0(q) + hy_1(q) \right] dq + O(h^2) \right\}, $$
%% where $\int_{(q_1-\varepsilon)_I}^{(q_2-\varepsilon)_I}$ means that the path of integration is chosen within the domain of definition of $\phi_+$ and determinations of the square root are taken as on the first sheet of the Riemann surface of the classical momentum. We have:
%% $$\frac{i}{h}\int_{(q_1-\varepsilon)_I}^{(q_2-\varepsilon)_I} \sqrt{E-(f')^2}dq
%% = \frac{i}{h}\int_{q_1-\varepsilon}^{q_2-\varepsilon} f'(q) (1-\frac{1}{2}\frac{E}{(f'(q))^2}) dq + O(E^2) $$ 
%% $$ \ = \ \frac{i}{h}\left( f(q_2-\varepsilon) - f(q_1-\varepsilon) - \frac{E}{2} \int_{q_1-\varepsilon}^{q_2-\varepsilon} \frac{dq}{f'(q)} + O(E^2) \right). $$

We will show the statement about $\Mbb^{(0)}_j$ for $j=1$.  
In the integral
$$ \Mbb_1^{(0)} \ = \ \int_{(q_1-\varepsilon)_I}^{(q_2-\varepsilon)_I} y_0(q) dq \ = \ \int_{(q_1-\varepsilon)_I}^{(q_2-\varepsilon)_I} \left[ \frac{f'f''}{2(E-(f')^2)} - \frac{f''}{2i\sqrt{E-(f')^2}}\right] dq $$ 
the first summand yields 
$$\int_{(q_1-\varepsilon)_I}^{(q_2-\varepsilon)_I} \frac{f'f''}{2(E-(f')^2)}dq \ = \ -\frac{1}{4}\Log \frac{E-[f'(q_2-\varepsilon)]^2}{E-[f'(q_1-\varepsilon)]^2} \ = \ $$
%and the $\Log$ is analytically continued in the domain with a cut located as on the first sheet. 
$$ \ = \ -\frac{1}{4}\left( \log \left\{\frac{E-[f'(q_2-\varepsilon)]^2}{E-[f'(q_1-\varepsilon)]^2}\right\} - 2\pi i \right) =  \frac{\pi i}{2} -\frac{1}{4} \log \left\{\frac{E-[f'(q_2-\varepsilon)]^2}{E-[f'(q_1-\varepsilon)]^2}\right\}, $$  
and the second summand 
$$ \int_{q_1-\varepsilon}^{q_2-\varepsilon} \frac{f''}{2i\sqrt{E-(f')^2}} dq \ = \  \int_{q_1-\varepsilon}^{q_2-\varepsilon} \frac{f''}{2f'} \left( 1+\frac{1}{2}\frac{E}{(f')^2} \right) dq \ + \ O(E^2)  \ = \ $$
$$ \ = \ \frac{1}{2} \Ln \left( \frac{f'(q_2-\varepsilon)}{f'(q_1-\varepsilon)} \right) \ 
- \left. \frac{E}{8[f'(q)]^2} \right|_{q=q_1-\varepsilon}^{q=q_2-\varepsilon} \ + \ O(E^2) \ = \ $$ 
$$ \ = \ \frac{1}{2}\ln\frac{f'(q_2-\varepsilon)}{[-f'(q_1-\varepsilon)]} - \frac{\pi i}{2} -  \frac{E}{8[f'(q_2-\varepsilon)]^2}  +  \frac{E}{8[f'(q_1-\varepsilon)]^2} + O(E^2).$$
%(It is $+f''/2f'$ in the second integral, because $\arg (E-(f')^2)=-\pi$ on the real line segment $(q_1^+,q_2^-)$, so $\arg \sqrt{E-(f')^2}=-\pi/2$, and, on the other hand $f'>0$ on that line segment)
If $j$ is even, the passage from complex $\Ln$ to the standard branch of logarithm will be slightly different.

% Calculation of the corresponding integrals of $y_1$ is contained in RDRW2.Appx47.tex

As for $\Mbb_j^{(1)}$, again for $j=1$, we use \eqref{y1is}, replace each summand by its limit for $E\to 0$, and obtain:
$$ \Mbb_1^{(1)} \ = \ \int_{(q=q_1-\varepsilon)_I}^{(q=q_2-\varepsilon)_I} y_1(q) dq \ = \ \int_{q=q_1-\varepsilon}^{q=q_2-\varepsilon} \left\{  - \frac{5(f')^2(f'')^2}{8 (f')^5}  -  \frac{f'(f'')^2}{2(f')^4}  
-  \frac{(f'')^2 }{8(-(f')^3)}  -   
\frac{f'f^{(3)}}{4(-(f')^3)}  - \frac{f^{(3)}}{4 (-(f')^2)} \right\} dq + O(E) \ = \ $$ 
%$$ \ = \ \int_{q=q_1-\varepsilon}^{q=q_2-\varepsilon} \left\{  - \frac{5(f'')^2}{8 (f')^3}  -  \frac{(f'')^2}{2(f')^3}  
%+  \frac{(f'')^2 }{8(f')^3}  +    
%\frac{f'f^{(3)}}{4(f')^3}  + \frac{f^{(3)}}{4 (f')^2} \right\} dq + O(E) \ = \ $$ 
$$ \ = \ \int_{q=q_1-\varepsilon}^{q=q_2-\varepsilon} \left\{  - \frac{(f'')^2}{(f')^3}   +    
\frac{f^{(3)}}{2(f')^2}  \right\} dq  + O(E) \ = \ 
\left.   \frac{f''}{2(f')^2}  \right|_{q=q_1-\varepsilon}^{q=q_2-\varepsilon} + O(E). $$ 
$\Box$

%% Hence  
%% $$ M_1 \ = \ -e^{\frac{[f(q_2-\varepsilon)-f(q_1-\varepsilon)]}{h}  - \frac{E}{2h} \int_{q_1-\varepsilon}^{q_2-\varepsilon} \frac{dq}{f'(q)} }\sqrt{\frac{[-f'(q_1-\varepsilon)]}{f'(q_2-\varepsilon)}} \sqrt[4]{\frac{[f'(q_1-\varepsilon)]^2-E}{[f'(q_2-\varepsilon)]^2-E}} \times \ \ \ \ $$ $$ \times \exp\left[  \frac{E}{8[f'(q_2-\varepsilon)]^2}-\frac{E}{8[f'(q_1-\varepsilon)]^2} + h( \frac{f''(q_2-\varepsilon)}{2(f'(q_2-\varepsilon))^2} - \frac{f''(q_1-\varepsilon)}{2(f'(q_1-\varepsilon))^2}) \right] (1+O(E^2/h)+O(E^2)+O(Eh)+O(h^2)),$$
%% which completes the proof for $j=1$. $\Box$

Analogously, we have:

\begin{Lemma} \label{LemmaMPrimj} We have:
$$ M'_j \ = \ \exp \left\{ \frac{1}{h}\tilde \Mbb^{(-1)}_j + \sum_{k=0}^\infty h^k \tilde \Mbb^{(k)}_j \right\} $$
where $\tilde \Mbb_j^{(k)}=\tilde \Mbb_j^{(k)}(E)$, $k\ge -1$, are analytic functions of $E$ at least for $0<E<<1$, such that when  $E\to 0$,
$$ \tilde \Mbb^{(-1)}_j \ = \ - [f(q_{j+1}-\varepsilon)-f(q_j-\varepsilon)] + E \int_{(q_j-\varepsilon)_{II}}^{(q_{j+1}-\varepsilon)_{II}} \frac{dq}{f'(q)} + O(E^2), $$
$$ \tilde \Mbb^{(0)}_j \ = \  \frac{1}{2}\ln \left| \frac{f'(q_{j+1}-\varepsilon)}{f'(q_{j}-\varepsilon)} \right| \ + \ \frac{1}{4}\ln \frac{[f'(q_j-\varepsilon)]^2-E}{[f'(q_{j+1}-\varepsilon)]^2-E} \ + \ \frac{E}{8[f'(q_{j}-\varepsilon)]^2}-\frac{E}{8[f'(q_{j+1}-\varepsilon)]^2} + O(E^2), $$
$$ \tilde \Mbb^{(1)}_j \ = \  O(E), $$
where $\int_{(q_j-\varepsilon)_{II}}^{(q_{j+1}-\varepsilon)_{II}}$ means that the integration path lies on the second sheet of the Riemann surface of $p(q)$ and $\ln$ stands for the branch of logarithm that is real for positive arguments.
\end{Lemma} 

% see RDRW2.Appx71.tex for a proof

%% \begin{Lemma} \label{LemmaMPrimj} We have:
%% \small
%% $$  M'_j \ = \ \exp \left\{-\frac{[f(q_{j+1}-\varepsilon)-f(q_j-\varepsilon)]}{h}  + \frac{E}{2h} \int_{(q_j-\varepsilon)_{II}}^{(q_{j+1}-\varepsilon)_{II}} \frac{dq}{f'(q)} \right\}\sqrt{\frac{|f'(q_{j+1}-\varepsilon)|}{|f'(q_j-\varepsilon)|}} \sqrt[4]{\frac{[f'(q_j-\varepsilon)]^2-E}{[f'(q_{j+1}-\varepsilon)]^2-E}}\times \ \ \ \ $$ $$ \times \exp\left[  \frac{E}{8[f'(q_j-\varepsilon)]^2}-\frac{E}{8[f'(q_{j+1}-\varepsilon)]^2}  \right] (1+O(E^2/h)+O(E^2)+O(Eh)+O(h^2)).$$
%% \normalsize
%% where $\int_{(q_j-\varepsilon)_{II}}^{(q_{j+1}-\varepsilon)_{II}}$ means that the integration path lies within the domain of definition of $\phi_-$, fig.\ref{RDRW2p32}, right.
%% \end{Lemma} 
% see RDRW2.Appx71.tex for a proof

%%%%%%%%%%%%%%%%%%%%%%%%%%%%%%%%%%%%%%%%%%%%%%%%%

% \newpage

\section{Passage from the equation $P\psi =E \psi$ to the equation $P\psi=hE_r\psi$.}  \label{secE2hEr} 

%Let $P$ be the Witten Laplacian \eqref{WLO}. So far we have studied formal WKB solutions $\psi_{\pm}(E,q,h)$ of the equation $\Psi \psi =E\psi$, cf.\eqref{PpsiEpsi} for a small, usually positive, number $E$. 

We saw in Sec.\ref{FormalhEr} that as long as we work with {\it formal} solutions of $P\psi=E\psi$ or $P\psi=hE_r\psi$, the passage from $E$ to $hE_r$ is straightforward. The situation is more subtle once we begin to study the correspondence between formal and actual resurgent solution of the equation $P\psi=E\psi$ or $P\psi=hE_r\psi$. This correspondence is expressed in terms of connection formulas and is the main topic of this section \ref{secE2hEr}.

We will repeat here the formal calculation of the connection coefficients across the double turning points of the equation $P\psi=hE_r\psi$ by the exact matching method, cf. our exposition in ~\cite[\S 7]{G} and references therein. Compared to ~\cite{G}, now we will push the calculation to one more order in $h$. 

Note that the connection coefficient called $c_j$ in ~\cite[\S 7]{G} will now be denoted $c'_j$, consistently with the notation of ~\cite[\S 8]{G}.

%Substituting $hE_r$ for $E$ also works without any problem for $s_{\gamma_k}$, $s_{\gamma'_k}$, $M_k$, $M'_k$, but the situation is not as simple for $s_{\sigma_k}$ and $s_{\sigma'_k}$. However, properly interpreting the last two expressions for $E=hE_r$ is crucial for the solution of the connection problem for the equation $P\psi=hE_r\psi$ as we shall see in a moment.

%The goal of this section is to derive the connection formula {\bf in the sense of section REF} for 

% \subsection{Connection formulas for the double turning point.} \label{ConnDouble}

\subsection{Stokes curves and Stokes regions for $P\psi=E\psi$ and $P\psi=hE_r\psi$.}
\label{SecStCu}

The purpose of this subsection \ref{SecStCu} is to give an extended literature reference for  the formulas \eqref{c1primE} and \eqref{c2E}.

Recall that following the setup of ~\cite{G} we assume $0<\arg h <<1$. In the complex plane of $q$ let us draw the Stokes curves -- the locus where a discontinuous change of exponential asymptotic expansions of solutions of our differential equation may happen.  Stokes regions are the domains into which the complex plane of $q$ is split by the Stokes curves. 

It is known, e.g. ~\cite{V83}, that for a Schr\"odinger equation 
$$ [ \, -h^2 \partial^2_q + V_0(q) + hV_1(q) \, ]\psi(q,h) \ = \  0, \ \ \arg h =\alpha, \ \ |h|\to 0 $$
with entire $V_0(q)$, $V_1(q)$, the Stokes curves are given by the condition $\int_{q_\star}^q \sqrt{V_0(q')} dq' \in e^{i\alpha} \R$, where $q_\star$ runs over the zeros of $V_0(q)$. 

\begin{figure}[h] \includegraphics{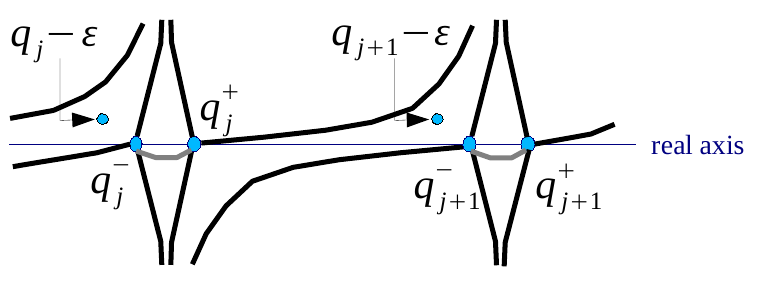} \caption{Stokes curves for the equation $P\psi=E\psi$, $0<E<<1$.} \label{Paper4p6} \end{figure}

\begin{figure}[h] \includegraphics{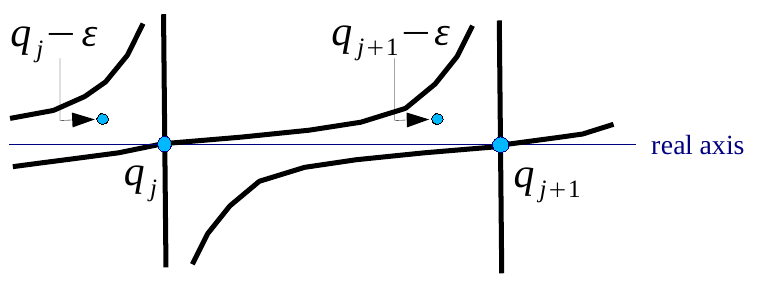} \caption{Stokes curves for the equation $P\psi=E\psi$, $E=0$, or for the equation $P\psi=hE_r\psi$, $E_r\in \C$.} \label{Paper4p7} \end{figure}

For the Witten Laplacian, this means the following. In the neighborhood of the real axis in the $q$ plane, the Stokes curves for the equation $P\psi=E\psi$, resp., $P\psi=hE_r\psi$, look as on fig. \ref{Paper4p6} and \ref{Paper4p7}. 

As we explained in sec.\ref{Method}, we are interested in solving the connection problem for the equation $P\psi = hE_r\psi$ between the intervals $(q_{j-1},q_j)$ and $(q_j,q_{j+1})$. Looking at the fig.\ref{Paper4p7}, this is the same as solving the connection problem between the Stokes regions containing points $q_j-\varepsilon$ and $q_{j+1}-\varepsilon$. 

For the equation $P\psi=E\psi$, $0<E<<1$, the answer to the connection problem between Stokes regions containing $q_j-\varepsilon$ and $q_{j+1}-\varepsilon$ is particularly simple, see ~\cite{V83}, ~\cite{DP99}, and our exposition in ~\cite[\S 7]{G}, and will be stated in \eqref{c1primE} and \eqref{c2E}.

%==========================

%%%%%%%%%%%%%%%%%%%%%%%%%%%%%%%%%%%%%%%%%%%%%%

%{\bf Proposed notation: $C_{q_1-\varepsilon}^{(q_1)}$ for the connection mtx in the basis of solutions normalized to one at $q_1-\varepsilon$ }

\subsection{Exact matching method around $q_1$.} \label{ExactMatchingq1}

 We are now going solve the connection problem across a double turning point $q_j$ of $P\psi=hE_r\psi$ in two representative cases -- for the double turning point $q_1$ where $f(q)$ has a local minimum in this subsection \ref{ExactMatchingq1}, and for the double turning point $q_2$ where $f(q)$ has a local maximum in the subsection \ref{ExactMatchingq2}; similar results will hold for other real local extrema of $f(q)$.

\subsubsection{Definition of the reduced connection coefficient $(c'_1)^{red}(E,h)$}

Let $E$ be a small positive number. Consider two formal solutions $\psi_+(E,q,h)$ and $\psi_-(E,q,h)$ of \eqref{WLcsb} corresponding to the first and to the second sheets of the Riemann surface of $p(q)$ and normalized in such a way that $\psi_+(q_1-\varepsilon)=\psi_-(q_1-\varepsilon)=1$, where $\varepsilon$ is a small complex number, $\arg \varepsilon \in [0,\frac{\pi}{2}]$, i.e. 
\begin{equation} \psi_{\pm}(E,q,h) =  \exp\left\{ \int_{q_1-\varepsilon}^q \frac{i}{h} \sqrt{E-[f'(q')]^2} + \sum_{j=0}^\infty y_j(q')dq' \right\} \label{mar4f1} \end{equation}

An actual resurgent function solution $\phi(E,q,h)$ of $P\psi=E\psi$, $0<E<<1$, whose exponential asymptotic representation on the interval $(q_{2n}^+-1,q_1^-)$ is given by $\phi(E,q,h)\sim \psi_+(E,q,h)$, has on $(q_1^+, q_2^-)$ an asymptotic representation 
\begin{equation}  \phi(E,q,h) \ \sim \ \psi_+(E,q,h) + c'_1(E,h) \psi_-(E,q,h) \label{mar4f2} \end{equation}
where 
\begin{equation} c'_1(E,h) \ = \ \exp ( 2\pi i s_{\sigma_1} ), \label{c1primE} \end{equation} 
cf. fig.\ref{Paper4p4} for the definition of $\sigma_1$ and sec.\ref{sigmaOdd} for the calculation of $s_{\sigma_1}$.

Since the contour $\sigma_1$ gets pinched by $q_1^-$ and $q_1^+$ when $E\to 0$, substituting $hE_r$ for $E$ directly into the expression of $c'_1(E)$ is problematic. Therefore, the following trick is used. 

Introduce the {\it reduced connection coefficient} $(c'_1)^{red}$ by
\begin{equation}  c'_1(E,h) = \frac{\sqrt{2\pi} h^{s_{\gamma_1} + \frac{1}{2}} }{\Gamma(-s_{\gamma_1})} (c'_1)^{red}(E,h). \label{c1redDefd} \end{equation}
Here $s_{\gamma_1}=s_{\gamma_1}(E,h)$ is defined on fig.\ref{Paper4p3} and computed in Sec.\ref{gammamon}.

It is believed that $(c'_1)^{red}(E,h)$ can be written in the form $\exp\{ \sum_{k=-1}^\infty h^k A_k(E) \}$ where $A_k(E)$ are holomorphic functions of $E$ near the origin; therefore, $(c'_1)^{red}(hE_r,h)$ is straightforward to define. Further, $s_{\gamma_1}(E,h)$ is representable by power series in $h$ with coefficients analytic with respect to $E$, therefore also $s_{\gamma_1}(hE_r,h)=\sum_{k=0}^\infty B_k(E_r)h^k $ makes sense as a power series in $h$ with coefficients analytic functions in $E_r$. Substitution of $t=s_{\gamma_1}(hE_r,h)$ into $h^{t+\frac{1}{2}}$ and $\frac{1}{\Gamma(-t)}$ is done by means of the Taylor series expansions of the latter two functions at $t=B_0(E_r)$.  Combining these definitions with \eqref{c1redDefd} allows us to define $c'_1(hE_r,h)$. Notice that the expression $h^{t+\frac{1}{2}}$ is representable as an $E_r$-dependent expansion in powers of $h$ and $h\ln h$.

%\begin{figure}[h]\includegraphics{RDRW2p36.pdf} \caption{Notation in the exact matching method around $q_1$} \label{nuthesisp14}
%\end{figure}

%For $E$ a positive real number,  the actual solution of \eqref{WLcsb} represented by $\psi_+$ in $R$ is represented by $\psi_+ +c'_1\psi_-$ in $R''$.   We know that $c'_1\psi_-$ corresponds to the analytic continuation of $\psi_+$ along a loop  $\sigma_1$ with base point $q_1-\varepsilon$  around the simple turning point $q_1^-(E)$, figure \ref{nuthesisp14}, i.e. the Stokes phenomenon transforms $\psi_+$ into 
%$$ \psi_+(q,E) \ + \ c'_1(E) \psi_-(q,E) \ = \ $$
%$$ \ = \ \psi_+(q,E) \ + \ \sqrt{2\pi} \frac{h^{s_{\gamma_1} + \frac{1}{2}} }{\Gamma(-s_{\gamma_1})} (c'_1)^{red}(E) \psi_-(q,E),$$
% where $c'_1$ is the monodromy of the formal solution along $\sigma_1$: 
% $$ c'_1  = \exp\left[ \int_{\sigma_1} dq \left\{ \frac{i}{h} S(q,E) + y_0(q) + hy_1(q)+h^2 y_2(q) +...\right\} \right] . $$
%and where we have denoted 
%$$ c'_1 = \frac{\sqrt{2\pi} h^{s_{\gamma_1} + \frac{1}{2}} }{\Gamma(-s_{\gamma_1})} (c'_1)^{red}.$$
%For $E=hE_r$ we will obtain
%$$ \psi_+(q,hE_r) \ + \ \sqrt{2\pi} \left.\frac{h^{s_{\gamma_1}+\frac{1}{2}}}{\Gamma(-s_{\gamma_1})}\right|_{E=hE_r}  (c'_1)^{red}(hE_r) \psi_-(q,hE_r). $$ 

\begin{Rmk} \label{SatoConj} {\rm This passage to the limit and replacing $E>0$ by $hE_r$ has been used in two papers ~\cite{DDP97} and ~\cite{DP99}, but ideally it would need a more solid mathematical justification. 
The first issue is purely algebraic: one has to show that the coefficients in the asymptotic expansion of $(c'_1)^{red}(E,h)$ with respect to $h$ are analytic functions of $E$ near the origin. %, or, equivalently, that all infinitely many coefficients $\Theta_j$, $j\in \Z$, $j\ge -1$, from \eqref{credTheta} are analytic in $E$ near the origin. 
This problem is known in the literature as the Sato's conjecture. See ~\cite{SS} for its solution for the Schr\"odinger equation with the harmonic oscillator potential. The case of a general potential may perhaps be proven using reduction of an arbitrary potential well to a harmonic oscillator using methods of ~\cite{AKT} and references therein. \footnote{We thank Shingo Kamimoto for pointing out to us both of these articles. We thank professors Aoki, Kawai, and Takei for explaining the result of ~\cite{AKT} and its significance. We must confess that as of this writing we have not carefully thought through the argument of ~\cite{SS}. } 
The other issue is analytical: in our setup, ~\cite[\S 2.1]{G}, resurgent functions are {\it equivalence classes} of analytic functions in $h$ and thus the notions of limit, parameter dependence, analytic continuation should be transferred from analytic functions to resurgent function with care. In particular, the fact that we can define $c'_1(hE_r,h)$ does not yet formally imply that this quantity gives the connection coefficient for the equation $P\psi=hE_r\psi$. We hope that all these technicalities will be resolved in due time. } \end{Rmk}

\subsubsection{Calculation of $(c'_1)^{red}(E,h)$ for $0<E<<1$.} \label{c1RedCalc}

% 
% Denote as in \eqref{DeltaSDeltay}
% $$\Delta S (E) \ = \ i\int_{\sigma_1} \sqrt{E-(f')^2} dq',  \ \ \ \Delta y_k(E) \ = \ \int_{\sigma_1} y_k(q) dq,  $$
% thus 
%$$ c'_1(E) \ = \ \exp \left\{ \frac{1}{h}\Delta S(E) + \Delta y_0(E) + h\Delta y_1(E) + O(h^2) \right\}. $$ 
Denote 
\begin{equation} (c'_1)^{red}(E,h) \ = \ \exp \{ \sum_{j=-1}^{\infty} h^j \Theta_{j}(E)  \} . 
% \ = \ \exp \{ \frac{1}{h} \Theta_{-1} + \Theta_{0} + h \Theta_1 + O(h^2) \}. 
\label{credTheta} \end{equation}

In this subsection \ref{c1RedCalc} we are going to evaluate $\Theta_{-1}$, $\Theta_0$, and $\Theta_1$ for $E\to 0$.

%Since
%$$ s_{\gamma_1} \ = \ \frac{\omega_{\gamma_1}}{2\pi  h}  - 1 + O(Eh) + O(h^2), $$
Formula \eqref{NextTermStirling1} shown is sec.\ref{AppliStir} gives: 
\begin{equation} \frac{\sqrt{2\pi}h^{s_{\gamma_1}+\frac{1}{2}}}{\Gamma(-s_{\gamma_1})} \ \sim \ % \frac{\exp ( -\frac{\omega_{\gamma_1}}{2\pi h} )}{\left(-\frac{\omega_\gamma}{2\pi} \right)^{-\frac{\omega_\gamma}{2\pi h}+\frac{1}{2}}   (1-\frac{\pi h}{6\omega_{\gamma_1}} )}(1+O_{E=fix}(h^2)+O(h)O(E\log E)) \ = \ $$
%$$ \ = \ 
\exp\left\{ \frac{1}{h}\left[ -\frac{\omega_{\gamma_1}}{2\pi} + \frac{\omega_{\gamma_1}}{2\pi} \Ln \left( -\frac{\omega_{\gamma_1}}{2\pi}\right) \right] - \frac{1}{2}\Ln \left( - \frac{\omega_{\gamma_1}}{2\pi} \right) + h\frac{\pi}{6\omega_{\gamma_1}} + O(h^2)+O(E\log E) h \right\}, \label{fe29e1} \end{equation} 
where the notation $O(h^2)+O(E\log E)h$  means an $E$-dependent power series $\sum_{k=1}^\infty \Gbb _k(E)h^k$ where $\Gbb_k(E)$ are analytic functions of $E$ for $0<E<<1$ and $\Gbb_1(E)=O(E\log E)$ when $E\to 0+$.  
 
%\underline{Dealing with $\Theta_{-1}$.} % old exposition, through ''difficult equality'', is in RDRW2.Appx53.tex

%Let us now calculate $\Theta_{-1}$, $\Theta_0$, and $\Theta_1$. 

 Combining \eqref{c1primE}, \eqref{c1redDefd}, \eqref{fe29e1}, and results of sec.\ref{sigmaOdd}, we have: \\
{\it a)}
$$ \Theta_{-1} = \Delta S +\frac{\omega_{\gamma_1}}{2\pi} - \frac{\omega_{\gamma_1}}{2\pi} \Ln\left[ -\frac{\omega_{\gamma_1}}{2\pi}\right]. $$
where $\Delta S$ was computed in \eqref{eq24feb}; more explicitly, 
$$ \Theta_{-1} = 2[f(q_1)-f(q_1-\varepsilon)] -a_0 E \Ln ( 2A )  + a_0 \frac{E^2}{8A^2}  
 \ - \   a_2  \frac{A^2}{2} E    \ - \ \sum_{j=1 \ \text{or} \ j\ge 3} a_j (-1)^j  \left\{   \frac{1}{j} A^j E  + \frac{1}{4(j-2)}E^2 A^{j-2}  \right\}   + $$
$$ \ - \  \frac{E}{2f''(q_1)} \Ln(2f''(q_1)) \ + \  \frac{a_2 E^2 }{16} \ + \ \frac{a_2 E^2}{8} \Ln(2f''(q_1))   \ - \ a_2 \frac{E^2}{4}\Ln 2A + o(E^2). $$
{\it b)}  
%\underline{Dealing with $\Theta_0$} % (The equality is exact, there is no error term in the first line) 
%The expressions for $\Theta_0$ and $\Theta_1$ are easier using formulas from sections \ref{MonoY0} and \ref{MonoY1}, we obtain:
$$ \Theta_0  \ = \ \arccosh \frac{A}{\sqrt{E}} \ - \ \frac{\pi i}{2} \ + \ \frac{1}{2}\Ln\left( -\frac{\omega_{\gamma_1}}{2\pi} \right) \ = \ $$
%$$ \ = \ \Ln 2A - \frac{1}{2}\Ln E - \frac{E}{4A^2} \ - \ \frac{\pi i}{2} + \frac{1}{2} \Ln (\frac{E}{2f''(q_1)} - \frac{a_2 E^2}{8}) + o(E) $$ 
$$ \ = \ \Ln 2A - \frac{E}{4A^2} \ - \ \frac{\pi i}{2} + \frac{1}{2} \left[ \Ln (\frac{1}{2f''(q_1)}) - \frac{a_2 f''(q_1) E}{4} \right] + o(E). $$ 
{\it c)}
%\underline{Dealing with $\Theta_1$} 
$$ \Theta_{1} \ = \ (\Delta y_1)(E) -  \frac{\pi}{6\omega_{\gamma_1}} + o(E^0) %\ = \ $$
%$$ \ = \ - \frac{1}{12} b_2   + \frac{b_0}{6E}   + \frac{1}{2}\sum_{j=0}^\infty b_{j}A^{j-2} \ - \ \frac{\pi}{6(-\frac{\pi E}{f''(q_1)} + \frac{a_2\pi E^2}{4})}  + o(E^0) $$
%$$ \ = \ - \frac{1}{12} b_2   - \frac{f''(q_1)}{6E}   + \frac{1}{2}\sum_{j=0}^\infty b_{j}A^{j-2} \ + \ \frac{f''(q_1)}{6E} (1+  \frac{a_2 f''(q_1) E}{4})  + o(E^0) $$
%$$ \ = \ - \frac{1}{12} b_2     + \frac{1}{2}\sum_{j=0}^\infty b_{j}A^{j-2} \ + \ \frac{f''(q_1)}{6E}  \frac{a_2 f''(q_1) E}{4}   + o(E^0) $$
%$$ \ = \ - \frac{1}{12} b_2     + \frac{1}{2}\sum_{j=0}^\infty b_{j}A^{j-2} \ + \ \frac{a_2 [f''(q_1)]^2 }{24}   + o(E^0) $$ $$
 \ = \ - \frac{1}{12} b_2     - \frac{1}{2}\frac{f''(q_1-\varepsilon)}{(f'(q_1-\varepsilon))^2} \ + \ \frac{a_2 [f''(q_1)]^2 }{24}   + o(E^0). $$
We remark that by lemma \ref{Theta1Analytic}, this last $o(E^0)$ is actually $O(E)$.

As we pointed out in Remark \ref{SatoConj}, there must be a conceptual way of proving analyticity of 
$\Theta_j(E)$ near $E=0$ for all $j\ge -1$; for now we will prove analyticity of $\Theta_{-1},\Theta_0,\Theta_1$ directly in section \ref{SomeProofs}.

% an intermediary proposition was here, see RDRW2.Appx36.tex

% see longer version in RDRW2.Appx118.tex
Note that the infinite sums appearing in the expression for $\Theta_{-1}$ for specific $f(q)$ can be evaluated by integration:  
\begin{equation} \sum_{j=1}^\infty a_j (-1)^j \frac{1}{j}A^j \ = \ \int_{q_1}^{q_1-\varepsilon}  (-\frac{1}{f'(q)} - a_0 \frac{f''(q)}{f'(q)}) dq, \label{SumAj} \end{equation}
\begin{equation}   \sum_{j=3}^\infty  a_j (-1)^j \frac{1}{j-2}A^{j-2} \ = \  
  \int_{q_1}^{q_1-\varepsilon} \left[ -\frac{1}{[f'(q)]^3}  - \frac{a_0 f''(q)}{[f'(q)]^3}  - \frac{a_1 f''(q)}{[f'(q)]^2}
- \frac{a_2f''(q)}{f'(q)} \right]  dq.  \label{SumAj2} \end{equation}

%applications of \eqref{SumAj2} to our situation: RDRW2.Appx119.tex

\subsubsection{Calculation of $c'_1(hE_r,h)$} 

Combining the above formulas, we are now able to to write down an expression for $c'_1(hE_r,h)$ to leading orders in $h$ and $h\ln h$ valid for any $E_r\in \C$. 

Since in the rest of the paper we will be substituting a small resurgent function for $E_r$, i.e. $E_r\in \C$ will be replaced by $E_r(h)\to 0$ for $h\to 0$, we will study the behavior of $c'_1(hE_r,h)$ for $\C\ni E_r\to 0$.

We have the following equality of formal power series in $h$ with coefficients holomorphic functions in $E_r$:
$$  s_{\gamma_1} (hE_r,h) = \left[ -\frac{E_r}{2f''(q_1)} - 1  \right] + O(E_r)h + O(h^2),$$
where $O(E_r)h + O(h^2)$ stands for $\sum_{k=1}^\infty A_k(E_r) h^k$ with $A_1(E_r)=O(E_r)$ for $E_r\to 0$. Therefore, 
$$ h^{s_{\gamma_1}(hE_r,h)+\frac{1}{2}} = h^{-\frac{E_r}{2f''(q_1)}-\frac{1}{2}} \exp[ \sum_{k=1}^\infty A_k(E_r)h^k\ln h ]  $$
and we will write the exponential as $1 + O(E_r) h\ln h + O(h^2\ln^2 h)$. Here we are obviously facing an $E_r$-dependent expansion in $h$ and $h\ln h$ for $h\to 0$. In all our formulas from here on we will treat $h\ln h$ as dominating $h$ and treat $h\ln^m h$ as dominated by $1$ for $h\to 0$. 

Further, $$ \frac{1}{\Gamma(1+t)} \ = \ % \frac{1}{t\Gamma(t)} \ = \ \frac{1}{t} (t + \underline{\gamma} t^2 + O(t^3)) = 
1 + \underline{\gamma} t + O(t^2), \ \ \text{as } \ t\to 0, $$
where $\underline{\gamma} = 0.5772...$ is the Euler-Mascheroni constant, hence the equality of power series in $h$ with coefficients analytic functions of $E_r$:
$$ \frac{1}{\Gamma(-s_{\gamma_1})} = \sum_{k=0}^\infty B_k(E_r) h^k, $$
$$ B_0(E_r) = 1 + \underline{\gamma} \frac{E_r}{2f''(q_1)} + O(E_r^2), \ \ \  B_1(E_r) = O(E_r). $$

Collecting all our calculations, we have:

%we get:
%$$ c'_1 |_{E=hE_r} \ = \ \frac{\sqrt{2\pi} h^{-\frac{E_r}{2f''(q_1)}-\frac{1}{2}+ O(E_r h)+O(h^2) } }{\Gamma(\frac{E_r}{2f''(q)}+1 + O(hE_r)+O(h^2)) } \exp \left( \frac{1}{h}\Theta_{-1} + \Theta_0 + h\Theta_{1} + O(h^2) \right)_{E=hE_r} \ = \ $$
%$$ \ = \ \frac{\sqrt{2\pi} h^{-\frac{E_r}{2f''(q_1)}-\frac{1}{2} } }{\Gamma(\frac{E_r}{2f''(q)}+1 ) } \exp \left( \frac{1}{h}\Theta_{-1} + \Theta_0 + h\Theta_{1}  \right)_{E=hE_r}(1+O(h^2)+O(E_r h\log h)+ O(E_r^2 h) + O(E_r^2))$$ % the last term of this formula used to be $O(E_r^3)$, I judged it was a typo and corrected it

\begin{Prop} \label{c1calculated} We have 
  \small
$$ c'_1(hE_r,h) = -i\sqrt{2\pi} h^{-\frac{E_r}{2f''(q_1)}-\frac{1}{2} } (1+\underline{\gamma}\frac{E_r}{2f''(q_1)})  \times $$
$$\times \exp \left[ \frac{2[f(q_1)-f(q_1-\varepsilon)]}{h} \ + 
 (1+\frac{E_r}{f''(q_1)})\ln \frac{2[-f'(q_1-\varepsilon)]}{\sqrt{2f''(q_1)}}   -  E_r\sum_{j=1}^\infty a_j   \frac{1}{j} [f'(q_1-\varepsilon)]^j    + \right. $$ 
$$ \ \ \ \ \ \ \ \left. +  h\left( -\frac{b_2}{12}-\frac{1}{2}\frac{f''(q_1-\varepsilon)}{(f'(q_1-\varepsilon))^2} + \frac{a_2(f''(q_1))^2}{24}\right) \right] $$ 
\begin{equation} \times ( [1+O(E_r^2)]+O(E_r) h\log h + O(E_r) h + O(h^2\log^2 h)), \label{mar4f3} \end{equation} % the last term of this formula used to be $O(E_r^3)$, I judged it was a typo and corrected it
\normalsize 
where $a_j=a_j(q_1)$, $b_2=b_2(q_1)$ were defined in section \ref{TaylorSer}.
\end{Prop}

% see RDRW2.Appx57.tex for the same thing calculated through the energy shift

Here the error term $ [1+O(E_r^2)]+O(E_r) h\log h + O(E_r) h + O(h^2\log^2 h)$ stands for an expansion $\sum_{k,\ell\ge 0} \BBB_{k,\ell}(E_r) h^k \log^\ell h$ with $\BBB_{00}(E_r)=1+O(E_r^2)$ for $E_r\to 0$, $\BBB_{11}=O(E_r)$, $\BBB_{10}=O(E_r)$, and for all other nonzero $\BBB_{k,\ell}$ either $k\ge 3$ or $k=2$ and $\ell=0,1,2$. 

%%%%%%%%%%%%%%%%%%%%%%%%%%%%%%%%%%%%%%%%%%%%%%%%%%%%%%%%%%%%%%%%%%%%%%%%%%%%%%%%%%%%%%

\subsection{Exact matching method around $q_2$.} \label{ExactMatchingq2}

%In this subsection, two formal solutions $\psi_+(q,h)$ and $\psi_-(q,h)$ of (\ref{WLcsb}) correspond to the first and to the second sheets of the Riemann surface of $p(q)$ and are normalized in such a way that $\psi_+(q_2-\varepsilon)=\psi_-(q_2-\varepsilon)=1$, where $\varepsilon$ is a   small complex number, $\arg \varepsilon \in (0,\frac{\pi}{2})$. 

%\begin{figure}[h]\includegraphics{RDRW2p35.pdf} \caption{Notation in the exact matching method around $q_2$} \label{nuthesisp15}
%\end{figure}

Let $0<E<<1$. Consider the following two formal solutions $\psi_+(E,q,h)$ and $\psi_-(E,q,h)$ of \eqref{WLcsb}  corresponding to the first and to the second sheets of the Riemann surface of $p(q)$
\begin{equation} \psi_{\pm}(E,q,h) =  \exp\{ \int_{q_2-\varepsilon}^q \frac{i}{h} \sqrt{E-[f'(q')]^2} + \sum_{j=0}^\infty y_j(q')dq' \} \label{mar4f4} \end{equation}
where $\varepsilon$ is a small complex number, $\arg \varepsilon \in [0,\frac{\pi}{2}]$.

An actual resurgent function solution $\phi(E,q,h)$ of $P\psi=E\psi$, $0<E<<1$, whose exponential asymptotic representation on the interval $(q_{1}^+,q_2^-)$ is given by $\phi(E,q,h)\sim \psi_-(E,q,h)$, has on $(q_2^+, q_3^-)$ an asymptotic representation 
\begin{equation} \phi(E,q,h) \ \sim \ c_2(E,h) \psi_+(E,q,h) + \psi_-(E,q,h) \label{mar4f5} \end{equation}
where 
\begin{equation} c_2(E,h) \ = \ \exp ( 2\pi i s_{\sigma'_2} ), \label{c2E} \end{equation} 
cf. fig.\ref{Paper4p4} for the definition of $\sigma'_2$ and sec.\ref{sigmaPEven} for the calculation of $s_{\sigma'_2}$.

%For $E$ a positive real number, the actual solution represented by $\psi_-$ in $R$ is represented by $\psi_- +c_2\psi_+$ in $R''$, where  $c_2\psi_+$ corresponds to the analytic continuation of $\psi_-$ along a contour $\sigma'_2$  with base point $q_2-\varepsilon$  around the simple turning point $q_2^-(E)$, figure \ref{nuthesisp15}.  Thus for $E=hE_r$ the connection coefficient $c_2$ is the limit of the formal monodromy along the path $\sigma'_2$. In other words,  the Stokes phenomenon transforms $\psi_-(q,E)$ into 

Introduce $c_2^{red}(E,h)$ by the equation
\begin{equation} c_2(E,h)  % \ = \ $$ $$ 
\ = \ \sqrt{2\pi} \frac{h^{s_{\gamma'_2} + \frac{1}{2}} }{\Gamma(-s_{\gamma'_2})} c_2^{red}(E,h). \label{c2redDefd} \end{equation}
It is believed that $c_2^{red}(E,h)$ analytically depends on $E$ in the sense that
$$ c_2^{red}(E,h) = \exp\{ \sum_{j=-1}^\infty h^j \Theta_j(E) \} $$
and $\Theta_j(E)$ are analytic functions of $E$ in the neighborhood of the origin. 

%in the same sense as $(c'_1)^{red}(E,h)$ from the previous section.

%
%where 
%$$ c_2 = \exp\left[ \int_{\sigma'_2} dq \left\{ \frac{1}{h} S(q,E) + y_0(q) + hy_1(q) + O(h^2) \right\}  \right] $$
%and where $c^{red}_2$ analytically and regularly depends on $E$. With the same caveat as in the previous subsection, we will substitute $E_r h$ for $E$ and obtain
%$$ c_2(hE_r) \psi_+(q,hE_r) \ + \  \psi_-(q,hE_r) \ = \ \sqrt{2\pi} \left.\frac{h^{s_{\gamma'_2}+\frac{1}{2}}}{\Gamma(-s_{\gamma'_2})}\right|_{E=hE_r}  c_2^{red}(hE_r)\psi_+(q,hE_r) \ + \  \psi_-(q,hE_r).  $$ 

%Together with calculations of the monodromy of the formal solution along the contour $\sigma'_2$, we will use the following asymptotic expansion valid for a fixed $E>0$ and derived from the Stirling formula:
The Stirling formula \eqref{NextTermStirling1} and \eqref{ggprim} give: 
\begin{equation} \frac{\sqrt{2\pi}h^{s_{\gamma'_2}+\frac{1}{2}}}{\Gamma(-s_{\gamma'_2})} \ \sim \ 
\exp\left\{ \frac{1}{h}\left[ \frac{\omega_{\gamma_2}}{2\pi} - \frac{\omega_{\gamma_2}}{2\pi} \Ln \left( \frac{\omega_{\gamma_2}}{2\pi}\right) \right] - \frac{1}{2}\Ln \left(  \frac{\omega_{\gamma_2}}{2\pi} \right) - h\frac{\pi}{6\omega_{\gamma_2}} +O(E\log E) h + O(h^2)\right\}. \label{ma23e1} \end{equation}
% longer exposition in RDRW2.Appx135.tex

%Introduce $\Theta_j$'s for $j\ge -1$ by formula
%$$  c_2^{red}(E,h) \ = \ \exp \{ \frac{1}{h}\Theta_{-1}(E) + \Theta_0(E) + h\Theta_1(E) + ... \} $$
%and so, 
Similarly to the previous subsection, 
with $A=-f'(q_2-\varepsilon)$, $a_j=a_j(q_2)$, $b_j=b_j(q_2)$, we have: %using \eqref{feb24f2} 
$$ \Theta_{-1} \ = \ i\int_{\sigma'_2} p(q,E) dq \ - \ \frac{\omega_{\gamma_2}}{2\pi} + \frac{\omega_{\gamma_2}}{2\pi} \Log \left( \frac{\omega_{\gamma_2}}{2\pi} \right) $$
$$  \ = \ -2[f(q_2)-f(q_2-\varepsilon)]+E a_0   \Log (-2A) -a_0\frac{E^2}{8A^2} + \left( \frac{Ea_0}{2} + \frac{E^2 a_2}{8} \right) \Log \frac{a_0}{2} $$
$$  - a_2\frac{A^2 E}{2} + a_2\frac{E^2}{16}  - a_2  \frac{E^2}{4}\Ln (-2A)  
 + E \sum_{j\ge 1, \ j\ne 2}^\infty a_j (-1)^j  \left( \frac{1}{j}A^j + \frac{1}{4(j-2)}A^{j-2}E \right) 
    + o(E^2); $$
$$ \Theta_0 \ = \ \int_{\sigma'_2} y_0(q,E) dq \ - \ \frac{1}{2}  \Log \left( \frac{\omega_{\gamma_2}}{2\pi} \right) \ = \ $$
$$ \ = \ -\Ln i   -  \Ln \frac{(-2A)}{\sqrt{2|f''(q_2)|}}  + \frac{E}{4A^2}  +  \frac{a_2 |f''(q_2)|\cdot E}{8}  + o(E);$$
$$ \Theta_1 \ = \ \int_{\sigma'_2} y_1(q,E) dq \ + \  \frac{\pi}{6\omega_{\gamma_2}} + o(E^0), $$
$$ \Theta_1 \ = \   \frac{1}{12} b_2  -  \frac{1}{2} \sum_{j=0}^\infty b_{j}  A^{j-2}  \ - \   \frac{a_2 [f''(q_2)]^2 }{24} + O(E^0) .$$

%=====================

% for $E=hE_r$, %cf detailed calculation in RDRW2.Appx45.tex

%Now we are almost ready combine these formulae  and calculate
%$$ c_2(hE_r) \ = \ \sqrt{2\pi} \left.\frac{h^{s_{\gamma'_2}+\frac{1}{2}}}{\Gamma(-s_{\gamma'_2})}\right|_{E=hE_r}  c_2^{red}(hE_r). $$ 
 % This implies that $1/\Gamma(-s_{\gamma'_2})$ is divisible by $E_r$. 
Using $$s_{\gamma'_2}(hE_r,h) = \frac{E_r}{2f''(q_2)}  + hO(E_r)+ O(h^2), $$
we can write
$$ \frac{1}{\Gamma(-s_{\gamma'_2})} \ = \ -\frac{E_r}{2f''(q_2)} + \underline{\gamma}\frac{E^2_r}{[2f''(q_2)]^2} + O(E_r^3) + h\times O(E_r)+O(h^2). $$

We combine the results as follows:

\begin{Prop} \label{c2calculated} We have  \small
$$ c_2(hE_r,h) =  -i\sqrt{2\pi} h^{ \frac{E_r}{2f''(q_2)} + \frac{1}{2}  }  (-\frac{E_r}{2f''(q_2)} + \underline{\gamma}\frac{E^2_r}{[2f''(q_2)]^2} + O(E_r^3) + hO(E_r)+O(h^2) ) \times $$
$$ \times \exp\left[ 2\frac{-f(q_2)+f(q_2-\varepsilon)}{h} - (1+\frac{E_r}{f''(q_2)})   \log \frac{2f'(q_2-\varepsilon)
}{\sqrt{-2f''(q_2)}}
 + E_r\sum_{j=1}^\infty a_j   \frac{1}{j}[f'(q_2-\varepsilon)]^j  \right. \ + \   $$
$$ \ + \  \left. h\left( \frac{1}{12} b_2  +  \frac{1}{2} \frac{f''(q_2-\varepsilon)}{(f'(q_2-\varepsilon))^2}   \ - \   \frac{a_2 [f''(q_2)]^2 }{24} + o(E^0) \right) + O(h^2) \right] \times $$ \begin{equation} \times ([1+O(E_r^2)]+O(E_r)h\ln h + O(E_r)h+ O(h^2\ln^2 h)), \label{mar4f6} \end{equation}
\normalsize
where $a_j=a_j(q_2)$, $b_2=b_2(q_2)$ were defined in section \ref{TaylorSer}.
\end{Prop}

{\bf Supplement.} Notice that if $E_r=0$, then $e^{-\frac{f(q)}{h}}$ is the formal solution of \eqref{WLcsb} corresponding to the second sheet determination of $p(q)$. But $e^{-\frac{f(q)}{h}}$ is also the actual solution of \eqref{WLcsb}, and therefore the connection coefficient vanishes: $$c_2(0,h)=0.$$ 

As a side note, the Supplement allows to collect the error estimate of Prof.\ref{c2calculated} similarly to how we did it in Prop.\ref{c1calculated}.

%cf RDRW2.Appx58.tex for the energy-shifted version of this calculation

%%%%%%%%%%%%%%%%%%%%%%%%%%%%%%%%%%%%%%

\subsection{An application of the Stirling formula} \label{AppliStir}

%the version relevant to the shifted energy Schroedinger can be found in RDRW2.Appx62.tex

In this subsection \ref{AppliStir} we will derive the formula \eqref{NextTermStirling1} which we used in \eqref{fe29e1} and \eqref{ma23e1}.

Let us first work with a fixed $0<E<<1$. 
We have calculated earlier that $s_{\gamma_1}=\frac{\omega_{\gamma_1}}{2\pi h}-1+\beta_1 h$, where $\beta_1=\beta_{1,1} + O(h)$ is a power series expansion in $h$.  and $\beta_{1,1}=O(E)$. Since $-s_{\gamma_1}$ has a positive real part which goes to infinity as $h\to 0+$, we can apply the Stirling formula to $\Gamma(-s_{\gamma_1})$ to get
 $$\frac{\sqrt{2\pi} h^{s_{\gamma_1}+\frac{1}{2}}}{\Gamma(-s_{\gamma_1})} \ \sim \  \frac{\sqrt{2\pi}h^{s_{\gamma_1}+\frac{1}{2}}}{(-s_{\gamma_1})^{-s_{\gamma_1}-\frac{1}{2}}\cdot e^{s_{\gamma_1}} \cdot \sqrt{2\pi}\cdot (1-\frac{1}{12}s_{\gamma_1}^{-1}+O(h^2))} .$$ 
%$$ \ = \  
%\frac{h^{s_\gamma+\frac{1}{2}}}{(-s_\gamma)^{-s_\gamma-\frac{1}{2}}e^{s_\gamma}(1-\frac{1}{12}s_\gamma^{-1}+O(h^2))}  
%\ = \ 
%\frac{e^{-s_\gamma}}{\left(-\frac{\omega_\gamma}{2\pi}+( h-\beta h^2)\right)^{-\frac{\omega_\gamma}{2\pi h}
%+\frac{1}{2}-\beta h} (1-\frac{1}{12}s_\gamma^{-1}+O(h^2)) } 
%\ = \ $$
%$$ \ = \  
%\frac{e^{-s_\gamma}}{\left(-\frac{\omega_\gamma}{2\pi} \right)^{-\frac{\omega_\gamma}{2\pi h}+\frac{1}{2}-\beta h}  \left(1-\frac{2\pi h (1 -\beta h)}{\omega_\gamma}\right)^{-\frac{\omega_\gamma}{2\pi h}+\frac{1}{2}-\beta h} (1-\frac{1}{12}s_\gamma^{-1}+O(h^2)) } 
%\ = \ $$ 
%$$\frac{e^{-\frac{\omega_\gamma}{2\pi h} + 1 -\beta h} }{\left(-\frac{\omega_\gamma}{2\pi} \right)^{-\frac{\omega_\gamma}{2\pi h}+\frac{1}{2}-\beta h}  \exp\left\{ 1 - \beta h  \right\}  (1-\frac{1}{12}s_\gamma^{-1}) }(1+O(h^2)) $$
A few routine steps of simplification bring us to
\begin{equation} \frac{\sqrt{2\pi} h^{s_{\gamma_1}+\frac{1}{2}}}{\Gamma(-s_{\gamma_1})} \ \sim \  
\frac{\exp\{-\frac{\omega_{\gamma_1}}{2\pi h}\}}{\left(-\frac{\omega_{\gamma_1}}{2\pi} \right)^{-\frac{\omega_{\gamma_1}}{2\pi h}+\frac{1}{2}-\beta_1 h}   (1-\frac{\pi h}{6\omega_{\gamma_1}} )}(1+O(h^2)).  
 \label{NextTermStirling} \end{equation}
In \eqref{NextTermStirling}, the LHS is a true function, and the RHS its asymptotic expansion valid for $E>0$ and for $h$ in a small sectorial neighborhood of zero in the positive real direction.

Now let us remember the $E$-dependence in \eqref{NextTermStirling}. Then $(1+O(h^2))$ on the RHS should be interpreted as a power series $1+ \sum_{k=2}^\infty \Gbb_k(E)h^k$ where $\Gbb_k(E)$ are analytic functions of $E$ for $0<E<<1$.

From sec.\ref{gammamon} we know that $\beta_1 =\beta_1(E,h)= \sum_{k=0}^\infty \BBB_k(E) h^k$ with $\BBB_0(E)=O(E)$. We can drop $\beta_1$ from the RHS of \eqref{NextTermStirling} at the expense of changing the error term as follows:
\begin{equation} \frac{\sqrt{2\pi} h^{s_{\gamma_1}+\frac{1}{2}}}{\Gamma(-s_{\gamma_1})} \ \sim \  
 \frac{\exp\{-\frac{\omega_{\gamma_1}}{2\pi h}\}}{\left(-\frac{\omega_{\gamma_1}}{2\pi} \right)^{-\frac{\omega_{\gamma_1}}{2\pi h}+\frac{1}{2}}   (1-\frac{\pi h}{6\omega_{\gamma_1}} )}(1+O(h^2)+O(h)O(E\log E)), 
 \label{NextTermStirling1} \end{equation}
where $1+O(h^2)+O(h)O(E\log E)$ stands for an $E$-dependent power series $1+ \sum_{k=1}^\infty \Gbb'_k(E)h^k$ where $\Gbb'_k(E)$ are analytic functions of $E$ for $0<E<<1$ and $\Gbb'_1(E)=O(E\log E)$ when $E\to 0+$. 

\subsection{Partial proof of analyticity of $(c'_1)_{red}$} \label{SomeProofs}

The purpose of this subsection \ref{SomeProofs} is not to give a full justification of the methods, but rather to prove some results confirming that our approach is consistent and makes sense. We are going to show that $\Theta_{-1}(E)$, $\Theta_0(E)$, $\Theta_1(E)$ from \eqref{credTheta} are indeed analytic functions of $E$.

\begin{Lemma} The quantity
$$  \Theta_{-1} \ = \ \Delta S + \frac{\omega_{\gamma_1}}{2\pi} - \frac{\omega_{\gamma_1}}{2\pi} \log[-\frac{\omega_{\gamma_1}}{2\pi}] $$
is analytic with respect to $E$ in the neighborhood of zero.  \label{SrIsAnalytic} \end{Lemma}

\textsc{Proof.} We need to show that the term containing $\log E$ in $\Delta S$ cancels $\frac{\omega_{\gamma_1}}{2\pi}\Log E$. Indeed, we have seen that 
$$  \Delta S = -2 \sum_{k=0}^\infty (-1)^k a_k \int_A^{\sqrt{E}} u^k \sqrt{u^2-E} du = -2 \sum_{k=0}^\infty (-1)^k a_k \int_{\arccosh A/\sqrt{E}}^0 E^{\frac{k}{2}+1} \cosh^k t \sinh^2 t dt$$
$$ \ = \  -2 \sum_{k=0}^\infty (-1)^k a_k \int_{\arccosh A/\sqrt{E}}^0 2^{-k-2} E^{\frac{k}{2}+1} (e^t+e^{-t})^k (e^t-e^{-t})^2 dt.$$
% paranoidal sign verification in RDRW2.Appx33.tex

Writing the integrand as the sum of exponents and integrating, we realize that only the summand $e^{0t}dt$ will eventually give rise to a logarithmic singularity for $E\to 0$. Writing $reg(E)$ for an arbitrary function that is analytic with respect to $E$ near the origin, we have:
$$ \Theta_{-1} \ = \ reg(E) \ - \ 2 \sum_{k=2j=0}^\infty  a_{2j} \int_{\arccosh A/\sqrt{E}}^0 2^{-2j-2} E^{j+1} (C^{j-1}_{2j} -2C^{j}_{2j}+C^{j+1}_{2j})  dt  $$
%$$ \ = \ reg(E) \ + \ 2 \sum_{k=2j=0}^\infty  a_{2j} [\arccosh \frac{A}{\sqrt{E}}] 2^{-2j-2} E^{j+1} (-2) (\frac{(2j)!}{j!j!}-\frac{(2j)!}{(j-1)!(j+1)!}) \ = \  $$ 
$$ \ = \ reg(E) \ - \  \sum_{k=2j=0}^\infty  a_{2j}   2^{-2j} E^{j+1} \frac{(2j)!}{j!j!}(1-\frac{j}{j+1}) \arccosh (\frac{A}{\sqrt{E}})\ = \    $$
$$ \ = \ % reg(E) \ - \  \sum_{k=2j=0}^\infty  a_{2j} \arccosh (\frac{A}{\sqrt{E}}) 2^{-2j} E^{j+1} \frac{(2j-1)!! 2^j j!}{j!j!} \frac{1}{(j+1)} \ = \ 
reg(E) \ + \ \sum_{k=2j=0}^\infty  a_{2j}  \frac{(2j-1)!!}{(2j+2)!!}  E^{j+1} \Log E . $$
The singularity that comes out of $\frac{\omega_{\gamma_1}}{2\pi} \Log E$ is the same by formula \eqref{Oct6}. $\Box$

% ----------------------

\begin{Lemma} $\Theta_0$ is analytic for $E$ around $0$. \label{Theta0Analytic} \end{Lemma}

\textsc{Proof} is obvious from \eqref{ArccoshAs} and proposition \ref{sgamma1calcd} $\Box$

% -----------------------

\begin{Lemma} $\Theta_1$ is analytic for $E$ around $0$. \label{Theta1Analytic} \end{Lemma}

\textsc{Proof.} The fact that $\Theta_1$ has no pole (i.e. $\frac{1}{E}$) singularity has been demonstrated in section \ref{ExactMatchingq1}. Now let us check that all logarithmic singularities $E^k \Ln E$ are absent in $\Theta_1$ as well. The question reduced to identifying the logarithmic singularity in the integral along the contour $\sigma$, fig.\ref{RDRW2p4}:
$$ \int_\sigma  \frac{\sum_{j=0}^\infty (j+1)[-jb_{j+2}] u^{j}}{24\sqrt{u^2-E}} dq \ = \ $$
$$ \ = \ \frac{1}{12} \sum_{j=0}^\infty E^{\frac{j}{2}}(j+1) [ - j b_{j+2}] \int_{\arccosh(A/\sqrt{E})}^0    \cosh^j(t) dt \ = \   $$
%Its singularity is the same as in 
$$ reg(E) \ + \  \frac{1}{12} \sum_{j=2k\ge 0} E^{\frac{j}{2}}(j+1) [ - j b_{j+2}]     
2^{-j} C^{(j/2)}_j (-\arccosh(A/\sqrt{E}))  $$
%the same as in  $$ \frac{1}{12} \sum_{k\ge 0} E^{k}(2k+1) [ - (2k) b_{2k+2}] 2^{-k} \frac{(2k)!}{2^k k!k!} \frac{1}{2}\Ln E  \ = \ $$
$$ \ = \ reg(E) \ - \ \frac{1}{24} \sum_{k\ge 0} E^{k}  b_{2k+2}   \frac{(2k+1)!!}{(2k-2)!!} \Ln E . $$

The logarithmic singularity in the $h^1$ term of $\Ln \frac{\sqrt{2\pi}h^{s_\gamma+\frac{1}{2}}}{\Gamma(-s_\gamma)}$ comes from $-\beta\Ln \left(-\frac{\omega_{\gamma_1}}{2\pi}\right)$, i.e. from $-\beta\Ln E$, where 
$$ \beta = \frac{1}{2\pi} \int_{\gamma_1} y_1 dq = \frac{1}{24} \sum_{k \ge 2}^\infty b_{2k} E^{k-1} \frac{(2k-1)!!}{(2k-4)!!}.  $$
That means that $E^k \Ln E$ terms in $\Theta_1$ cancel for all $k$. $\Box$
% old proof in RDRW2.Appx113.tex

\subsection{Remarks on more precise calculation of $c'_1(hE_r,h)$, $c_2(hE_r,h)$.} \label{FiniteTime}

Let us discuss what needs to be done to calculate more terms of $c'_1(hE_r,h)$, both with respect to $h$ and with respect to $E_r$; similar ideas apply to $c_2(hE_r,h)$.

In order to determine $c'(hE_r,h)$, we had to perform three pieces of somewhat laborious calculations -- computing $s_{\sigma_1}(E,h)$, $s_{\gamma_1}(E,h)$, and the expansion \ref{NextTermStirling}; all other steps were more or less substitution and cancelation of terms.  An analog of \eqref{NextTermStirling} is available by the Stirling formula to any order in $h$. Computations of $s_{\sigma_1}(E,h)$ and $s_{\gamma_1}(E,h)$ are parallel to each other, so let us discuss the former. 

The $j$-th term $y_j$ of the series \eqref{yjdefd} solving the equation \eqref{mar4eq7} is a linear combination of fractions of the type $\frac{f^{(\iota_1)}(q)...f^{(\iota_\ell)}(q) }{[E-(f'(q))^2]^{k/2}}$ where $f^{(\iota_\nu)}(q)$ denotes the $\iota_\nu$-th derivative of $f(q)$, and $k\in \N$. 

Similarly to \eqref{DDfB}, we can write $f^{(\iota_\nu)}(q)$ as a power series in $f'(q)$. Thus, the integral $\Delta y_j$, cf. \eqref{DeltaSDeltay}, can be transformed by substitution $u=-f'(q)$ into an integral of the type $\int_{\sigma} \frac{\sum_{m=1}^\infty C_m u^m}{(E-u^2)^{k/2}} du$, where $\sigma$ is the contour on fig.\ref{RDRW2p4}. If $k$ is even, this integral can be taken by the residue formula. If $k$ is odd, this integral can be calculated using several steps of integration by parts to reduce the denominator to $\sqrt{u^2-E}$ and then using substitution $u=\sqrt{E}\cosh t$, similarly to sec.\ref{DeltaY1Q1}. In terms of the coefficients $C_m$, the formula for $\Delta y_j(E)$ can be written precisely as a function of $E$.

% \newpage

\section{Ingredients of the quantization condition for the equation $P\psi=hE_r\psi$.}
\label{ingredients}

The purpose of this section \ref{ingredients} is to calculate the ingredients of the quantization condition mentioned in Step 3 of sec.\ref{SecOutline}.

\subsection{Calculation of $\mu_j(hE_r,h)$} \label{mujss}

Define: %In the sequel we will also need the following microfunctions $\mu_j$ that will be by abuse of language identified with their asymptotic expansions with respect to $h$:
$$ \mu_j(E,h) \ = \ 1 - e^{2\pi i s_{\gamma_j}(E,h)}, \ \text{$j$ odd}; \ \ \ \mu_j \ = \ 1+e^{-2\pi i s_{\gamma_j}(E,h)}, \text{$j$ even} $$
for $E\in \C$, $|E|$ sufficiently small; $\mu_j(hE_r,h)$ is then defined by substitution of $hE_r$ in place of $E$.

% {\bf Calculation of $\mu_j$} ---> RDRW2.Appx134.tex

Using Prop.\ref{sgamma1calcd} and performing routine simplifications, we obtain  %for odd $j$, % verified with Mathematica RDRW2.Math11; full version and old formulas cf. RDRW2.Appx67.tex
\begin{equation} \mu_j(hE_r,h)  \ = \ \frac{E_r \pi i }{f''(q_j)}  (1 - \frac{a_2(q_j)f''(q_j)}{4} hE_r  - \frac{\pi i}{2f''(q_j)} E_r )([1+O(E_r^2)]+O(h^2)), \ \ \ j  \ \text{odd}, \label{mu1} \end{equation}
%and for even $j$
\begin{equation} \mu_j(hE_r,h)  \ = \ \frac{E_r \pi i }{|f''(q_j)|}  (1 + \frac{a_2(q_j)|f''(q_j)|}{4} hE_r  - \frac{\pi i}{2|f''(q_j)|} E_r )([1+O(E_r^2)]+O(h^2)) \ \ \ j  \ \text{even} \label{mu2} \end{equation}
(in the sense that the $h^0$ term of the power series with respect to $h$ is given up to $O(E_r^2)$ for $E_r\to 0$, $h^1$ term is given precisely, and $h^2$ and smaller terms are neglected).

\subsection{Calculation of $\tau_j(hE_r,h)$} \label{taujss}

We define for $E\in \C$ 
$$ \text{for odd $j$ : } \  \tau_j \ = \ \tau_j(E,h) \ = \ c'_j c_{j+1} M^{-1}_j M'_j, $$
$$ \text{for even $j$ : } \ \tau_j \ = \ \tau_j(E,h) \ = \ c_j c'_{j+1} M_j  (M'_j)^{-1}, $$
and the definition is then extendable to $\tau_j(hE_r,h)$ by means of Props.\ref{c1calculated} and \ref{c2calculated}. 

Let us now evaluate two representative cases of $\tau_j$

{\bf Calculation of $\tau_1$.} Using lemmas \ref{LemmaMj}, \ref{LemmaMPrimj} and propositions \ref{c1calculated}, \ref{c2calculated}, and inserting $E=hE_r$ into the corresponding formulas, obtain after routine simplification:
% cf RDRW2.Appx96.tex for what I hope is a correct calculation
$$\tau_1(hE_r,h)  \ = \ e^{2\frac{f(q_1)-f(q_2)}{h}}  \frac{E_r \pi }{\sqrt{|f''(q_2)|\cdot f''(q_1)}} h^{-\frac{E_r}{2f''(q_1)} + \frac{E_r}{2f''(q_2)}   } (1+\underline{\gamma}\frac{E_r}{2f''(q_1)} +  \underline{\gamma}\frac{E_r}{2|f''(q_2)|})  \times $$
$$\times \exp \left[  \frac{E_r}{f''(q_1)}\log \frac{2(-f'(q_1-\varepsilon))}{\sqrt{2f''(q_1)}}   -  E_r\sum_{j=1}^\infty a_j(q_1)    \frac{1}{j} (f'(q_1-\varepsilon))^j  + h\left(   -\frac{b_2(q_1)}{12} + \frac{a_2(q_1)(f''(q_1))^2}{24}   \right)   \right] $$  
$$ \times \exp\left[ - \frac{E_r}{f''(q_2)}  \log \frac{2f'(q_2-\varepsilon)}{\sqrt{-2f''(q_2)}}
 + E_r\sum_{j=1}^\infty a_j(q_2)    \frac{1}{j}(f'(q_2-\varepsilon))^j + h\left( \frac{b_2(q_2)}{12}   -    \frac{a_2(q_2) [f''(q_2)]^2 }{24} \right)  \right] \times  $$
 $$ \times \exp\left[ E_r  \int_{(q_1-\varepsilon)_I}^{(q_2-\varepsilon)_I} \frac{dq}{f'(q)} \ + \ \frac{\pi i E_r}{ f''(q_1)}  \right] ([1+ O(E_r^2)]+ O(E_r) h\log h + O(E_r)h +O(h^2\ln^2 h)).$$
Here $\int_{(q_1-\varepsilon)_I}^{(q_2-\varepsilon)_I}$ denotes the integral along a path lying within the domain of definition of $\phi_+$, fig.\ref{Paper4p2}. Formulas for other $\tau_j$ with odd $j$ are analogous.

% old calculation of $\tau_1$ is contained in RDRW2.Appx68.tex

{\bf Calculation of $\tau_2$.} Analogously, % cf RDRW2.Appx72.tex for details,
$$ \tau_2(hE_r,h) = e^{ 2\frac{f(q_3)-f(q_2)}{h} } \frac{E_r \pi}{\sqrt{ |f''(q_2)| \cdot f''(q_3)}}  h^{ \frac{E_r}{2f''(q_2)}  -\frac{E_r}{2f''(q_3)} } 
  ( 1 +  \underline{\gamma}\frac{E_r}{2|f''(q_2)|} +\underline{\gamma}\frac{E_r}{2f''(q_3)}) \times $$
$$ \times \exp\left[  - \frac{E_r}{f''(q_2)}   \log \frac{2f'(q_2-\varepsilon)}{\sqrt{-2f''(q_2)}}
 + E_r\sum_{j=1}^\infty a_j(q_2) \frac{1}{j}(f'(q_2-\varepsilon))^j  + h\left( \frac{1}{12} b_2(q_2)    \ - \   \frac{a_2(q_2) [f''(q_2)]^2 }{24}  \right) \right] \times $$
$$   \exp \left[ 
 \frac{E_r}{f''(q_3)}\log \frac{-2f'(q_3-\varepsilon)}{\sqrt{2f''(q_3)}}   -  E_r \sum_{j=1}^\infty a_j(q_3)    \frac{1}{j} (f'(q_3-\varepsilon))^j     +  h\left( -\frac{b_2(q_3)}{12}+ \frac{a_2(q_3)(f''(q_3))^2}{24}\right) \right] $$ 
$$ \times  \exp \left\{ - E_r\int_{(q_2-\varepsilon)_I}^{(q_3-\varepsilon)_I} \frac{dq}{f'(q)}  \ - \ \frac{\pi i E_r}{f''(q_2)} \right\} ([1+ O(E_r^2)]+ O(E_r) h\log h + O(E_r)h +O(h^2\ln^2 h)),$$
and similarly for other $\tau_j$ with even $j$. 

In a calculation of these monodromies for a specific $f(q)$ we can use formula \eqref{SumAj}.

\subsection{Formula for the product of $M'_j$s}

Let us calculate the expression that we called $1+E_r\kappa(E_r,h)$ in ~\cite[Lemma 8.2]{G}:
$$ 1+ E_r \kappa(E_r,h) \ \stackrel{\rm def}{=} \ M'_1(hE_r,h) M'_2(hE_r,h)...M'_{2n}(hE_r,h) \ = \ $$
(by lemma \ref{LemmaMPrimj})
%$$ M'_1 M'_2 ... M'_{2n} \ = \ \exp \left( \frac{E}{2h}\oint_{\text{2nd sheet}} \frac{dq}{f'(q)} \right) (1+O(E^2/h)+O(E^2)+O(Eh)+O(h^2)), $$
%or
\begin{equation} % M'_1 M'_2 ... M'_{2n} |_{E=hE_r} 
\ = \ \exp \left( \frac{E_r}{2}\oint_{\text{2nd sheet}} \frac{dq}{f'(q)} \right) (1+hO(E_r^2)+O(h^2)), \label{coeffk} \end{equation}
where $\oint_{\text{2nd sheet}}$ means that the integral is taken along a loop from some $q_0$ to $q_0+1$ along a loop lying within the domain of definition of $\phi_-$.
%\end{Cor}

%\begin{equation}  M'_1 M'_2 ... M'_{2n} |_{E=hE_r} \ = \ 1 +  \frac{E_r}{2}\oint_{\text{2nd sheet}} \frac{dq}{f'(q)}  + O(E_r)^2 + O(E_r^2 h)+O(h^2) \label{jan20} \end{equation}

%\subsection{One more formula}
%{\bf Do we need it? - rewrite this subsection anyway}
%
%\begin{Cor} We have 
%$$ \left. \prod_{j=1}^{2n} M_j^{-1} M'_j \right|_{E=hE_r} \ = \ \exp \left\{  \sum_{j=1}^{2n} \frac{\pi i E_r}{|f''(q_j)|} 
%+ E_r \oint_{\text{1st sheet}} \frac{dq}{f'(q)} \right\} (1+O(E_r^2 h)+O(h^2)),$$
%where $\oint_{\text{1st sheet}}$ means that the integral is taken along a loop from some $q_0$ to $q_0+1$ along a loop lying within the domain of definition of $\phi_+$.
%\end{Cor}

%%%%%%%%%%%%%%%%%555%%%%%%%%%%%%%%%%%%%%%%%%%%%%%%%%%%%%%%%%%5

% \newpage

\section{Quantization condition and formulas for eigenfunctions.}
\label{procedure}

In this section we give explicit details and formulas to execute the method of sec.\ref{Method} of calculating low-lying eigenvalues and corresponding eigenfunctions of the Witten Laplacian

\subsection{Quantization condition} \label{QCPs}

The explicit form of the transfer matrix $F(E_r,h)$ from \eqref{transferMtxDefd} was discussed in ~\cite[\S 8]{G}; the quantization condition for the Witten Laplacian will be written in terms of a related $2\times 2$ matrix $G_0=G_0(E_r,h)$ whose coefficients are $E_r$-dependent resurgent symbols in $h$, or, by abuse of notation, $E_r$-dependent exponential asymptotic expansions in $h$ of the form
$$ \sum_j e^{\frac{c_j}{h}} (\sum A_{j,k,\ell}(E_r) h^k \ln^\ell h) $$
 with $A_{j,k,\ell}(E_r)$ analytic functions of $E_r$ and $c_j\in \R$ independent of $E_r$. 

The quantization condition is: 
\begin{equation}   \ker (  G_0 - \frac{1}{1+E_r \kappa(E_r,h)} Id ) \ \ne \{ 0 \} , \label{QCondMtxForm} \end{equation}
or equivalently
%Rewrite the quantization condition \eqref{QCond} as
\begin{equation} 
-\frac{1}{1+E_r \kappa}  + \Tr G_0  - (1+E_r \kappa )\det G_0 \ = \ 0, \label{QCondPrim} \end{equation}
where the coefficient $\kappa(E_r,h)$ was defined and computed in \eqref{coeffk}.
In the case when $f(q)$ has $n$ local minima and $n$ local maxima on the period $q\in [0,1)$, the matrix $G_0=G_0(E_r,h)$  has the following explicit form:  
$$ G_0 \ = \ \prod_{k=n,...,1} \left( \begin{array}{cc} \tau_{2k} &  \\ & 1 \end{array} \right)
\left( \begin{array}{cc} \tau_{2k-1}^{-1} +1 & \mu_{2k-1} \tau_{2k-1}^{-1} +1 \\ \mu_{2k} \tau_{2k-1}^{-1} +1 & \mu_{2k-1} \mu_{2k} \tau_{2k-1}^{-1} + 1 \end{array} \right), $$
where the factors in the product are arranged, from left to right, in the order $k=n,n-1,...,1$.
%$$ \ = \ 
%\left( \begin{array}{cc} \tau_4(\tau_3^{-1} +1) & \tau_4(\mu_3 \tau_3^{-1} +1) \\ \mu_4 \tau_3^{-1} +1 & \mu_3 \mu_4 \tau_3^{-1} + 1 \end{array} \right)  
%\left( \begin{array}{cc} \tau_2(\tau_1^{-1} +1) & \tau_2(\mu_1 \tau_1^{-1} +1) \\ \mu_2 \tau_1^{-1} +1 & \mu_1 \mu_2 \tau_1^{-1} + 1 \end{array} \right)  \ = \ $$
In particular, if $n=2$,
\footnotesize
\begin{equation} G_0 \ = \ 
\left( \begin{array}{cc} \tau_4(\tau_3^{-1} +1)\tau_2(\tau_1^{-1} +1) +  \tau_4(\mu_3 \tau_3^{-1} +1)(\mu_2 \tau_1^{-1} +1) & 
\tau_4(\tau_3^{-1} +1)\tau_2(\mu_1 \tau_1^{-1} +1) + \tau_4(\mu_3 \tau_3^{-1} +1)(\mu_1 \mu_2 \tau_1^{-1} + 1 )
\\
(\mu_4 \tau_3^{-1}+1)\tau_2(\tau_1^{-1} +1) + (\mu_3 \mu_4 \tau_3^{-1} + 1) (\mu_2 \tau_1^{-1}+1)&
(\mu_4 \tau_3^{-1}+1)\tau_2(\mu_1 \tau_1^{-1} +1) + (\mu_3 \mu_4 \tau_3^{-1} + 1)(\mu_1 \mu_2 \tau_1^{-1} + 1) 
\end{array} \right), \label{G0Explicit} \end{equation} \normalsize
where $\mu_j=\mu_j(hE_r,h)$ and $\tau_j=\tau_j(hE_r,h)$ were calculated in sec.\ref{mujss} and \ref{taujss}, respectively.

\subsection{Formulas for the eigenfunctions}

Assume $E_r(h)$ is an exponentially small solution of the quantization condition \eqref{QCondPrim}. %low-lying resurgent eigenvalue of our Witten Laplacian.

Let $\phi_{\pm}(hE_r(h),q,h)$ be formal WKB solutions obtained from $\phi_\pm(hE_r,q,h)$ of sec.\ref{FormalSolSec} by substituting $E_r(h)$ in place of $E_r\in \C$ into the expression of $\phi_{\pm}(hE_r,q,h)$. As $E_r(h)$ has an exponential asymptotic expansion with several, usually countably many term, $E_r(h) \sim \sum_j e^{-\frac{b_j}{h}} a_j(h,h\ln h)$, the $q$-dependent exponential asymptotic expansions of $\phi_\pm(hE_r(h),q,h)$ will be of the same nature.

Let $q_0$ be as in sec.\ref{FormalSolSec}. Denote  $B_0(E,h)=\phi_+(E,q_1-\varepsilon,h)$, $B'_0(E,h)=\phi_-(E,q_1-\varepsilon,h)$.

Suppose a vector of resurgent symbols $(Z^{(0)}_+, Z^{(0)}_-)^T$ belongs to the kernel \eqref{QCondMtxForm}. Then, cf. ~\cite[\S 8.2]{G}, the vector 
$$ \left( \begin{array}{c} D^{(0)}_+ \\ D^{(0)}_- \end{array} \right) \ = \ \left( \begin{array}{c} B'_0 B_0^{-1} (c'_1)^{-1} Z^{(0)}_+ \\ Z^{(0)}_- \end{array} \right) $$
belongs to $\ker (F(E_r,h)-Id) = 0 $  and thus the eigenfunction corresponding to the resurgent eigenvalue $hE_r(h)$ will be representable, for $q\in (0,q_1)$, by an exponential asymptotic expansion
$$ D^{(0)}_+(h)\phi_+ (hE_r(h),q,h) + D^{(0)}_-(h) \phi_-(hE_r(h),q,h).$$

%where $\phi_+$ and $\phi_-$ are formal solutions of the Witten Laplacian normalized by $\phi_+(q_0)=\phi_-(q_0)=1$ for some $q_0\in \C$, $f'(q_0)\ne 0$.

The connection formulas for the equation $P\psi = hE_r\psi$ can be rephrased as follows. For $q\in (q_j,q_{j+1})$, $j=1,...,2n$, $q_{2n+1}:=q_1+1$, the same eigenfunction is representable by the following exponential asymptotic expansion
$$ D^{(j)}_+(h)\phi_+ (hE_r(h),q,h) + D^{(j)}_-(h) \phi_-(hE_r(h),q,h),$$
$$ D^{(j)}_+(h) = Z_+^{(j)}(h) \times \frac{B'_0(hE_r(h),h)}{B_0(hE_r(h),h)} \prod_{\ell=1}^{j-1} \frac{M'_\ell(hE_r(h),h)}{M_\ell(hE_r(h),h)} \times \left\{ \begin{array}{cc} (c'_j(hE_r(h),h))^{-1}, & \text{$j$ odd} \\ c_j(hE_r(h),h) & \text{$j$ even} \end{array} \right. \ \ , $$ $$ D^{(j)}_-(h)  = Z_-^{(j)}(h), $$
and $Z_\pm^{(j)}$ are defined recursively based on $Z_{\pm}^{(0)}$:
\begin{equation} \left( \begin{array}{c} Z^{(j)}_+ \\ Z^{(j)}_- \end{array} \right) 
\ = \ \left( \begin{array}{c} Z^{(j-1)}_+  + \mu_j Z^{(j-1)}_- \\ Z^{(j-1)}_+  + Z^{(j-1)}_- \end{array} \right) \ \ \text{$j$ odd}; \ \ \ \ 
\ \ \ \left( \begin{array}{c} Z^{(j)}_+ \\ Z^{(j)}_- \end{array} \right) 
\ = \ 
 \left( \begin{array}{c}   \tau_{j-1}^{-1} Z^{(j-1)}_+ +  Z^{(j-1)}_-  \\   \mu_j \tau_{j-1}^{-1} Z^{(j-1)}_+ +  Z^{(j-1)}_- \end{array} \right), \ \ \text{$j$ even}. \label{ZjFla} \end{equation}

\label{tildebasis}
The same formulas will look nicer if we introduce the following basis of formal WKB solutions.  
Let $\tilde \psi_-(E,q,h)=\phi_-(E,q,h)$, in particular, $\tilde \psi_-(E,h,q_0)=1$.  Suppose $E$ is in a neighborhood of the positive real axis; accordingly to our conventions, we think of $\tilde \psi_-(E,h,q)$ as defined on the second sheet of the Riemann surface of $p(q)$. Then $\tilde \psi_-(E,q,h)$ can be analytically continued (as a formal WKB solution) across the cut
connecting $q_1^-(E)$ and $q_1^+(E)$ along the path $\sigma'_1$, fig.\ref{Paper4p4}; we thus obtain a formal WKB solution $\tilde \psi_+(E,q,h)$ on the first sheet of $p(q)$. After that, substitute $hE_r$, where $E_r\in \C$, or $hE_r(h)$, where $E_r(h)$ is a small resurgent function, to obtain $\tilde \psi_{\pm}(hE_r,q,h)$, resp. $\tilde \psi_\pm(hE_r(h),q,h)$.

We can write the exponential asymptotics of the eigenfunction corresponding to the eigenvalue $hE_r(h)$ as
$$ \tilde D^{(j)}_+(h)\tilde \psi_+ (hE_r(h),q,h) + \tilde D^{(j)}_-(h) \tilde \psi_-(hE_r(h),q,h)\ \ \ \ \text{on} \ (q_j,q_{j+1}), \ \  j=1,...,2n,$$
where
\begin{equation} \tilde D^{(j)}_+=  Z_+^{(j)} \times \prod_{\ell=1}^{j-1} \tau_{\ell}^{(-1)^{\ell+1}}, \ \ \ \tilde D^{(j)}_-=Z_-^{(j)}. \label{tldDj} \end{equation}

%$$ \tilde D^{(2)}_+=\tau_1 Z_+^{(2)}$, $\tilde D^{(2)}_-=Z_-^{(2)}$ \\
%etc

Remark that  
$$\tilde \psi_-(hE_r,q,h)\hat\sim e^{\frac{-f(q)+f(q_0)}{h}},    \ \ \ \tilde \psi_+(hE_r,q,h) \hat\sim e^{\frac{f(q)-2f(q_1)+f(q_0)}{h}}$$
in the sense of the notation $\hat\sim$ introduced on p.\pageref{hatsimN}.

We also see that removing the condition $\tilde \psi_-(E,q_0,h)=1$ will only rescale both solutions $\tilde \psi_{\pm}(E,q_0,h)$. Thus, up to rescaling the vector $\tilde (\psi_+(E,q,h),\psi_-(E,q,h))$, we can make our formulas independent of the choice of $q_0$.

\vskip2.5pc

Remark also that if $E_r$ satisfies the quantization condition, then the corresponding eigenfunction of the Witten Laplacian is periodic and 
%is an eigenvalue of the Witten Laplacian and if for $j=0,...,4$ the hyperasymptotic expansions $D^{(j)}_+\phi_+(q) + D^{(j)}_-\phi_-(q)$ define its eigenfunction $\psi(q)$ satisfying $\psi(q)=\psi(q+1)$, then 
we must have
$$ [ M_1 M_2 M_3 M_4 ]^{-1} D^{(0)}_+ \ = \ D^{(4)}_+ ; \ \ [ M'_1 M'_2 M'_3 M'_4 ]^{-1} D^{(0)}_- \ = \ D^{(4)}_-. $$
Rewriting this condition in terms of $Z^{(j)}_\pm$, we arrive at 
%$$ D^{(0)}_+ \ = \ \frac{B'_0}{B_0 c'_1} Z^{(0)}_+; \ \ \ D^{(4)}_+ \ = \ \frac{c_4 B'_0 M'_1 M'_2 M'_3}{B_0 M_1 M_2 M_3} Z^{(4)}_+ $$
%then 
%$$ [M_1 M_2 M_3 M_4]^{-1} \frac{B'_0}{B_0 c'_1} Z^{(0)}_+ \ = \ \frac{c_4 B'_0 M'_1 M'_2 M'_3}{B_0 M_1 M_2 M_3} Z^{(4)}_+ $$
%$$   Z^{(0)}_+ \ = \ (1+E_r \kappa) c'_1 c_4 (M'_4)^{-1} M_4 Z^{(4)}_+ $$
%i.e. 
%$$ Z^{(0)}_+ \ =\ (1+E_r \kappa)\tau_4 Z^{(4)}_+ $$
%We have $D^{(j)}_-=Z^{(j)}_-$, so we must have
%$$ Z^{(0)}_- \ =\ (1+E_r \kappa) Z^{(4)}_- $$
\begin{equation} Z^{(0)}_+ \ = \ (1+E_r \kappa) c'_1 c_4 (M'_4)^{-1} M_4 Z^{(4)}_+; \ \ \ Z^{(0)}_- \ =\ (1+E_r \kappa) Z^{(4)}_- .\label{closure} \end{equation}
If $Z^{(j)}_{\pm}$ are calculated without algebraic mistakes, they must satisfy the formulas \eqref{closure}. 

% verification of equalities connecting $Z^{(0)}$ and $Z^{(4)}$ can be found in RDRW2.Appx104.tex

%%%%%%%%%%%%%%%%%%%%%%%%%%%%%%%%%%%%%%%%%%%%%%%

% \newpage

\section{Example 1.} \label{SectionEx1}

%{\bf Notation.} For a resurgent symbol $\phi$ we will write $\phi \hat\sim e^{\frac{\alpha}{h}}$ if $-\alpha$ is the location of the left-most nonzero microfunction in the decomposition of $\phi$, or, informally, if  $e^{\frac{\alpha}{h}}$ is the leading exponential in $\phi$. We will denote by ${\cal E}^{a}$ the set of resurgent symbols or corresponding resurgent functions of exponential type $\le a$ in $h$, i.e. of those whose majors have no singularities left of the vertical line $\Re \xi = -a$.

%In order to solve the quantization condition \eqref{QCondPrim} 
%\begin{equation} \left(\frac{1}{1+E_r \kappa}\right)^2 - {\rm Tr}G_0  \frac{1}{1+E_r \kappa} + \det G_0 = 0 \label{QCond} \end{equation}
%for the rescaled energy $E_r$, it is important to understand the determinant and the trace of the matrix $G_0$.

Let  
\begin{equation} f(q) = \frac{1}{2\pi}\left[ \sin 2\pi (q + \frac{1}{8}) + \cos 4\pi (q + \frac{1}{8}) \right], \label{fEx1} \end{equation}
$$f'(q) = \cos 2\pi (q + \frac{1}{8}) - 2 \sin 4\pi (q + \frac{1}{8})  .$$ 
The summand $\frac{1}{8}$ in the argument assures that $f'(0)\ne 0$.

% RDRW2.Appx63.tex contains the detailed calculation of location of  the turning points as well as the Stokes pattern

The critical points of $f(q)$ are:
$$ \begin{array}{cccccc} q_1 = \frac{1}{8} &&& f(q_1)= 0  && f''(q_1)=6\pi,\\  
q_2  = \frac{3}{8} - \frac{1}{2\pi}\arcsin \frac{1}{4} &&& f(q_2)  = \frac{9}{16\pi} && f''(q_2)=-7.5\pi, \\
q_3 = \frac{5}{8} &&& f(q_3)  = -\frac{1}{\pi} && f''(q_3)=10\pi, \\ 
q_4 = \frac{7}{8} + \frac{1}{2\pi}\arcsin \frac{1}{4} &&& f(q_4) =  \frac{9}{16\pi} && f''(q_4)=-7.5\pi.
 \end{array}$$

% RDRW2.Appx124.tex contains an adapted exposition corresponding roughly to Paper2

% actually calculating $\tau$s by taking appropriate integrals: cf RDRW2.Appx120.tex

% see RDRW2.Appx76.tex for direct calculation of $\tau_3$ and $\tau_4$

Now we are going to exploit the symmetry of our choice of $f(q)$. 

\begin{Lemma} \label{SymmetryMonodr} Suppose $f(q)$ has two local minima $q_1$, $q_3$ and two local maxima $q_2,q_4$ and satisfies $f(q)=f(2q_3-q)$. Then  $$\tau_1=\tau_4, \ \ \ \tau_2=\tau_3,  \ \ \ \mu_2=\mu_4.$$ \end{Lemma}

\textsc{Proof.} {\it a)}
Observe that for $E$ real, the equation $P\psi = E\psi$ has two real solutions, i.e. those satisfying 
\begin{equation} \psi({\bar q}) = \overline{\psi(q)}, \label{PsiIsReal} \end{equation} 
therefore the same equality must be satisfied by any solution of this equation.
Thus: \\ {\it reflecting a contour $\delta$ on the Riemann surface of $p(q)$ with respect to the (preimage of the) real axis of the complex plane of $q$, while keeping the contour on the same sheet of the Riemann surface of $p(q)$ changes the monodromy of a formal solution by complex conjugation.} \\
 Indeed, the monodromy changes from $\frac{\psi(\delta(1))}{\psi(\delta(0))}$ to $\frac{\psi(\overline{\delta(1)})}{\psi(\overline{\delta(0)})}=\frac{\overline{\psi(\delta(1))}}{\overline{\psi(\delta(0))}}$, where $\delta(0)$ and $\delta(1)$ are the endpoints of $\delta$.

\begin{figure}[h] \includegraphics{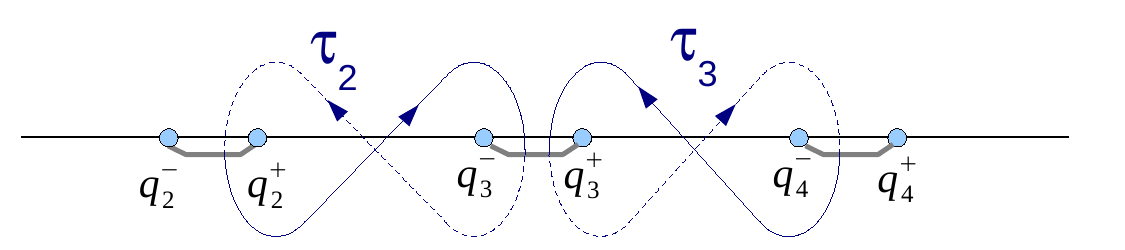} \caption{Contours formal monodromies along which equal to  $\tau_j(E,h)$ for $0<E<<1$. } \label{ContsForTaus} \end{figure}

\begin{figure}[h] \includegraphics{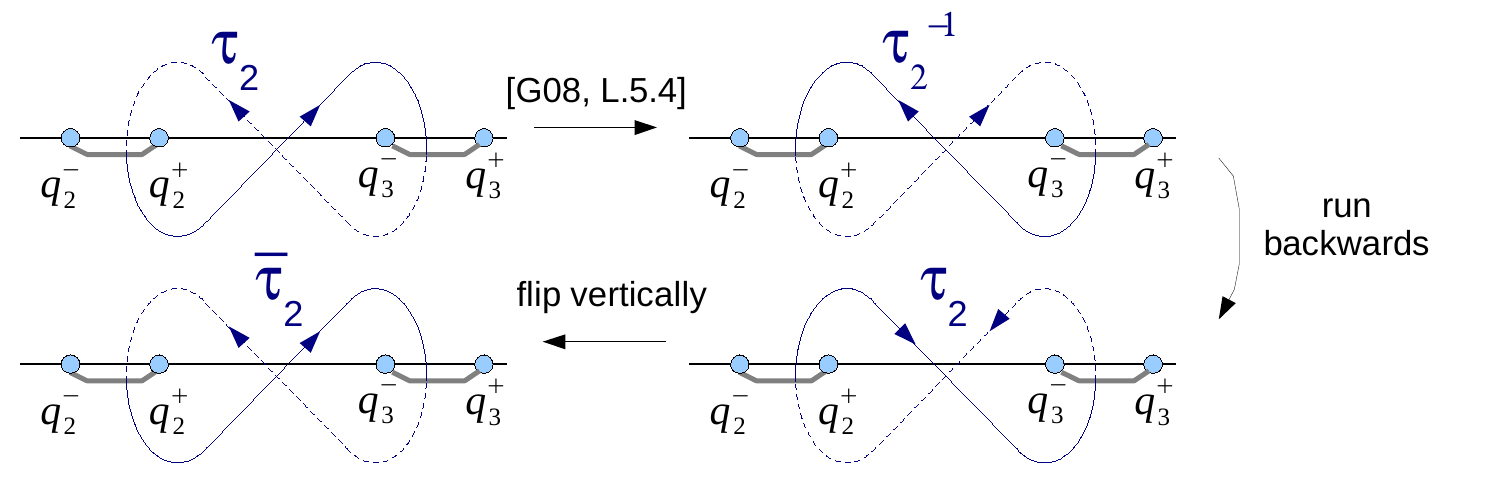} \caption{Formal monodromy $\tau_2=\tau_2(E,h)$ and the related monodromies -- step b) in the proof of lemma \ref{SymmetryMonodr}.}
\label{Paper4p8}
\end{figure}

{\it b)} The coefficients $\tau_j(E,h)$ can be equivalently defined as monodromies of formal solutions along contours shown on fig.\ref{ContsForTaus}. Let us show that for $E>0$ and $h$ real, $\tau_j(E,h)$ is real. E.g., for $\tau_2(E,h)$ perform the transformation of the contour shown on fig.\ref{Paper4p8} : the first transformation is done using ~\cite[Lemma 5.4]{G}, the second simply reverses the direction of the contour, and the third reflects the contour with respect to the real axis, cf. part a) of this proof. Since we returned to the same contour, we conclude that $\overline{\tau_2(E,h)}=\tau_2(E,h)$.

{\it c)} The equation $P\psi=E\psi$ is preserved if we replace $q$ by $2q_3-q$, and so are the determinations of $p(q)$ on the sheets of the Riemann surface of $p(q)$. Hence, a formal monodromy along a contour $\delta$ coincides with a formal monodromy along a contour obtained from $\delta$ by $q\mapsto 2q_3-q$. But the contour for $\tau_3$ is obtained from the contour for $\tau_2$ by means of $q\mapsto 2q_3-q$ followed by reflection with respect to the real axis. Using steps a) and b), we conclude that $\tau_3=\tau_2$.

{\it d)} Similarly, $s_{\gamma_2}=s_{\gamma_4}$ and hence $\mu_2=\mu_4$.  $\Box$

% When $f$ is a real trigonometric polynomial, $E>0$, $h>0$, then $\tau_j$'s are also real {\bf WHY?}, and so flipping the contours (see fig. \ref{ContsForTaus}) defining them with respect to the real axis will give rise to the same formal monodromies. 

%Now if we reflect an integration contour $\delta$ with respect to the point $q_3$, we obtain an integration path that we will denote $2q_3-\delta$. Notice that if $\psi(q)$ is a formal WKB solution of $P\psi=E\psi$, then so is $\psi(2q_3-q)$, and both solutions correspond to the same sheet of the Riemann surface of the classical momentum (because both are either exponentially growing or exponentially decreasing in the direction away from $q_3$ along the real line). Therefore, if we reflect a contour $\delta$ with respect to $q_3$ while keeping it on the same sheet, the formal monodromies along that contour will remain the same.

%The lemma follows from the fact that the contour defining $\tau_3$ can be obtained from that defining $\tau_2$ by reflecting it with respect to $q_3$ and then reflecting it with respect to the real axis. Analogously for $\tau_1$ and $\tau_4$.

% calculation of $k$ see in RDRW2.Appx114.tex

\vskip2.5pc

Solving \eqref{QCondPrim} for $E_r$ begins by writing out the $Tr G_0$ term. Using \eqref{G0Explicit}, we have:
%
%Recall that 
%$$ G_0 = \left( \begin{array}{cc} \tau_4(\tau_3^{-1} + 1) & \tau_4(\mu_3 \tau_3^{-1} +1) \\ \mu_4 \tau_3^{-1} +1 & \mu_3\mu_4\tau_3^{-1} \end{array} \right)
%\left( \begin{array}{cc} \tau_2(\tau_1^{-1} + 1) & \tau_2(\mu_1 \tau_1^{-1} +1) \\ \mu_2 \tau_1^{-1} +1 & \mu_1\mu_2\tau_1^{-1} \end{array} \right)  $$
%and
\begin{equation} \begin{array}{ccrrrr} Tr G_0 &  = &  \tau_4 \tau_3^{-1} \tau_2 \tau_1^{-1} & +  \tau_4 \tau_2 \tau_3^{-1} & 
+ \tau_4\tau_2\tau_1^{-1} & + \tau_2 \tau_4 \\ 
&& + \mu_2\mu_3 \tau_4 \tau_3^{-1} \tau_1^{-1} & + \mu_3 \tau_4 \tau_3^{-1} & + \mu_2\tau_4\tau_1^{-1} & + \tau_4 \\
&& + \mu_1\mu_4 \tau_2 \tau_3^{-1} \tau_1^{-1} & + \mu_4 \tau_2 \tau_3^{-1} & + \mu_1\tau_2\tau_1^{-1} & 
+ \tau_2 \\
&& + \mu_1 \mu_2 \mu_3 \mu_4 \tau_1^{-1} \tau_3^{-1} & + \mu_1 \mu_2 \tau_1^{-1} & + \mu_3 \mu_4 \tau_3^{-1} & 
+ 1 .
\end{array} \label{G0expl} \end{equation}

%%%%%%%%%%%%%%%%%%%%%%%%%%%555
% the old version based on $f(q_3)=-1/2\pi$ and accordingly on different exponents in $\tau_2$ and $\tau_3$ can be found in RDRW2.Appx49.tex

Using the formulas for $\tau_k$, we find
\begin{equation}
 \tau_1 = \tau_4 \hat\sim  e^{-\frac{9}{8\pi h}} E_r , \ \ \  
 \tau_2 = \tau_3 \hat\sim e^{-\frac{25}{8\pi h}} E_r , \label{mr5f1} \end{equation} 
and therefore we can write, loosely,
$$ \begin{array}{ccrrrr} Tr G_0 &  \hat\sim &  1 & +  E_r e^{-\frac{9}{8\pi h}}  & 
+ E_r e^{-\frac{25}{8\pi h}}  & + E_r^2 e^{-\frac{34}{8\pi h}}  \\ 
&& + E_r e^{\frac{25}{8\pi h}}  & + E_r e^{\frac{16}{8\pi h}} & + E_r & + E_r e^{-\frac{9}{8\pi h}} \\
&& + E_r e^{\frac{9}{8\pi h}}  & + E_r  & + E_r e^{-\frac{16}{8\pi h}} & 
+ E_r e^{-\frac{25}{8\pi h}} \\
&& + E_r^2 e^{\frac{34}{8\pi h}}& + E_r e^{\frac{9}{8\pi h}} & + E_r e^{\frac{25}{8\pi h}} & 
+ 1, 
\end{array} $$
by which we mean that, e.g., the exponential type of the summand $\tau_4\tau_2\tau_3^{-1}$ is the same as the exponential type of $E_r e^{-\frac{9}{8\pi h}}$ and that, therefore, this summand contributes to the points corresponding to $E_r^k e^{-\frac{9}{8\pi h}}$, $k\ge 1$, in the Newton polygon, cf. step 4 of sec.\ref{SecOutline}, of the quantization condition \eqref{QCondPrim}. 

By ~\cite[\S 8]{G}, the first and the third summands on the LHS of \eqref{QCondPrim} contribute only terms of type $E_r^k$, $k\ge 1$ (degree zero with respect to $e^{\frac{1}{h}}$).

%---------------

\subsection{Solving the quantization condition} \label{SolvingQCex1}

Since $f(q)$ has two real local minima and two local maxima on its period, the Witten Laplacian will have two resurgent exponentially small eigenvalues, one of them equal to zero; from now on $hE_r(h)$ will denote the other, nonzero exponentially small eigenvalue.  In this subsection we are going to calculate the beginning of the exponential asymptotic expansion of $E_{r}(h)$ using the Newton polygon, cf. step 4 of sec.\ref{SecOutline} and ~\cite[\S 9]{G}.

\begin{figure} \includegraphics{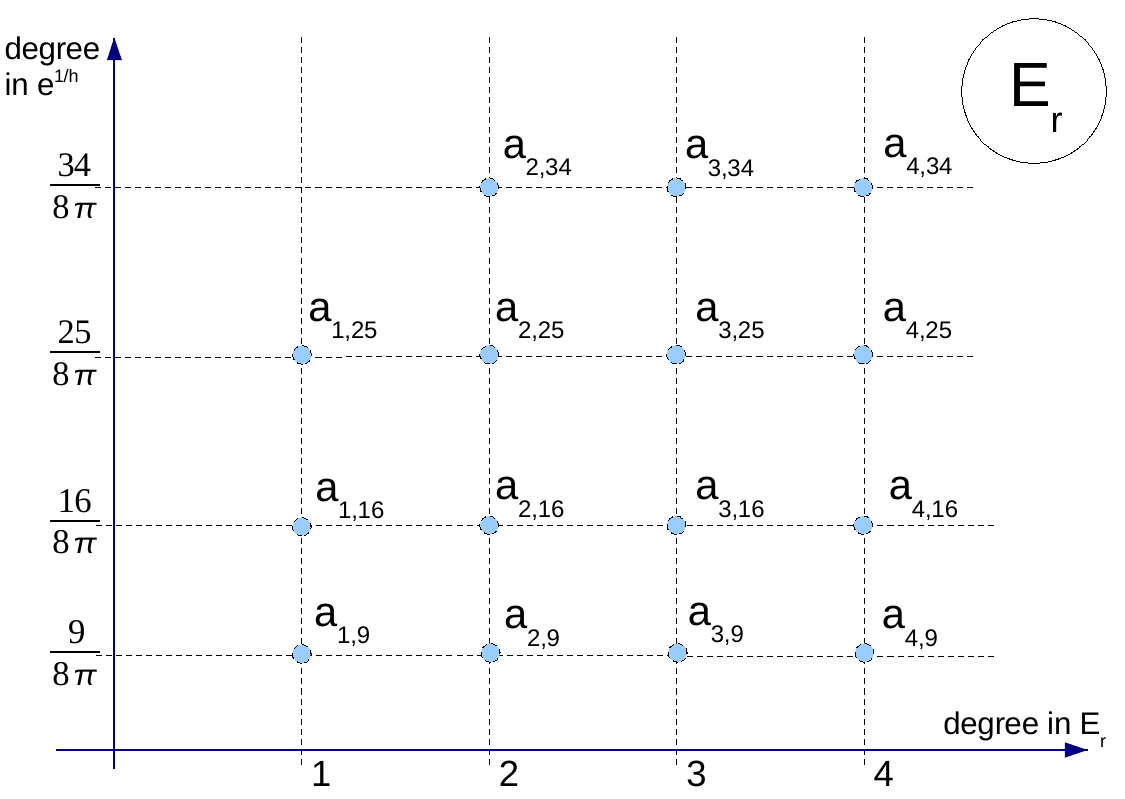} % previously it was RDRW2p11.pdf
\caption{Newton polygon of the quantization condition \eqref{QCondPrim} for the Example 1.} \label{Paper4p9}
\end{figure}

The upper left portion of the Newton polygon of the quantization condition \eqref{QCondPrim} for our example is shown on fig.\ref{Paper4p9}. It is convenient to modify the notation of \eqref{NPGenForm} by putting $\frac{1}{8\pi}$ under the exponent; let us write 
\begin{equation} {\cal F}(E_r,h) := \sum_{k,c} E_r^k e^{\frac{c}{8\pi h}} a_{k,c}(h) \ = \ 0. \label{NPGF8pi} \end{equation}
% ; the coefficients $a_{k,c}=a_{k,c}(h)$ are as in \eqref{NPGenForm}.

% \begin{figure} \includegraphics{RDRW2p12.pdf} \end{figure}

%% 
%% Represent the l.h.s. of the quantization condition \eqref{QCondPrim} as a sum of powers of $E_r$ and $e^{\frac{1}{h}}$, namely,
%% $$-\frac{1}{1+E_r k}  + \Tr G_0  - (1+E_r k )\det G_0 \ = \ \sum_{j,\omega} a_{j\omega} E_r^j e^{\frac{\omega}{8\pi h}}, $$ 
%% and draw the Newton polygon of  \eqref{QCondPrim}  on figure \ref{Paper4p9} as explained in ~\cite{G}. For 
%% the calculation that we are going to carry out, only terms of exponential order $\ge \frac{8}{9\pi h}$ will be important; in particular contributions from the first and the third term on the l.h.s. of \eqref{QCondPrim} are of exponential order $\le 0$ and therefore need not be considered in detail.  % but those terms have been discussed in RDRW2.Appx110.tex

In ~\cite{G} we explained that the leading exponential summand of the exponential asymptotic expansion of $E_{r}(h)$ is obtained by looking the the north-west edges of the Newton polygon, in our case that means -- by solving for $E_0=E_0(h)$ the equation
$$ a_{1,25}E_0 e^{\frac{25}{8\pi h}} + a_{2,34} E_0^{2} e^{\frac{34}{8\pi h}}  \ = \ 0; $$
then $E_r$ will be equal to $E_0$ up to terms of smaller exponential type with respect to $h$, i.e.,
 $E_{r}(h)\  \approx \ -\frac{a_{1,25}(h)}{a_{2,34}(h)} e^{-\frac{9}{8\pi h}}$. To find the next term in the exponential asymptotic expansion of $E_r(h)$, let us make a substitution 
$$E_r(h) =  \left( -\frac{a_{1,25}}{a_{2,34}} + E_1(h) \right) e^{-\frac{9}{8\pi h}} $$
and solve the quantization condition for $E_1(h)$ under additional requirement that $E_1(h)$ should be exponentially small. 

In other words, if the quantization condition as an equation on $E_r(h)$ was written in the form \eqref{NPGF8pi}, we rewrite it in the form
\begin{equation} {\cal F}_1(E_1,h) := \sum_{k,c} E_1^k e^{\frac{c}{8\pi h}} b_{k,c}(h) \ = \ 0 \label{E1NewtonPoly} \end{equation}
where
$$ {\cal F}_1(E_1,h) = e^{-\frac{16}{8\pi h}}{\cal F} (-\frac{a_{1,25}}{a_{2,34}} + E_1,h). $$

\begin{figure} \includegraphics{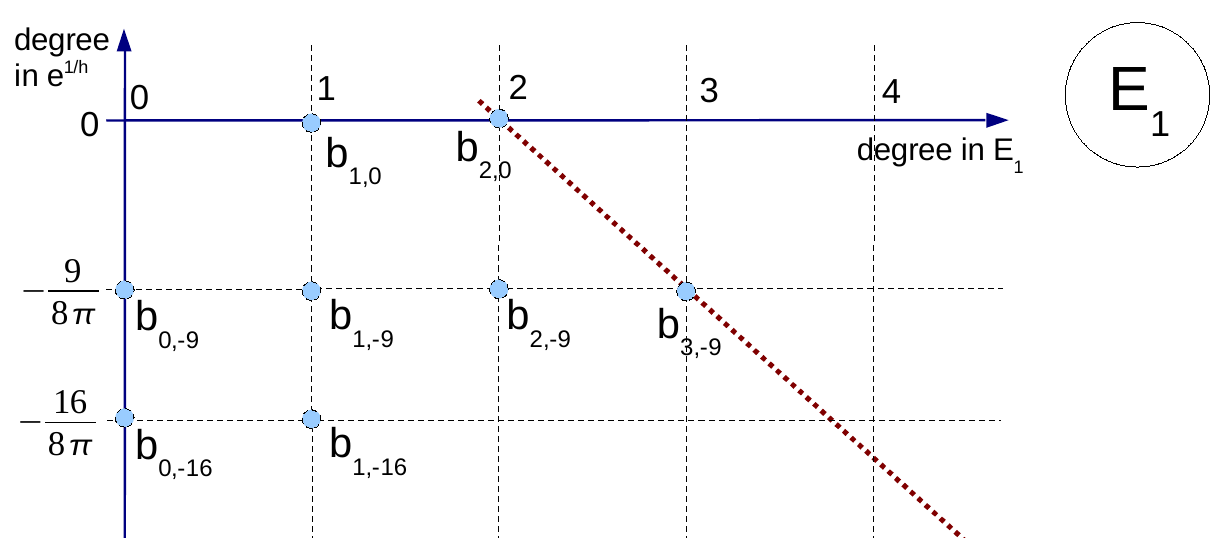} \caption{Newton polygon for $E_1$.} \label{E1polygon} \end{figure}

Let us plot the (possibly) nonzero summands of the RHS of \eqref{E1NewtonPoly} on fig.\ref{E1polygon}; there are no nonzero terms to the right of the slanted dotted line through $b_{2,0}$ and $b_{3,-9}$. 

%We re-express the quantization condition  in terms of $E_1$, 
%$$ \sum_{j,\omega} b_{j,\omega} E_1^j e^{\frac{\omega}{8\pi h}} \ = \ e^{-\frac{16}{8\pi h}} \sum_{j,\omega} a_{j,\omega} E_r^j e^{\frac{\omega}{8\pi h}} $$
%and plot the summands on the figure \ref{E1polygon}, where:

We have:
 $$ b_{1,0} = -a_{1,25} , \ \ \ b_{2,0}=a_{2,34} ,$$
$$b_{0,-9} =  -\frac{a_{1,25}a_{1,16}}{a_{2,34}} + a_{2,25} (\frac{a_{1,25}}{a_{2,34}})^2 - a_{3,34}(\frac{a_{1,25}}{a_{2,34}})^3 .$$
% same thought with more words is presented in RDRW2.Appx123.tex 

Analogously to the first step, from the Newton polygon on fig.\ref{E1polygon} we infer that
$$ E_1 \ = \ e^{-\frac{9}{8\pi h}} \left( -\frac{b_{0,-9}}{b_{1,0}} + E_2 \right) $$
where $E_2$ has to be exponentially small. The next step of this procedure and a Newton polygon for $E_2$ (which we will not draw here) yields $E_2\in{\cal E}^{-\frac{7}{8\pi }}$.

%Substitute $E_1 = e^{-\frac{9}{8\pi h}} E_2$. Draw a polygon for $E_2$:
%\begin{figure} \includegraphics{RDRW2p14.pdf} \caption{Newton polygon for $E_2$} \end{figure}

Let calculate the coefficients $a_{2,34}, a_{3,34}, a_{1,25}, a_{2,25}$, and  $a_{1,16}$. Only the following four summands in $\Tr G_0$ contribute to these coefficients:

% A calculation carried out in RDRW2.Appx52.tex
$$\mu_1\mu_2\mu_3\mu_4\tau_1^{-1} \tau_3^{-1}  
 \ = \ e^{\frac{34}{8\pi h} }  \frac{ E_r^2\pi^2  }{\sqrt{f''(q_1)\cdot |f''(q_2)|\cdot f''(q_3)\cdot |f''(q_4)|}} \times $$ $$ \times h^{\frac{E_r }{2}\left(\frac{1}{f''(q_1)} +\frac{1}{|f''(q_2)|} +\frac{1}{f''(q_3)} +\frac{1}{|f''(q_4)|}\right) } ([1+O(E_r)]+O(h\ln h)); $$
$$ \mu_2\mu_3 \tau_4 \tau_3^{-1} \tau_1^{-1} \ = \ - e^{ \frac{25}{8\pi h} } \frac{E_r \pi }{\sqrt{f''(q_3) |f''(q_2)|}} 
 h^{\frac{E_r}{2} \left( \frac{1}{|f''(q_2)|} + \frac{1}{f''(q_3)}  \right)}([1+O(E_r)]+O(h\ln h));
$$
$$ 
\mu_3\mu_4   \tau_3^{-1}   \ = \  - e^{\frac{25}{8\pi h}} \frac{E_r \pi }{ \sqrt{f''(q_3)|f''(q_4)|}}   h^{\frac{E_r }{2} \left( \frac{1}{f''(q_3)}+ \frac{1}{|f''(q_4)|}  \right)} 
    ([1+O(E_r)]+O(h\ln h));  $$
$$ 
\mu_3 \tau_4 \tau_3^{-1}  \ = \  e^{\frac{16}{8\pi h}}\frac{E_r \pi i }{\sqrt{f''(q_1) f''(q_3)}}      h^{\frac{E_r}{2} \left( \frac{1}{f''(q_3)} - \frac{1}{f''(q_1)}  \right)}  (1+O_{h=fix}(E_r)+O(h)), $$
where the notation $(1+O(E_r)) +O(h\ln h)$ means an expression of the form $\sum \BBB_{k\ell}(E_r) h^k \ln^\ell h$ with $\BBB_{00}(E_r)=1+O(E_r)$ for $E_r\to 0$ and all other $B_{k\ell}=0$ if $h^k\ln^\ell h$ dominates $h\ln h$.  
%The notation $O_{h=fix}(E_r)$ means terms that contain factors of degree $\ge 1$ with respect to $E_r$, regardless of their degree with respect to $h$ or $\log h$. 

We have:
$$ a_{2,34} = \frac{ \pi^2  }{\sqrt{f''(q_1)\cdot |f''(q_2)|\cdot f''(q_3)\cdot |f''(q_4)|}} (1+O(h)) \ = \ \frac{1}{15\sqrt{15}};
$$
\footnotesize 
$$ a_{3,34} = \frac{ \pi^2  }{\sqrt{f''(q_1)\cdot |f''(q_2)|\cdot f''(q_3)\cdot |f''(q_4)|}} \left[\frac{1}{f''(q_1)} +\frac{1}{|f''(q_2)|} +\frac{1}{f''(q_3)} +\frac{1}{|f''(q_4)|}\right] \frac{\log h + O(h^0)}{2}  \ = \ $$
$$ \ = \  \frac{1}{15\sqrt{15}}  \left[\frac{1}{6\pi } +\frac{2}{7.5\pi} +\frac{1}{10\pi}\right] \frac{\log h + O(h^0)}{2} \ = \ \frac{4}{15^2\sqrt{15} \pi}  \log h + O(h^0);  $$ 
\normalsize
$$ a_{1,25} = - \frac{ \pi }{\sqrt{f''(q_3) |f''(q_2)|}}  - \frac{ \pi }{\sqrt{f''(q_3) |f''(q_4)|}} + O(h) \ = \ -\frac{2}{\sqrt{75}} + O(h); 
$$
\footnotesize $$ a_{2,25} = \left\{- \frac{ \pi }{\sqrt{f''(q_3) |f''(q_2)|}}\left( \frac{1}{f''(q_3)}+ \frac{1}{|f''(q_2)|} \right)  - \frac{ \pi }{\sqrt{f''(q_3) |f''(q_4)|}}\left( \frac{1}{f''(q_3)}+ \frac{1}{|f''(q_4)|} \right) \right\} \frac{\log h + O(h^0)}{2} $$
$$ \ = \ -\frac{2}{\sqrt{75}}[ \frac{1}{10\pi} + \frac{2}{15\pi} ] \frac{\log h + O(h^0)}{2} \ = \ -\frac{1}{5\sqrt{3}}\cdot\frac{7}{30\pi} \log h + O(h^0) \ = \ -\frac{7}{150\sqrt{3}\pi} \log h + O(h^0); $$
\normalsize 
$$ a_{1,16} = \frac{\pi i }{\sqrt{f''(q_1) f''(q_3)}} + O(h) \ = \ \frac{i}{\sqrt{60}} + O(h). $$
We think that the imaginary value of $a_{1,15}$ is attributable to our choice of $\arg h >0$. Passage to the case $\arg h=0$ would involve techniques studied, e.g., in ~\cite{M99}.

Proceed with the calculation:
$$ r: = -\frac{a_{1,25}}{a_{2,34}} % =  \frac{\frac{ \pi }{\sqrt{f''(q_3) |f''(q_2)|}}  +  \frac{ \pi }{\sqrt{f''(q_3) |f''(q_4)|}}}{\frac{ E_r^2\pi^2  }{\sqrt{f''(q_1)\cdot |f''(q_2)|\cdot f''(q_3)\cdot |f''(q_4)|}}}+O(h) $$ $$ 
\ = \  \frac{1}{\pi} ( \sqrt{f''(q_1)\cdot |f''(q_4)|}   +  \sqrt{f''(q_1)\cdot |f''(q_2)|}) + O(h) \ = \ 6\sqrt{5} + O(h). $$
% RDRW2.Appx129.tex has a calculation of the rest with wrong signs
$$ b_{0,-9} = a_{1,16} r + a_{2,25} r^2 + a_{3,34} r^3 = $$
%\textcolor{red}{Wrong signs! Actually, $b_{0,-9} = a_{1,16} r + a_{2,25} r^2 + a_{3,34} r^3$. Redo the rest!}
%\footnotesize (ignore the first term because it gives something of order $h^0$)
$$ \ = \  \left\{ -\frac{7}{150\sqrt{3}\pi} [6\sqrt{5}]^2 + \frac{4}{15^2\sqrt{15}\pi}[6\sqrt{5}]^3  \right\}\log h + O(h^0) \ = \ \frac{18\sqrt{3}}{5\pi}\log h + O(h^0);$$
% an attempt of a more conceptual treatment is given in RDRW2.Appx111.tex
 $$b_{1,0} \ = \ -a_{1,25} = \frac{2}{\sqrt{75}}+O(h).$$

Collecting the results, we find the following formula for the nonzero low-lying eigenvalue:
$$ hE_r(h) \ = \ h e^{-\frac{9}{8\pi h}} \left(- \frac{a_{1,25}}{a_{2,34}} + E_1 \right) \ = \ $$
%$$ \ = \ e^{-\frac{9}{8\pi h}} \left(- \frac{a_{1,25}}{a_{3,24}} + e^{-\frac{9}{8\pi h}}E_2 \right) \ = \ $$
$$ \ = \ h e^{-\frac{9}{8\pi h}} \left(- \frac{a_{1,25}}{a_{3,24}} + e^{-\frac{9}{8\pi h}}\left[ -\frac{b_{0,-9}}{b_{1,0}} + {\cal E}^{-\frac{7}{8\pi }} \right] \right) \ = \ $$
\begin{equation} \ = \  e^{-\frac{9}{8\pi h}} (6\sqrt{5}h+ o(h^1)) +   e^{-\frac{18}{8\pi h}} (-\frac{27}{\pi}h \log h + O(h^1)) + {\cal E}^{-\frac{25}{8\pi }} . \label{Apr1st} \end{equation}

%{\bf Remark.} Calculation of the next term in $h$ in the above asymptotic expansions multiplying $e^{-\frac{9}{8\pi h}}$ or $e^{-\frac{18}{8\pi h}}$ requires taking the integrals as in formulas \eqref{SumAj}, \eqref{SumAj2}, which in our example can be done by hand. Performing this calculation, however, did not bring the author any new insight.

%---------------------------

\subsection{Asymptotic expansion of the eigenfunction corresponding to the nonzero low-lying eigenvalue.} \label{eigenfunEx1}

Let us calculate the eigenfunction corresponding to the eigenvalue \eqref{Apr1st}. 

As explained in sec.\ref{procedure}, we will start with $(Z_+^{(0)},Z_-^{(0})^T \in Ker( G_0 - \frac{1}{1+E_rk} Id)$. We can take 
$$ Z_+^{(0)} = \lambda \cdot [G_0]_{12}, \ \ \ Z_-^{(0)}=-\lambda\cdot ([G_0]_{11} - \frac{1}{1+E_r \kappa}) $$
where $\lambda=\lambda(h)$ is any nonzero resurgent symbol which we will choose as $\lambda=\frac{1}{\tau_3^{-1}\tau_4\mu_3}$ to simplify the formulas.

Using \eqref{mr5f1},
$$\tau_1 =\tau_4 \hat \sim e^{-\frac{9}{8\pi h}} E_r \hat\sim e^{-\frac{18}{8\pi h}}; \ \ \ \ \ \tau_2 = \tau_3 \sim e^{-\frac{25}{8\pi h}} E_r \hat\sim e^{-\frac{34}{8\pi h}}. $$

Let us write down the exponential orders of the various summands in $[G_0]_{11} - (1+E_r \kappa)^{-1}$ and in $[G_0]_{12}$. Namely,
% old treatment in RDRW2.Appx131.tex
$$ [G_0]_{11} - (1+E_r \kappa)^{-1} 
\ = \ \overbrace{ ( \tau_4 \tau_2 \tau_1^{-1} \tau_3^{-1} - (1+E_rk)^{-1}) }^{\hat\sim e^{-\frac{9}{8\pi h}} }  
+ \overbrace{ \tau_4 \tau_2 \tau_3^{-1} }^{ \hat\sim e^{-\frac{18}{8\pi h}} } 
+ \overbrace{ \tau_4 \tau_2 \tau_1^{-1} }^{ \hat\sim e^{-\frac{34}{8\pi h}} }
+ \overbrace{ \tau_4 \tau_2 }^{ \hat\sim e^{-\frac{52}{8\pi h}} } $$
$$ \ \ \ 
+ \underbrace{ \tau_4 \tau_1^{-1} \tau_3^{-1} \mu_3\mu_2 }_{ \hat\sim e^{\frac{16}{8\pi h}} } 
+ \underbrace{ \tau_4 \tau_1^{-1} \mu_2 }_{ \hat\sim  e^{-\frac{9}{8\pi h}} }
+ \underbrace{ \tau_3^{-1} \tau_4 \mu_3 }_{ \hat\sim  e^{\frac{7}{8\pi h}} } 
+ \underbrace{ \tau_4 }_{ \hat\sim  e^{-\frac{18}{8\pi h}} } ,
$$
hence
%%Thus, we have two main terms  $\tau_4 \tau_1^{-1} \tau_3^{-1} \mu_3\mu_2$ and $\tau_3^{-1} \tau_4 \mu_3$.\\
%% Furthermore,
%\footnotesize
%$$ \frac{[G_0]_{11} - (1+E_r \kappa)^{-1}}{\tau_3^{-1} \tau_4 \mu_3} \ = \  \frac{\tau_4 \tau_1^{-1} \tau_3^{-1} \mu_3\mu_2 + \tau_3^{-1} \tau_4 \mu_3 +  \tau_4 \tau_1^{-1} \mu_2 + ( \tau_4 \tau_2 \tau_1^{-1} \tau_3^{-1} - (1+E_rk)^{-1}) + {\cal E}^{-\frac{18}{8\pi}}}{\tau_3^{-1} \tau_4 \mu_3} \ = \ $$
%$$ \ = \ 
% \frac{[G_0]_{11} - (1+E_r \kappa)^{-1}}{\tau_3^{-1} \tau_4 \mu_3} \ = \  \frac{\tau_4 \tau_1^{-1} \tau_3^{-1} \mu_3\mu_2 + \tau_3^{-1} \tau_4 \mu_3 +  \tau_4 \tau_1^{-1} \mu_2 + ( \tau_4 \tau_2 \tau_1^{-1} \tau_3^{-1} - (1+E_rk)^{-1}) }{\tau_3^{-1} \tau_4 \mu_3} + {\cal E}^{-\frac{25}{8\pi}} \ = \ $$
$$  \frac{[G_0]_{11} - (1+E_r \kappa)^{-1}}{\tau_3^{-1} \tau_4 \mu_3} \ = \   \underbrace{\tau_1^{-1} \mu_2}_{\hat\sim e^{\frac{9}{8\pi h}} } + 1 +  \underbrace{\left(\frac{\tau_4 \tau_1^{-1} \mu_2 + ( \tau_4 \tau_2 \tau_1^{-1} \tau_3^{-1} - (1+E_rk)^{-1}) }{\tau_3^{-1} \tau_4 \mu_3} \right)}_{\hat\sim e^{-\frac{16}{8\pi h}}}+ {\cal E}^{-\frac{25}{8\pi}}. $$
%and the third summand $\sim e^{-\frac{16}{8\pi h}}$.

Further,
$$[G_0]_{12} \ = \ 
  \overbrace{ \mu_1 \tau_1^{-1} \tau_2\tau_3^{-1}\tau_4 }^{\hat\sim  e^{-\frac{9}{8\pi h}} }
+ \overbrace{ \tau_2\tau_3^{-1}\tau_4 }^{ \hat\sim e^{-\frac{18}{8\pi h}} }
+ \overbrace{ \mu_1 \tau_1^{-1} \tau_2 \tau_4 }^{\hat\sim e^{-\frac{43}{8\pi h}} }
+ \overbrace{ \tau_2\tau_4 }^{\hat\sim  e^{-\frac{52}{8\pi h}} } $$
$$ \ \ \ \ 
+ \underbrace{ \mu_1 \mu_2 \mu_3  \tau_1^{-1} \tau_3^{-1}\tau_4 }_{\hat\sim e^{\frac{7}{8\pi h}} }
+ \underbrace{ \mu_3 \tau_3^{-1} \tau_4 }_{  e^{\frac{7}{8\pi h}} }
+ \underbrace{ \mu_1 \mu_2 \tau_1^{-1} \tau_4 }_{\hat\sim e^{-\frac{18}{8\pi h}} }
+ \underbrace{ \tau_4 }_{ \hat\sim  e^{-\frac{18}{8\pi h}} }, $$ 
hence
%Main terms are $\mu_1 \mu_2 \mu_3  \tau_1^{-1} \tau_3^{-1}\tau_4$ and $\mu_3 \tau_3^{-1} \tau_4$, and we have
$$ 
\frac{[G_0]_{12}}{\mu_3 \tau_3\tau_4^{-1}} 
%= \frac{\mu_1 \mu_2 \mu_3  \tau_1^{-1} \tau_3^{-1}\tau_4 + \mu_3 \tau_3^{-1} \tau_4+ \mu_1 \tau_1^{-1} \tau_2\tau_3^{-1}\tau_4 + {\cal E}^{-\frac{18}{8\pi h}}}{\mu_3 \tau_3^{-1}\tau_4} \ = \ $$
%$$ = \frac{\mu_1 \mu_2 \mu_3  \tau_1^{-1} \tau_3^{-1}\tau_4 + \mu_3 \tau_3^{-1} \tau_4+ \mu_1 \tau_1^{-1} \tau_2\tau_3^{-1}\tau_4 }{\mu_3 \tau_3^{-1}\tau_4} + {\cal E}^{-\frac{25}{8\pi h}}\ = \ $$ $$ 
\ = \ 1+ \underbrace{\mu_1 \mu_2 \tau_1^{-1}}_{\hat\sim e^0} + 
\underbrace{ \frac{\mu_1 \tau_1^{-1} \tau_2 }{\mu_3} }_{\hat\sim e^{-\frac{16}{8\pi h}}} + {\cal E}^{-\frac{25}{8\pi h}}.
$$

Thus,
$$ \left( \begin{array}{c} Z^{(0)}_+ \\ Z^{(0)}_- \end{array} \right) \ = \
 \left( \begin{array}{c}  1+ \mu_1 \mu_2 \tau_1^{-1} + 
\frac{\mu_1 \tau_1^{-1} \tau_2 }{\mu_3} + {\cal E}^{-\frac{25}{8\pi h}}  
\\ 
- (\tau_1^{-1} \mu_2 + 1 +  \frac{\tau_4 \tau_1^{-1} \mu_2 + ( \tau_4 \tau_2 \tau_1^{-1} \tau_3^{-1} - (1+E_rk)^{-1}) }{\tau_3^{-1} \tau_4 \mu_3} + {\cal E}^{-\frac{25}{8\pi}} ) \end{array} \right). $$

%IMPORTANT: There are two summands  contributing to the vertex $e^{\frac{25}{8\pi h}} E_r$, namely, $\mu_2 \mu_3\tau_4 \tau_3^{-1} \tau_1^{-1}$ and $\mu_3 \mu_4 \tau_3^{-1}$.
% That means that for $E_r$ satisfying the quantization condition, we have (see equation \eqref{QCondCorr1203})
% see also RDRW2.Appx130.tex
Before writing down the explicit expressions for $Z^{(j)}_\pm$, let us derive the following consequence of the quantization condition. Using the explicit form of $\Tr G_0$, assuming $E_r$ satisfies \eqref{QCondPrim}, and keeping only the largest terms in \eqref{QCondPrim}, we obtain
$$ (\mu_2 \mu_3 \tau_4 \tau_3^{-1} \tau_1^{-1} + \mu_3\mu_4\tau_3^{-1}) + \mu_1\mu_2\mu_3\mu_4 \tau_1^{-1} \tau_3^{-1} = {\cal E}^{\frac{7}{8\pi}} $$
which, taking into account $\tau_1=\tau_4$ and $\mu_2=\mu_4$, simplifies to
%$$ \mu_2 \tau_4 \tau_1^{-1} + \mu_4  + \mu_1\mu_2 \mu_4 \tau_1^{-1} =  {\cal E}^{-\frac{18}{8\pi}}  $$
%or, since $\tau_1=\tau_4$ and $\mu_2=\mu_4$,
%$$ 2\mu_2   + \mu_1\mu_2^2 \tau_1^{-1}  =  {\cal E}^{-\frac{18}{8\pi}}   $$
\begin{equation} 2   + \mu_1\mu_2 \tau_1^{-1}  = {\cal E}^{-\frac{9}{8\pi}}.  \label{QCsimp1203} \end{equation}
%---------------

% RDRW2.Appx108.tex contains a more detailed version, before it was shortened on March 12th.
Now, applying formulas \eqref{ZjFla}, we obtain successively:
\small $$  \left( \begin{array}{c} Z^{(1)}_+ \\ Z^{(1)}_- \end{array} \right) 
\ = \ 
\left( \begin{array}{c} 1 - \mu_1 +  
\frac{\mu_1 \tau_1^{-1} \tau_2 }{\mu_3}  + {\cal E}^{-\frac{25}{8\pi}} 
 \\ 
 -\tau_1^{-1} \mu_2 + \mu_1 \mu_2 \tau_1^{-1} + 
\frac{\mu_1 \tau_1^{-1} \tau_2 }{\mu_3}   
 -  \frac{\tau_4 \tau_1^{-1} \mu_2 + ( \tau_4 \tau_2 \tau_1^{-1} \tau_3^{-1} - (1+E_rk)^{-1}) }{\tau_3^{-1} \tau_4 \mu_3} + {\cal E}^{-\frac{25}{8\pi}} \end{array} \right) \ = \ 
 $$
\normalsize
% some calculations skipped, cf. RDRW2.Appx95.tex
% which is convenient to rewrite as
$$ 
 \ = \ \left( 1 - \mu_1 +  
\frac{\mu_1 \tau_1^{-1} \tau_2 }{\mu_3}  + {\cal E}^{-\frac{25}{8\pi}}  \right)
\left( \begin{array}{c} 1
 \\ 
 \tau_1^{-1} \mu_2 \left[ -1 + \frac{\mu_1 \tau_1^{-1} \tau_2 }{\mu_3} 
\right] (1  + {\cal E}^{-\frac{25}{8\pi}}) \end{array} \right);
 $$
% \normalsize

$$ \left( \begin{array}{c} Z^{(2)}_+ \\ Z^{(2)}_- \end{array} \right) 
\ = \ 
 \left( \begin{array}{c}   \tau_1^{-1} Z^{(1)}_+ +  Z^{(1)}_-  \\   \mu_2 \tau_1^{-1} Z^{(1)}_+ +  Z^{(1)}_- \end{array} \right) \ = \ % $$
%$$ \ = \  Z^{(1)}_+  \left( \begin{array}{c}   \tau_1^{-1}  +  \tau_1^{-1} \mu_2 \left[ -1 + \frac{\mu_1 \tau_1^{-1} \tau_2 }{\mu_3} 
%\right] (1  + {\cal E}^{-\frac{25}{8\pi}})  \\   \mu_2 \tau_1^{-1}  +  \tau_1^{-1} \mu_2 \left[ -1 + \frac{\mu_1 \tau_1^{-1} \tau_2 }{\mu_3} 
%\right]   + {\cal E}^{-\frac{16}{8\pi}} \end{array} \right) % $$
% $$ \ = \ 
 Z^{(1)}_+  \left( \begin{array}{c}   \tau_1^{-1}  +  \tau_1^{-1} \mu_2 \left[ -1 + \frac{\mu_1 \tau_1^{-1} \tau_2 }{\mu_3} 
\right]  + {\cal E}^{-\frac{16}{8\pi}}  \\     \tau_1^{-1} \mu_2  \frac{\mu_1 \tau_1^{-1} \tau_2 }{\mu_3} 
  + {\cal E}^{-\frac{16}{8\pi}} \end{array} \right). $$

\begin{Rmk} \label{ResurgImp1} {\rm It is interesting to note that in the calculation of $Z^{(2)}_-$ the contributions from the leading exponential orders in $Z^{(1)}_+$ and $Z^{(1)}_-$ cancel and the nonzero value of $Z^{(2)}_-$ is due purely to subdominant exponentials in $Z^{(1)}$. Neglecting these subdominant terms would make the rest of the calculation impossible. This little algebraic detail is philosophically  important: it shows that constructing asymptotic expansions of an eigenfunction of the Witten Laplacian on all intervals $(q_j,q_{j+1})$ must be difficult without methods of resurgent analysis.} \end{Rmk}

% a wrong -- too naive -- calculation is in RDRW2.Appx115.tex

Let us continue:
$$\left( \begin{array}{c} Z^{(3)}_+ \\ Z^{(3)}_- \end{array} \right) 
\ = \   
% RDRW2.Appx107.tex contains the missing calculation
\tau_2 \tau_1^{-1}Z^{(1)}_+
\left( \begin{array}{cc} (1 + \tau_1^{-1} \mu_2  \mu_1  )(1 + {\cal E}^{-\frac{9}{8\pi}} )
\\  
 \tau_1^{-1} \mu_2  \frac{\mu_1}{\mu_3}(1 +  {\cal E}^{-\frac{9}{8\pi}})  \end{array} \right) %$$
% $$
\ = \   \tau_2 \tau_1^{-1}Z^{(1)}_+
\left( \begin{array}{cc} -1 + {\cal E}^{-\frac{9}{8\pi}} 
\\  
 \tau_1^{-1} \mu_2  \frac{\mu_1}{\mu_3} +  {\cal E}^{0}  \end{array} \right), $$
where we have used \eqref{QCsimp1203} in the last step.
%(where the bottom is $\sim e^{\frac{9}{8\pi h}}$.)
%\normalsize
Finally,
$$\left( \begin{array}{c} Z^{(4)}_+ \\ Z^{(4)}_- \end{array} \right) \ = \  
% \left( \begin{array}{c}   \tau_3^{-1} Z^{(3)}_+ +  Z^{(3)}_-  \\   \mu_4 \tau_3^{-1} Z^{(3)}_+ +  Z^{(3)}_- \end{array} \right) \ = \ %$$
% $$ \ = \ 
 \tau_2 \tau_1^{-1} Z^{(1)}_+
 \left( \begin{array}{c}   \tau_3^{-1} (-1 + {\cal E}^{-\frac{9}{8\pi}}) +  \tau_1^{-1} \mu_2  \frac{\mu_1}{\mu_3} +  {\cal E}^{0}   \\   \mu_4 \tau_3^{-1} (-1 + {\cal E}^{-\frac{9}{8\pi}}) +  \tau_1^{-1} \mu_2  \frac{\mu_1}{\mu_3} +  {\cal E}^{0} \end{array} \right) \ = \  
-\tau_2 \tau_1^{-1} Z^{(1)}_+
 \left( \begin{array}{c}   \tau_3^{-1}  +  {\cal E}^{\frac{25}{8\pi }}   \\  \mu_2 \tau_3^{-1}  +  {\cal E}^{\frac{16}{8\pi}} \end{array} \right).  $$
\normalsize

We finish this section by computing the coefficients $\tilde D_\pm^{(j)}$ from \eqref{tldDj}:
\begin{equation} 
\begin{array}{lcl}
\tilde D^{(1)}_+ = 1+{\cal E}^{-\frac{9}{8\pi }}, && \tilde D^{(1)}_- = \underbrace{\tau_1^{-1}\mu_2}_{\hat\sim e^{\frac{9}{8\pi }}} (1+{\cal E}^{-\frac{9}{8\pi }}), \\ 
\tilde D^{(2)}_+ = 1+{\cal E}^{-\frac{9}{8\pi }}, && \tilde D^{(2)}_- = \underbrace{\tau_1^{-2}\tau_2 \mu_1\mu_2\mu_3^{-1}}_{\hat\sim e^{-\frac{7}{8\pi h}}} (1+{\cal E}^{-\frac{9}{8\pi }}), \\ 
\tilde D^{(3)}_+ = -1+{\cal E}^{-\frac{9}{8\pi }}, && \tilde D^{(3)}_- = \underbrace{\tau_1^{-2}\tau_2 \mu_1\mu_2\mu_3^{-1}}_{\hat\sim e^{-\frac{7}{8\pi h}}} (1+{\cal E}^{-\frac{9}{8\pi }}), \\
\tilde D^{(4)}_+ = -1+{\cal E}^{-\frac{9}{8\pi }}, && \tilde D^{(4)}_- = -\underbrace{\mu_2\tau_1^{-1}}_{\hat\sim e^{\frac{9}{8\pi h}}} (1+{\cal E}^{-\frac{9}{8\pi }}).  
\end{array}
\label{efunEx1}
\end{equation}
% calculation in Paper4.Scan1.pdf
To the leading term, % Paper4.Scan2.pdf
 $$ \tau_1^{-1}\mu_2 = ie^{\frac{9}{8\pi h}}(\frac{2}{\sqrt{5}}+O(h)), \ \ \ \ \tau_1^{-2}\tau_2\mu_1\mu_2\mu_3^{-1}=ie^{-\frac{7}{8\pi h}}(\frac{2}{\sqrt{3}}+O(h)), $$ and we obtain the result claimed in \eqref{ansEx1}.

%===============
%
%The closure condition would be {\bf Roughly!}
%$$ (1+E_r \kappa) \tau_4\left( 1 - \mu_1 +  
%\frac{\mu_1 \tau_1^{-1} \tau_2 }{\mu_3}  + {\cal E}^{-\frac{25}{8\pi}}  \right) \tau_2 \tau_1^{-1} 
% \tau_3^{-1} (-1 + {\cal E}^{-\frac{9}{8\pi}}) +  \tau_1^{-1} \mu_2  \frac{\mu_1}{\mu_3} +  {\cal E}^{0})  \ = \ 1+%\mu_1\mu_2 \tau_1^{-1} $$
%At this precision level both sides are roughly $-1$. The bottom works out as well!

% RDRW2.Appx116.tex contains a wrong matrix -- the matrix that has only the leading, boundary of the Newton polygon, terms for each entry. I used this matrix in the earlier version of my calculation

% RDRW2.Appx117.tex contains a more precise calculation of various products of $tau$ and $mu$.

%%%%%%%%%%%%%%%%%%%%%%%%%%%%%%%%%%%%%%5

% \newpage

\section{Example 2.} \label{SectionEx2}

\begin{figure}[h] \includegraphics{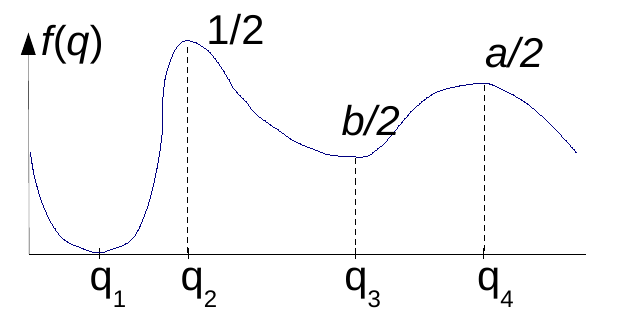} \caption{Graph of $f(q)$ in Example 2.} \label{genericf} \end{figure}

Let $f(q)$ be a trigonometric polynomial with two local minima and two local maxima $q_1,q_3$ and two local maxima $q_2,q_4$ on the period $[0,1]$, where $0<q_1<q_2<q_3<q_4<1$. Up to shifting $q$ by a constant we can assume that $q_1$ is the global minimum of $f$, and up to changing $f(q)$ into $f(2q_1-q)$, that $q_2$ is its global maximum. Changing further $f(q)$ by an affine linear transformation $f\mapsto Af+B$, we can assume $f(q_1)=0$, $f(q_2)=\frac{1}{2}$, $f(q_3)=\frac{b}{2}$, $f(q_4)=\frac{a}{2}$, where $0\le b < a \le 1$, figure \ref{genericf}. All these transformations of $f$ produce easily controllable changes in the eigenvalues and eigenfunctions of the Witten Laplacian. 

We will actually assume that the inequalities are strict:
\begin{equation} \text{Assume: \ \ } 0 \ < \ b \ < \ a < \ 1, \label{ZeroABOne} \end{equation}
and will gradually put more restrictions on $a$ and $b$ more specific as we progress through this section.

In our situation $$ \tau_1 \hat \sim E_r e^{-\frac{1}{h}} ; \ \ \tau_2 \hat\sim E_r e^{-\frac{1-b}{h}} ; \ \ \tau_3 \hat\sim  E_r e^{-\frac{a-b}{h}}; \ \ \tau_4 \hat\sim E_r e^{-\frac{a}{h}}.$$ 

In order to find the two low-lying eigenvalues of the Witten Laplacian (one of which equals, as we know already, to zero), we need to solve the same  quantization condition \eqref{QCondPrim} and will use the formula \eqref{G0expl} for $\Tr G_0$.
%\begin{equation} -\frac{1}{1+E_r \kappa}  \ + \  \Tr G_0 \ -  \ (1+E_r \kappa) \det G_0 \ = \ 0 .\label{QCondFeb16} \end{equation}
%where $\Tr G_0$ has the same form as \eqref{G0expl}.

%$$ \begin{array}{ccrrrr} Tr G_0 &  = &  \tau_4 \tau_3^{-1} \tau_2 \tau_1^{-1} & +  \tau_4 \tau_2 \tau_3^{-1} & 
%+ \tau_4\tau_2\tau_1^{-1} & + \tau_2 \tau_4 \\ 
%&& + \mu_2\mu_3 \tau_4 \tau_3^{-1} \tau_1^{-1} & + \mu_3 \tau_4 \tau_3^{-1} & + \mu_2\tau_4\tau_1^{-1} & + \tau_4 \\
%&& + \mu_1\mu_4 \tau_2 \tau_3^{-1} \tau_1^{-1} & + \mu_4 \tau_2 \tau_3^{-1} & + \mu_1\tau_2\tau_1^{-1} & 
%+ \tau_2 \\
%&& + \mu_1 \mu_2 \mu_3 \mu_4 \tau_1^{-1} \tau_3^{-1} & + \mu_1 \mu_2 \tau_1^{-1} & + \mu_3 \mu_4 \tau_3^{-1} & 
%+ 1 
%\end{array} $$

% RDRW2.Appx125.tex contains the two intermediary steps
In the loose sense explained in the Example 1, we have now
$$ \begin{array}{ccrrrr} Tr G_0 &  \hat \sim &  1  & 
+  E_r e^{ -\frac{1}{h} } & 
+ E_r e^{\frac{b-a}{h}  } & 
+ E_r^2 e^{\frac{b-a-1}{h} } \\ 
&& + E_r e^{\frac{1-b}{h}} & 
+  E_r e^{-\frac{b}{h}} & 
+  E_r e^{\frac{1-a}{h}} & 
+ E_r e^{-\frac{a}{h}} \\
&& + E_r e^{\frac{a}{h} } & 
+  E_r e^{\frac{a-1}{h}} & 
+  E_r e^{\frac{b}{h}} & 
+ E_r e^{\frac{b-1}{h}} \\
&& + E_r^2  e^{\frac{1+a-b}{h}} & + E_r e^{\frac{1}{h}} & +  E_r e^{\frac{a-b}{h}} & 
+ 1 .
\end{array} $$

\begin{figure} %\includegraphics{RDRW2p16.pdf} 
\includegraphics{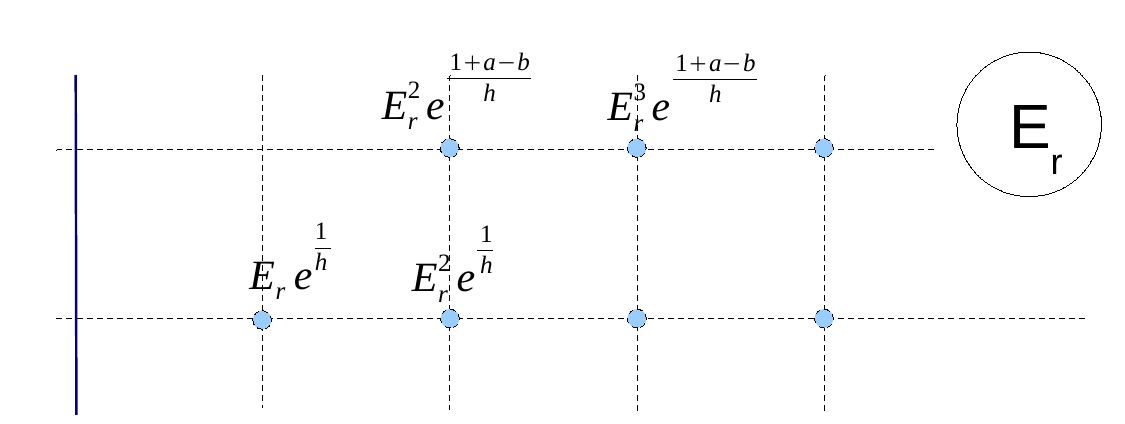} 
\caption{The Newton polygon of \eqref{QCondPrim}, situation of Example 2.} \label{NPolyBdry} \end{figure}

The Newton polygon corresponding to \eqref{QCondPrim} is as shown on fig.\ref{NPolyBdry}, with the $E_r^2 e^{\frac{1+a-b}{h}}$-term coming from the $\mu_1\mu_2\mu_3\mu_4\tau_1^{-1}\tau_3^{-1}$ summand, and the $E_r e^{\frac{1}{h}}$ term -- from the $\mu_1\mu_2\tau_1^{-1}$ summand. We conclude that the nonzero low-lying eigenvalue will have the exponential type $E_r \hat\sim e^{\frac{b-a}{h}}$.

% calculation of different elements of the Newton polygon in terms of $f$ (i.e. not just in terms of $\mu$ and $\tau$ can be found in RDRW2.Appx112.tex 
% RDRW2.Appx127.tex contains some words toward computing the $E_0$ Newton polygon.
% RDRW2.Appx126.tex -- when we were trying to classify the situations in cases 1,2,3.
% RDRW2.Appx97.tex contains a long/wrong version of this computation.

As $G_0 - (1+E_r \kappa)^{-1} Id$ is a $2\times 2$ matrix of rank $1$, a nonzero vector in its kernel is proportional to 
$([G_0-(1+E_rk)^{-1}]_{12}, -[G_0-(1+E_rk)^{-1}]_{11})^T$, i.e. to $([G_0]_{12}, -[G_0]_{11}+(1+E_rk)^{-1})^T$.

For $E_r \hat\sim e^{\frac{b-a}{h}}$, the exponential types of various summands in $[G_0-(1+E_rk)^{-1}]_{11}$, $[G_0]_{12}$ are as follows:
$$ [G_0-(1+E_rk)^{-1}]_{11} =  
\underbrace{( \tau_4 \tau_2 \tau_1^{-1} \tau_3^{-1} - 1)}_{{\cal E}^{b-a} } + 
\underbrace{\tau_4 \tau_2 \tau_3^{-1} }_{\hat\sim e^{\frac{-1+b-a}{h}}} + 
\underbrace{\tau_4 \tau_2 \tau_1^{-1}}_{\hat\sim e^{\frac{2b-2a}{h}}} + 
\underbrace{\tau_4 \tau_2}_{\hat\sim e^{\frac{-1+3b-3a}{h}}} + $$
$$ \ \ \ + 
\underbrace{\tau_4 \tau_1^{-1} \tau_3^{-1} \mu_3\mu_2}_{\hat\sim e^{\frac{1-a}{h}}}  + 
\underbrace{\tau_4 \tau_1^{-1} \mu_2}_{\hat\sim e^{\frac{1+b-2a}{h}}} + 
\underbrace{\tau_3^{-1} \tau_4 \mu_3}_{\hat\sim e^{-\frac{a}{h}}}  + 
\underbrace{\tau_4}_{\hat\sim e^{\frac{b-2a}{h}}}.  $$
(The first summand should typically be $\hat \sim e^{\frac{b-a}{h}}$, but it is conceivable that its exponential type is actually smaller for a special choice of $f$.)
$$ [G_0]_{12} \ = \ \underbrace{\mu_1 \tau_1^{-1} \tau_2\tau_3^{-1}\tau_4}_{\hat\sim e^{\frac{b-a}{h}} } + 
\underbrace{\tau_2\tau_3^{-1}\tau_4}_{\hat\sim e^{\frac{-1+b-a}{h}}} +
\underbrace{\mu_1 \tau_1^{-1} \tau_2 \tau_4}_{\hat\sim e^{\frac{3b-3a}{h}}}+
\underbrace{\tau_2\tau_4}_{\hat\sim e^{\frac{-1+3b-3a}{h}}}+$$
$$ \ \ \ + 
\underbrace{\mu_1 \mu_2 \mu_3  \tau_1^{-1} \tau_3^{-1}\tau_4}_{\hat\sim e^{\frac{1+b-2a}{h}}} + 
\underbrace{\mu_3 \tau_3^{-1} \tau_4}_{\hat\sim e^{\frac{2b-3a}{h}}} + 
\underbrace{\mu_1 \mu_2 \tau_1^{-1} \tau_4}_{\hat\sim e^{\frac{1+2b-3a}{h}} }+
\underbrace{\tau_4}_{\hat\sim e^{\frac{b-2a}{h}}} .$$
 
% cd RDRW2.Appx99.tex for a longer calculation of this 

The two largest summands in the above formulas are thus $\mu_2\mu_3\tau_1^{-1}\tau_3^{-1}\tau_4+ \mu_2\tau_1^{-1}\tau_4$ and $\mu_1\mu_2\mu_3\tau_1^{-1}\tau_3^{-1}\tau_4+\mu_1\mu_2\tau_1^{-1}\tau_4$, respectively; so it is reasonable to take 
$$ \left( \begin{array}{c} Z^{(0)}_+ \\ Z^{(0)}_- \end{array} \right) \ = \  
\frac{1}{\mu_2\tau_1^{-1}\tau_3^{-1}\tau_4(\mu_3+\tau_3)} \left( \begin{array}{c} [G_0]_{12} \\ -[G_0]_{11}+(1+E_rk)^{-1} \end{array} \right). $$

There are too many summands in the entries of $G_0$ for us to be able to get an enlightening exposition, so we will artificially impose additional assumptions on $(a,b)$. These assumptions will help us select the dominant exponential, the first subdominant, the second subdominant, etc, terms in every exponential asymptotic expansion we are going to write down in a moment. There might be a combinatorial structure to various inequalities between $(a,b)$ we are going to introduce, but we are not ready to comment on it at the present time. 

%\begin{figure} \includegraphics{RDRW2p18C3.pdf} \caption{Values of $(a,b)$ satisfying \eqref{adassCase3} }  \label{Case3ab} \end{figure}

\begin{figure} \includegraphics{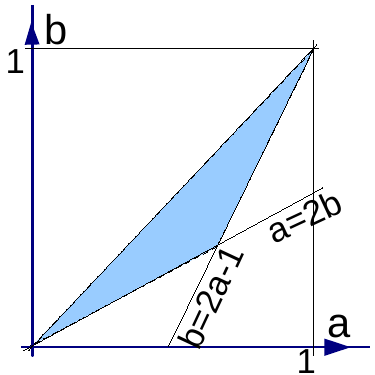} \includegraphics{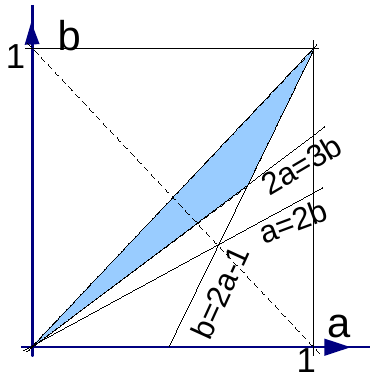} \includegraphics{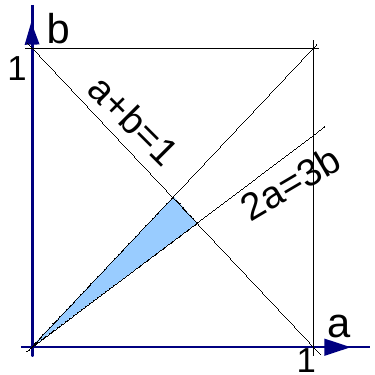} \includegraphics{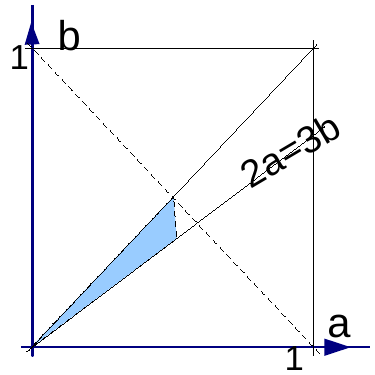} 
\caption{Values of $(a,b)$ satisfying, from left to right, \eqref{adassCase3}, \eqref{adass2a3b}, \eqref{ApBle1}, \eqref{AleHalf}.  }  \label{AllAdAss} \end{figure}

Under 
\begin{equation} \text{additional assumptions: }    a<2b ; \ \ b>2a-1 \label{adassCase3} \end{equation}
(first part of the figure \ref{AllAdAss}), we can write
$$\frac{[G_0]_{11}-(1+E_r \kappa)^{-1}}{\mu_2 \tau_1^{-1} \tau_3^{-1} \tau_4 (\mu_3 + \tau_3)} =  1+ \frac{( \tau_4 \tau_2 \tau_1^{-1} \tau_3^{-1} - (1+E_rk)^{-1})} { \mu_2 \tau_1^{-1} \tau_4\mu_3\tau_3^{-1} } + {\cal E}^{-1+2b-a}, $$
% see RDRW2.Appx100.tex for the calculation
$$\frac{[G_0]_{12}}{\mu_2 \tau_1^{-1} \tau_3^{-1} \tau_4 (\mu_3 + \tau_3)} =  1+ \underbrace{\text{\fbox{$\frac{\tau_1}{\mu_1\mu_2}$}}}_{\hat\sim e^{\frac{-1+a-b}{h}}} + \frac{\tau_2}{\mu_2\mu_3} - \frac{\tau_2 \tau_3}{\mu_2\mu_3^2} +  {\cal E}^{-1+2b-a}. $$
 
% see RDRW2.Appx101.tex for a detailed calculation.

Restricting further to
\begin{equation} \text{additional assumptions: }    2a<3b ; \ \ b>2a-1, \label{adass2a3b} \end{equation}
see the second part of the figure \ref{adass2a3b}, we can absorb the boxed term into the error ${\cal E}^{-1+2b-a}$. 

%\begin{figure} \includegraphics{RDRW2p28.pdf} \caption{Values $(a,b)$ satisfying \eqref{adass2a3b}.} \label{Fig2a3b} \end{figure}

We conclude that 
$$ \left( \begin{array}{c} Z^{(0)}_+ \\ Z^{(0)}_- \end{array} \right) 
\ = \  
\left( \begin{array}{c} \mu_1 \left(  1+ \frac{\tau_2}{\mu_2\mu_3} - \frac{\tau_2 \tau_3}{ \mu_2  \mu_3^2  }  + {\cal E}^{-1+2b-a} \right) \\
- \left( 1+ \frac{( \tau_4 \tau_2 \tau_1^{-1} \tau_3^{-1} - (1+E_rk)^{-1})} { \mu_2 \tau_1^{-1} \tau_4\mu_3\tau_3^{-1} } + {\cal E}^{-1+2b-a} \right)
\end{array} \right).
$$

%$$ \left( \begin{array}{c} D^{(0)}_+ \\ D^{(0)}_- \end{array} \right) \ \approx \
% const\cdot \left( \begin{array}{c}
%B'_0 B_0^{-1} (c'_1)^{-1}\mu_1  \left(  1+ \frac{\tau_2}{\mu_2\mu_3} \right) \\
%- \left( 1+ \frac{( \tau_4 \tau_2 \tau_1^{-1} \tau_3^{-1} - (1+E_rk)^{-1})} { \mu_2 \tau_1^{-1} \tau_4\mu_3\tau_3^{-1} }\right)
%\end{array} \right)  $$

Using \eqref{ZjFla}, 
$$ \left( \begin{array}{c} Z^{(1)}_+ \\ Z^{(1)}_- \end{array} \right)\ = \ \left( \begin{array}{c}  \mu_1 
\left(   \frac{\tau_2}{\mu_2\mu_3} - \frac{( \tau_4 \tau_2 \tau_1^{-1} \tau_3^{-1} - (1+E_rk)^{-1})} { \mu_2 \tau_1^{-1} \tau_4\mu_3\tau_3^{-1} } - \frac{\tau_2 \tau_3}{ \mu_2  \mu_3^2  } + {\cal E}^{-1+2b-a} \right)  
\\ 
-1 + \mu_1  + {\cal E}^{-1+b} \end{array} \right),  $$
with the following exponential orders of the ingredients of $Z^{(1)}$: 
 $$ \frac{\tau_2}{\mu_2\mu_3} \hat\sim e^{\frac{-1+a}{h}} ; \ \  \frac{( \tau_4 \tau_2 \tau_1^{-1} \tau_3^{-1} - (1+E_rk)^{-1})} { \mu_2 \tau_1^{-1} \tau_4\mu_3\tau_3^{-1} } \hat\sim e^{\frac{-1+b}{h}} ; \ \ \frac{\tau_2 \tau_3}{ \mu_2  \mu_3^2  } \hat\sim e^{\frac{-1+b}{h}}. $$

%there were some words here before, see RDRW2.Appx103.tex

We have
$$  \frac{Z^{(1)}_-}{\tau_1^{-1} Z^{(1)}_+} \ = \ \underbrace{ -\frac{\tau_1 \mu_2\mu_3}{ \tau_2 \mu_1  }}_{\hat\sim e^{-\frac{a}{h}}} + {\cal E}^{b-2a}, $$ 
% the division was carried out in RDRW2.Appx102.tex 
hence by \eqref{ZjFla}
\small $$ 
\left( \begin{array}{c} Z^{(2)}_+ \\ Z^{(2)}_- \end{array} \right) 
\ = \ \frac{\mu_1}{\tau_1} 
\left(   \frac{\tau_2}{\mu_2\mu_3} - \frac{( \tau_4 \tau_2 \tau_1^{-1} \tau_3^{-1} - (1+E_rk)^{-1})} { \mu_2 \tau_1^{-1} \tau_4\mu_3\tau_3^{-1} } - \frac{\tau_2 \tau_3}{ \mu_2  \mu_3^2  } + {\cal E}^{-1+2b-a} \right) 
 \left( \begin{array}{c}  1+{\cal E}^{-a}   \\  \underbrace{\mu_2}_{\hat\sim e^{\frac{b-a}{h}}} -\underbrace{\left(\frac{\tau_1 \mu_2\mu_3}{ \tau_2 \mu_1  }\right)}_{\hat\sim e^{-\frac{a}{h}}} + {\cal E}^{b-2a}    \end{array} \right)
$$ \normalsize
where the coefficient in front of this vector nothing but $\tau_1^{-1} Z^{(1)}_+$.

The formula \eqref{ZjFla} gives 
%Next level: 
%$$\left( \begin{array}{c} D^{(3)}_+ \\ D^{(3)}_- \end{array} \right) 
% \ = \ 
% B'_0  M'_1 M'_2 \left( \begin{array}{cc} B_0 M_1 M_2 & 0 \\ 0 & B'_0 M'_1 M'_2 \end{array} \right)^{-1}
%\left( \begin{array}{cc}  (c'_3)^{-1}  [\tau_2 X +  \mu_3  Y] \\  \tau_2  X +   Y \end{array} \right)
%  $$
\footnotesize
$$ 
\left( \begin{array}{c} Z^{(3)}_+ \\ Z^{(3)}_- \end{array} \right) 
\ = \ \frac{\mu_1}{\tau_1} 
\left(   \frac{\tau_2}{\mu_2\mu_3} - \frac{( \tau_4 \tau_2 \tau_1^{-1} \tau_3^{-1} - (1+E_rk)^{-1})} { \mu_2 \tau_1^{-1} \tau_4\mu_3\tau_3^{-1} } - \frac{\tau_2 \tau_3}{ \mu_2  \mu_3^2  } + {\cal E}^{-1+2b-a} \right) 
 \left( \begin{array}{c}  \tau_2+{\cal E}^{-1+2b-2a} +\mu_3\left[ \mu_2 - \frac{\tau_1 \mu_2\mu_3}{\tau_2 \mu_1} \right] +{\cal E}^{2b-3a}   \\  \tau_2 + {\cal E}^{-1+2b-2a} + \mu_2 -\frac{\tau_1 \mu_2\mu_3}{ \tau_2 \mu_1  } + {\cal E}^{b-2a}    \end{array} \right).
$$
\normalsize

%(we are using $\tau_2\sim E_r e^{\frac{b-1}{h}} \sim e^{\frac{-1+2b-a}{h}}$ in the error estimates)\\

Under 
\begin{equation} \text{additional assumption: \ \ }  b+a<1 , \ \ 2a< 3b \label{ApBle1} \end{equation}
we have $\tau_2\in{\cal E}^{b-2a}$ and therefore
\footnotesize
$$ 
\left( \begin{array}{c} Z^{(3)}_+ \\ Z^{(3)}_- \end{array} \right) 
\ = \  \frac{\mu_1}{\tau_1} 
\left(   \frac{\tau_2}{\mu_2\mu_3} - \frac{( \tau_4 \tau_2 \tau_1^{-1} \tau_3^{-1} - (1+E_rk)^{-1})} { \mu_2 \tau_1^{-1} \tau_4\mu_3\tau_3^{-1} } - \frac{\tau_2 \tau_3}{ \mu_2  \mu_3^2  } + {\cal E}^{-1+2b-a} \right) 
 \left( \begin{array}{c}    \overbrace{\mu_3\mu_2}^{\hat\sim e^{\frac{2b-2a}{h}}} - \overbrace{\left(\frac{\tau_1 \mu_2\mu_3^2}{\tau_2 \mu_1}\right)}^{\hat\sim e^{\frac{b-2a}{h}}}  + \overbrace{\tau_2}^{\hat\sim e^{\frac{-1+2b-a}{h}}} + {\cal E}^{2b-3a}  \\   \mu_2 -\frac{\tau_1 \mu_2\mu_3}{ \tau_2 \mu_1  } + {\cal E}^{b-2a}    \end{array} \right).
$$
\normalsize
%More crudely we can write $Z_{+}^{(3)} = \mu_3\mu_2+{\cal E}^{b-2a} $.
%
%Still the next step:
%$$\left( \begin{array}{c} D^{(4)}_+ \\ D^{(4)}_- \end{array} \right) 
% \ = \ 
% B'_0  M'_1 M'_2 M'_3 \left( \begin{array}{cc} B_0 M_1 M_2 M_3& 0 \\ 0 & B'_0 M'_1 M'_2 M'_3 \end{array} \right)^{-1}
% \left( \begin{array}{c}  c_4 \left[ \tau_3^{-1} X +  Y \right] \\   \mu_4 \tau_3^{-1} X +   Y \end{array} \right)$$
Finally, use \eqref{ZjFla} to obtain:
\footnotesize
$$ 
\left( \begin{array}{c} Z^{(4)}_+ \\ Z^{(4)}_- \end{array} \right) 
\ = \  \frac{\mu_1}{\tau_1} 
\left(   \frac{\tau_2}{\mu_2\mu_3} - \frac{( \tau_4 \tau_2 \tau_1^{-1} \tau_3^{-1} - (1+E_rk)^{-1})} { \mu_2 \tau_1^{-1} \tau_4\mu_3\tau_3^{-1} } - \frac{\tau_2 \tau_3}{ \mu_2  \mu_3^2  } + {\cal E}^{-1+2b-a} \right) \times $$ $$ \ \ \ \ \ \times 
 \left( \begin{array}{c}   \tau_3^{-1}(\mu_2\mu_3+{\cal E}^{b-2a})+\mu_2+{\cal E}^{-a}     \\  \mu_4\tau_3^{-1}(\tau_2 +\mu_3\left[ \mu_2 - \frac{\tau_1 \mu_2\mu_3}{\tau_2 \mu_1} \right] +{\cal E}^{2b-3a}) +  \mu_2 -\frac{\tau_1 \mu_2\mu_3}{ \tau_2 \mu_1  } + {\cal E}^{b-2a}  \end{array} \right).
$$
\normalsize
Under one more 
\begin{equation} \text{additional assumption: } a<\frac{1}{2}; \ \ \ 2a<3b, \label{AleHalf} \end{equation} 
we have % $\mu_4 \tau_3^{-1} \tau_2  \sim E_r e^{\frac{a-1}{h}} \sim e^{\frac{b-1}{h}}$. 
 $\mu_4 \tau_3^{-1} \tau_2  \hat\sim e^{\frac{b-1}{h}} \in {\cal E}^{b-2a}$, and the expression simplifies:
\footnotesize
$$ 
\left( \begin{array}{c} Z^{(4)}_+ \\ Z^{(4)}_- \end{array} \right) 
\ = \  \frac{\mu_1}{\tau_1} 
\left(   \frac{\tau_2}{\mu_2\mu_3} - \frac{( \tau_4 \tau_2 \tau_1^{-1} \tau_3^{-1} - (1+E_rk)^{-1})} { \mu_2 \tau_1^{-1} \tau_4\mu_3\tau_3^{-1} } - \frac{\tau_2 \tau_3}{ \mu_2  \mu_3^2  } + {\cal E}^{-1+2b-a} \right) 
 \left( \begin{array}{c}   \tau_3^{-1}\mu_2\mu_3+\mu_2+{\cal E}^{-b}     \\  \left[ \mu_4\tau_3^{-1}\mu_3 + 1 \right] \cdot \left[ \mu_2 - \frac{\tau_1 \mu_2\mu_3}{\tau_2 \mu_1} \right]   + {\cal E}^{b-2a}    \end{array} \right).
$$
\normalsize

We will see now that the bracket $[\mu_4\tau_3^{-1}\mu_3 + 1]$ in the expression for $Z^{(4)}_-$  is not $\hat\sim e^{\frac{0}{h}}$ as would appear from the first glance, but is of a smaller exponential type.
Indeed, the quantization condition \eqref{QCondPrim} and the explicit form \eqref{G0expl} of $\Tr G_0$ imply
%full derivation RDRW2.Appx128.tex  
$$ (\mu_3  \tau_3^{-1}  + 1) \mu_2\tau_4\tau_1^{-1} + \mu_1 \mu_2  \tau_1^{-1}(1+  \mu_3 \mu_4 \tau_3^{-1}) = {\cal E}^{a}, $$
or
%$$ (\mu_3  \tau_3^{-1}  + 1) \tau_4 + \mu_1 (1+  \mu_3 \mu_4 \tau_3^{-1}) = {\cal E}^{-1+a} $$
%$$ (\mu_3  \tau_3^{-1}  + 1) \frac{\tau_4}{\mu_1} +  (1+  \mu_3 \mu_4 \tau_3^{-1}) = {\cal E}^{-1+2a-b} $$
$$   (1+  \mu_3 \mu_4 \tau_3^{-1}) = - (\mu_3  \tau_3^{-1}  + 1) \frac{\tau_4}{\mu_1} +  {\cal E}^{-1+2a-b} \ \hat\sim e^{-\frac{b}{h}}. $$

\begin{Rmk} \label{ResurgImp2} {\rm Here we observe again the cancelation of the leading exponential terms and stress again the importance of subdominant exponentials for the calculation of the asymptotics of eigenfunctions on all intervals $(q_j,q_{j+1})$.} \end{Rmk}

%Let us put an additional assumption that  AND $a<1/2$ (hence also $1-b>a$), that will restrict our variables further, figure \ref{Case3March5}.
%\begin{figure} \includegraphics{RDRW2p27.pdf} \caption{Case 3 restricted on March 5th by $e^{\frac{3b-3a}{h}}\in{\cal E}^{\frac{b-a-\min\{b,1-a\}}{h}}$ AND $a<1/2$} \label{Case3March5} \end{figure}

We finish this section by computing the coefficients $\tilde D_\pm^{(j)}$ from \eqref{tldDj}. Under assumptions \eqref{AleHalf}, we have
% cf Paper4.Scan3.pdf
\begin{equation} 
\begin{array}{lcl}
\tilde D^{(1)}_+ = \underbrace{\left( \frac{\mu_1\tau_2}{\mu_2\mu_3}\right)}_{\hat\sim e^{\frac{b-1}{h}}} (1+{\cal E}^{b-a}), && \tilde D^{(1)}_- = -1+{\cal E}^{b-a}, \\ 
\tilde D^{(2)}_+ =   \frac{\mu_1\tau_2}{\mu_2\mu_3} (1+{\cal E}^{b-a}), && \tilde D^{(2)}_- = \underbrace{\left( \frac{\mu_1\tau_2}{\tau_1\mu_3}\right)}_{\hat\sim e^{\frac{b}{h}}} (1+{\cal E}^{b-a}), \\ 
\tilde D^{(3)}_+ = \underbrace{\mu_1}_{\hat\sim e^{\frac{b-a}{h}}} (1+{\cal E}^{b-a}), && \tilde D^{(3)}_- = \underbrace{\mu_2}_{\hat\sim e^{\frac{b-a}{h}}} (1+{\cal E}^{b-a}), \\
\tilde D^{(4)}_+ = \mu_1 (1+{\cal E}^{b-a}), && \tilde D^{(4)}_- = -1+{\cal E}^{b-a} .
\end{array}
\label{efunEx2}
\end{equation}

\begin{Rmk} \footnote{The material contained in this remark was explained to the author by Prof. A.Gabrielov.} {\rm There exist trigonometric polynomials $f$ satisfying assumptions of figure \ref{AllAdAss}, i.e. having two local minima and two local maxima satisfying inequalities \eqref{adassCase3}, or \eqref{adass2a3b}, or \eqref{ApBle1}, or \eqref{AleHalf}. Indeed, one should take any Morse $C^{\infty}$ function $f_0$ with two local minima and two local maxima satisfying the inequalities, say, 
\small 
\begin{equation} f_0(q_1)<f_0(q_3)<f_0(q_4)<f_0(q_2); \ \ 2[f_0(q_4)-f_0(q_1)]<3[f_0(q_3)-f_0(q_1)]; \ \ 
2[f_0(q_4)-f_0(q_1)]<f_0(q_2)-f_0(q_1), \label{OpenCondF} \end{equation} 
\normalsize
that are, up to shift and rescaling, correspond to the conjunction of \eqref{AleHalf} and \eqref{ZeroABOne}. Then the Fourier series of $f_0$ will converge to $f_0$ uniformly together with all derivatives, and an $n$-th partial sum $f_n$ of that Fourier series for sufficiently large $n$ will have critical points and critical values arbitrarily close to those of $f$. Since our conditions \eqref{OpenCondF} are open, $f_n$ will satisfy them for $n$ large enough. With a little more work one can produce a trigonometric polynomial with exactly prescribed critical points and critical values. Alternatively, one can generate examples of trigonometric polynomials satisfying \eqref{AleHalf} and \eqref{ZeroABOne} using a computer algebra system.} \end{Rmk}

% in RDRW2.Appx98.tex I started, out of desperation, recalculating the whole thing using the second row of the matrix G_0

% RDRW2.Appx105.tex contains an version from March 11th when I hadn't yet discovered a contradiction in assumptions on $(a,b)$.

\appendix

\section{Useful formulas}

For $A>0$ and $E>0$ and $E\to 0+$ we have the following asymptotics of various integrals:

\begin{equation} \arccosh (A/\sqrt{E}) = \Ln 2A - \frac{1}{2}\Ln E - \frac{E}{4A^2} + o(E) \label{ArccoshAs} \end{equation}

\begin{equation} 
\int_{\arccosh(\frac{A}{\sqrt{E}})}^0 \cosh^k t dt  =  
\left\{ \begin{array}{lcl}
-\arccosh(A/\sqrt{E}) && \text{if $k=0$} \\
-\frac{A^2}{2 E} +\frac{1}{4} + \frac{1}{16}\frac{E}{A^2} -  \frac{1}{2}\arccosh(A/\sqrt{E})  + o(E) && \text{if $k=2$} \\
 -\frac{A^4}{4E^2} -\frac{A^{2}}{4 E}+\frac{7}{32}  - \frac{3}{8}\arccosh(A/\sqrt{E}) + o(E^{0}) && \text{if $k=4$} \\
 -\frac{1}{k}A^k E^{-\frac{k}{2}} - \frac{1}{2(k-2)}A^{k-2}E^{1-\frac{k}{2}} 
- \frac{3}{8(k-4)}A^{k-4}E^{2-\frac{k}{2}} + o(E^{2-\frac{k}{2}}) && \text{if $k=1,3$ or $\ge 5$}
\end{array} \right.
\label{IntCosh} \end{equation}

 % details in RDRW2.Appx14
%for $k=2$:
%$$   \frac{A^2}{2 E} -\frac{1}{4} - \frac{1}{16}\frac{E}{A^2} +  \frac{1}{2}\arccosh(A/\sqrt{E})  + o(E)  \ = \  $$
%for $k=4$:
%$$   \frac{A^4}{4E^2} +\frac{A^{2}}{4 E}-\frac{7}{32}  + \frac{3}{8}\arccosh(A/\sqrt{E}) + o(E^{2-\frac{4}{2}}).  $$
%for $k=1,3$, or for $k\ge 5$:
% $$   \frac{1}{k}A^k E^{-\frac{k}{2}} + \frac{1}{2(k-2)}A^{k-2}E^{1-\frac{k}{2}} + 
%\frac{3}{8(k-4)}A^{k-4}E^{2-\frac{k}{2}} + o(E^{2-\frac{k}{2}}) . $$
% the result was checked on Maple; cf. also RDRW2.Appx13

\begin{equation} 
\int_{\arccosh(\frac{A}{\sqrt{E}})}^0 \sinh^2 t \cosh^k t dt  =  
\left\{ \begin{array}{lcl}
  -\frac{A^2}{2E}+\frac{1}{4}+\frac{E}{16A^2} + \frac{1}{2}\arccosh(A/\sqrt{E})+o(E) && \text{if $k=0$} \\
-\frac{A^4}{4E^2} +\frac{A^2}{4E} -  \frac{1}{32} +  \frac{1}{8}\arccosh(A/\sqrt{E})  + o(E^0) && \text{if $k=2$} \\
E^{-1-\frac{k}{2}}\left( -\frac{A^{k+2}}{k+2} + \frac{A^k E}{2k} + \frac{A^{k-2}E^2}{8(k-2)} + o(E^2) \right) && \text{if $k=1$ or $\ge 3$}
\end{array} \right.
\label{IntSinhCosh} \end{equation}
%{\bf The third case of this formula has not been verified}
% verification in RDRW2.Math2

%-------------

%If $\cosh t=x$ with $x>>1$, then 
%$$e^t = x\pm \sqrt{x^2-1} = x\left( 1+ 1 - \frac{1}{2x^2}+ \frac{\frac{1}{2}\cdot\left( -\frac{1}{2}\right)}{1\cdot 2 x^4}+...\right) = 2(x-\frac{1}{4x}-\frac{1}{16x^3})+O(\frac{1}{x^5})$$
%hence
%$$ e^{nt} \ = \ 2^nx^n\left( 1-n(\frac{1}{4x^2}+\frac{1}{16x^4}) + \frac{n(n-1)}{2}(\frac{1}{4x^2})^2 + O(\frac{1}{x^6})\right) \ = \ $$
%$$ \ = \  2^n (x^n - \frac{n}{4} x^{n-2} + ( \frac{n(n-1)}{32}-\frac{n}{16} ) x^{n-4}) + O(x^{n-6}) \ = \ $$
%$$ \ = \  2^n(x^n - \frac{n}{4} x^{n-2} +  \frac{n(n-3)}{32}  x^{n-4}) + O(x^{n-6}) \ = \ $$
%
%{\bf This formula has not been verified}
%
%=============

The following formulae are simple integration by parts used in sec.\ref{MonoFS}.
% old version RDRW2.Appx132.tex
\begin{equation} 
\int \frac{du}{(u^2-E)^{3/2}} \ = \  -\frac{1}{E} \frac{u}{\sqrt{u^2-E}}  \ ; \ \ \ \ \  
\int \frac{udu}{(u^2-E)^{3/2}}  \ = \ - \frac{1}{\sqrt{u^2-E}} \label{sqrt32deg0} \end{equation}
\begin{equation} \int \frac{u^k du}{(u^2-E)^{3/2}} \ = \ 
 -\frac{u^{k-1}}{\sqrt{u^2-E}} + (k-1) \int \frac{u^{k-2}}{\sqrt{u^2-E}} du, \ \ \ k\ge 2 \label{sqrt32deg2} \end{equation}
\begin{equation}  \int \frac{du}{(u^2-E)^{\frac{5}{2}}} \ = \ \frac{1}{E^{2}} \frac{u}{\sqrt{u^2-E}} - \frac{1}{E^{2}} \frac{u^3}{3(u^2-E)^{\frac{3}{2}}} \ ; \ \ \ \ 
\int \frac{udu}{(u^2-E)^{\frac{5}{2}}}  \ = \ - \frac{1}{3(u^2-E)^{\frac{3}{2}}} \label{sqrt52deg1} \end{equation}
\begin{equation} \int \frac{u^k du}{(u^2-E)^{\frac{5}{2}}} 
\ = \ -\frac{u^{k-1} }{3(u^2-E)^{\frac{3}{2}}} + \frac{k-1}{3} \int \frac{u^{k-2}}{(u^2-E)^{\frac{3}{2}}} du,  \ k\ge 2 \label{sqrt52deg2} \end{equation}

{\Large \bf Acknowledgements. } The author would like to thank Takashi Aoki, Andrei Gabrielov, Stavros Garoufalidis, Shingo Kamimoto, Takahiro Kawai, Yoshitsugu Takei, and Boris Tsygan for their kind help during the work on this paper. This work was supported by World Premier International Research Center Initiative (WPI Initiative), MEXT, Japan.


\begin{thebibliography}{99}
\bibitem[AKT09]{AKT} Aoki T., Kawai T., Takei Y., {\it The Bender-Wu analysis and the Voros theory. II.}  Algebraic analysis and around,  19--94, Adv. Stud. Pure Math., 54, Math. Soc. Japan, Tokyo, 2009. 
 \bibitem[CNP]{CNP} 
 B.Candelpergher, J.-C. Nosmas, F.Pham, {\it Approche de la r\'esurgence.}  Actualités Math\`ematiques.  Hermann, Paris, 1993.
% \bibitem[D92]{D92} E.Delabaere, {\it Introduction to the Écalle theory.}  Computer algebra and differential equations (1992),  59--101, London Math. Soc. Lecture Note Ser., 193, Cambridge Univ. Press, Cambridge, 1994.
% \bibitem[DDP93]{DDP93} E.Delabaere, H.Dillinger, F.Pham, {\it R\'esurgence de Voros et p\'eriodes des courbes hyperelliptiques.} 
% Ann. Inst. Fourier (Grenoble) 43 (1993), no. 1, 163--199.  
 \bibitem[DDP97]{DDP97} 
 E.Delabaere, H.Dillinger, F.Pham, {\it Exact semiclassical expansions for one-dimensional quantum oscillators.} J.Math.Phys, 38 (1997)
\bibitem[DP99]{DP99} 
E.Delabaere, F.Pham, {\it Resurgent methods in semi-classical asymptotics.} Ann. Inst. Poincar\'e Phys. Th\'eor. 77 (1999)
\bibitem[F05]{Fuk} 
K.Fukaya, {\it Multivalued Morse theory, Asymptotic Analysis, and Mirror Symmetry}.  Graphs and patterns in mathematics and theoretical physics,  p.205--278,
Proc. Sympos. Pure Math., 73, Amer. Math. Soc., Providence, RI, 2005. 
%\bibitem[E]{E} 
%J.\'Ecalle, {\it Les fonctions r\'esurgentes.} Publications Math. d'Orsay, preprint, 1981 
\bibitem[G11]{G} A.Getmanenko, {\it Resurgent analysis of the Witten Laplacian in one dimension.}  Funkcialaj Ekvacioj, 54 (2011), p.383-438.
\bibitem[G12]{G12} A.Getmanenko, {\it On eigenfunctions corresponding to a small resurgent eigenvalue}, Asymptotic Analysis, 76(2), 2012.
%\bibitem[G09]{G09} A.Getmanenko,  {\it Shatalov-Sternin's construction of complex WKB solutions and the assosicated Riemann surface.} {\tt arXiv:0907.2934}
\bibitem[HKN04]{HeKlNi} 
B.Helffer, M.Klein, F. Nier, {\it  Quantitative analysis of metastability in reversible diffusion processes via a Witten complex approach.}  Mat. Contemp.  26  (2004), 41--85.
\bibitem[M99]{M99} F.Menous, {\it 
Les bonnes moyennes uniformisantes et une application \`a la resommation r\'eelle.} Ann. fac. sciences de Toulouse, S\'er. 6, 8 no. 4 (1999), p.579-628. 
%\bibitem[J94]{J94} A.O.Jidoumou, {\it Mod\`eles de r\'esurgence param\'etrique: fonctions d'Airy et cylindro-paraboliques.} J. Math. Pures Appl. (9) 73 (1994), no. 2, 111--190. 
%\bibitem[Sch]{Sch} P.Schapira, {\it Microdifferential systems in the complex domain.}  Grundlehren der Mathematischen Wissenschaften, 269. Springer, 1985. 
\bibitem[ShSt]{ShSt} 
B.Yu. Sternin, V.E. Shatalov, {\it Borel-Laplace transform and asymptotic theory.  Introduction to resurgent analysis.} CRC Press, Boca Raton, FL, 1996.
\bibitem[SS06]{SS} H.Shen, H.J.Silverstone. {\it Observations on the JWKB treatment of the quadratic barrier.} 
-- Algebraic Analysis of Differential Equations, 2006 -- Springer
\bibitem[V83]{V83}
A.Voros, {\it Return of the quatric oscillator. The complex WKB method.} Ann. Inst. H.Poincar\'e Phys. Th\'eor. 39 (1983)
\end{thebibliography}
\end{document}